\documentclass[10pt,letterpaper]{article}

\usepackage[margin=1.0in]{geometry}
\usepackage{amsmath,amssymb}
\usepackage{graphicx}
\usepackage{booktabs}
\usepackage{textcomp,marvosym}
\usepackage{cite}
\usepackage{nameref,hyperref}
\usepackage[protrusion=true,expansion=true]{microtype}
\usepackage[table]{xcolor}
\usepackage{array}
\usepackage{multirow}
\usepackage[ruled,vlined,linesnumbered]{algorithm2e}

\usepackage{setspace} 
\usepackage[aboveskip=1pt,labelfont=bf,labelsep=period,justification=raggedright,singlelinecheck=off]{caption}

\makeatletter
\renewcommand{\@biblabel}[1]{\quad#1.}
\makeatother

\RestyleAlgo{ruled}
\SetKwComment{Comment}{/* }{ */}
\renewcommand{\Return}[1]{\textbf{return} #1}

\begin{document}
\begin{center}
{\Large\textbf{Reliable mechanistic operator recovery with biologically-informed neural networks: principles for architecture and optimisation design}}

\bigskip
Rebecca M. Crossley \textsuperscript{*1\ddag},
Yuan Yin\textsuperscript{1\ddag},
Sarah L. Waters\textsuperscript{1} and
Ruth E. Baker\textsuperscript{1}

\medskip
1 Mathematical Institute, University of Oxford, Oxford, United Kingdom, OX2 6GG

\ddag \, These authors contributed equally to this work.

* rebecca.crossley@maths.ox.ac.uk
\end{center}

\section*{Abstract}

Many biological processes are governed by complex dynamical mechanisms that remain incompletely understood despite increasing volumes of experimental data. Biologically-informed neural networks (BINNs) seek to address this challenge by embedding mechanistic differential equations into neural network training, enabling interpretable constitutive operators to be recovered directly from sparse and noisy observations. However, the extent to which reliable operator recovery depends on architectural design, optimisation strategy, and the information contained within the available data, is not yet well understood. Here, we present a systematic empirical study of how these factors influence mechanistic inference using BINNs applied to canonical one-dimensional advection--diffusion--reaction partial differential equation models. Across a suite of benchmark problems, we investigate how network expressivity, learning rate, loss weighting, and batch size influence optimisation behaviour, reconstruction accuracy, and operator recovery. We show that successful mechanistic inference is governed by balancing competing objectives rather than maximising any single aspect of the model or optimisation. Moderately expressive architectures outperform overly complex networks, intermediate learning rates balance efficient exploration of the loss landscape with optimisation stability, accurate operator recovery requires an appropriate balance between data-fitting and PDE residual losses, and intermediate batch sizes provide the best compromise between efficient parameter space exploration, computational efficiency, and reproducibility. We further identify practical diagnostics for recognising common failure modes, including over-fitting, unstable optimisation, and poor mechanistic recovery when the ground truth is unavailable. Together, these findings establish evidence-based guidelines for deploying BINNs as credible tools for biological model discovery and demonstrate that reliable mechanistic inference is achieved not by maximising network complexity or optimisation effort, but by appropriately balancing model expressivity, optimisation, physical consistency, and data informativeness.


\section*{Author summary}

Mathematical models are widely used throughout biology to describe processes ranging from cell migration and tissue growth to ecological spread. As biological datasets become increasingly abundant, there is growing interest in learning the mechanisms underlying these processes directly from experimental observations. However, this remains challenging because biological measurements are often sparse, noisy, and irregularly sampled. Biologically-informed neural network (BINN) models address this challenge by combining neural networks with mechanistic differential equations, enabling unknown biological processes to be inferred directly from data. Although promising, the performance of BINNs depends on user-defined choices such as network architecture, optimisation settings, and the balance between fitting data and satisfying the governing equations, none of which have previously been widely documented in the literature.

In this study, we systematically investigate how these design choices influence the ability of BINNs to recover the correct underlying biological mechanisms. Using a suite of benchmark models spanning increasing levels of mechanistic complexity, we show that reliable operator recovery depends on balancing competing objectives rather than simply increasing network complexity or optimisation effort. We identify practical diagnostics for recognising common failure modes, including over-fitting, unstable optimisation, and poor mechanistic recovery, and provide evidence-based guidance for selecting network architectures, learning rates, loss weightings, and batch sizes. Together, our findings provide practical recommendations and a conceptual framework for applying BINNs to mechanistic model discovery from biological data.



\enlargethispage{0.5cm}

\section*{Introduction}

Mechanistic partial differential equation (PDE) models are central to mathematical biology because they provide interpretable descriptions of processes ranging from bacterial dispersal and tumour invasion to ecological range expansion and tissue growth~\cite{murray1989mathematical,murray2001mathematical}. Rather than simply reproducing observed behaviour, these models seek to explain the biological mechanisms governing system dynamics through constitutive operators describing processes such as diffusion, advection, and growth. Calibrating such models from experimental data is therefore fundamentally an inverse problem: the objective is to recover the underlying mechanisms from observations that are typically sparse, noisy, irregularly sampled, and often limited to partial observations of the system state. Classical inference methods can struggle under these conditions, particularly when the mathematical form of one or more governing operators is unknown. For example, although Bayesian approaches provide rigorous uncertainty quantification~\cite{falco2023quantifying,martina2021bayesian,mcgurk2024data}, they often become computationally demanding when data are sparse or irregularly sampled. These challenges have motivated the development of hybrid methods that combine mechanistic modelling with machine learning.

Physics-informed neural networks (PINNs) provide one such framework: it embeds governing differential equations directly into the loss function of a neural network~\cite{Raissi2019physics,Pang2020,Fan2026}. By minimising PDE residuals throughout the spatio-temporal domain~\cite{Lagergren:2020:BIN}, PINNs use known physical structure to regularise the recovery of PDE solutions and unknown parameters from limited observations. This approach has been successfully applied across a wide range of scientific disciplines, including weather forecasting, wave propagation, medical imaging, epidemiology, and fluid mechanics~\cite{ahmadi2025physics,cai2021physics,inam2025ai,millevoi2024physics,movahhedi2023predicting,rasht2022physics,rodrigues2024using,soto2024physics}. In biological applications, PINNs have evolved into biologically-informed neural networks (BINNs), which tailor network architectures and operator parameterisations to biological processes such as density-dependent diffusion and nonlinear reaction kinetics~\cite{aronson1980density,Lagergren:2020:BIN,lavery2026physicsinformedneuralnetworksbiological,stokes2024speed,crossley2024phenotypic}. 
Unlike the original PINNs paper~\cite{Raissi2019physics}, 
which assume that the governing PDE is known and seeks only to recover the solution or unknown parameters, applications of BINNs have also demonstrated an ability to infer interpretable constitutive operators directly from sparse biological datasets~\cite{Lagergren:2020:BIN,Nardini:2020:LEF}. Here, the inference problem can be separated into two complementary tasks. First, a neural network is trained to recover a smooth approximation of the latent solution from noisy observations. Next, the learned solution is supplied as input into additional neural networks representing the unknown constitutive operators, allowing both the solution and the governing mechanisms to be learned simultaneously.

More formally, consider a dynamical system governed by
\[
\frac{\partial u}{\partial t}=\mathcal{F}(u;\boldsymbol{p}),
\]
where $u(\boldsymbol{x},t)$ is the system state at location $\boldsymbol{x}$ at time $t$, and the differential operator $\mathcal{F}$ contains model parameters $\boldsymbol{p}$. In a classical PINN, the form of $\mathcal{F}$ is assumed known and the latent solution is approximated from observations $u_o(\boldsymbol{x}_i,t_i)$, $i=1,\ldots,N_{\mathrm{data}}$, using a neural network $u_\theta(\boldsymbol{x},t)$ with parameters $\boldsymbol{\theta}$~\cite{baker2025modelling} through a combined data-fitting and PDE-constrained optimisation. Any unknown parameters of the model, $\boldsymbol{p}$, can be learned at the same time as part of the optimisation process. Rather than treating $\mathcal{F}$ as fully known, BINNs on the other hand represent any unknown components of the governing operator $\mathcal{F}$ using additional neural networks, whose parameters are learned jointly with those of the solution network, $u_\theta(\boldsymbol{x},t)$. Throughout this work we will consider one-dimensional advection--diffusion--reaction equations of the form
\begin{equation}
    \frac{\partial u}{\partial t} = \frac{\partial }{\partial x}\left(D\left(u\right)\frac{\partial u}{\partial x} - uV(u)\right) + u\,G(u), 
    \label{Eq:PDE}
\end{equation}
where the constitutive functions $D(u)$, $V(u)$, and $G(u)$ may be partially or entirely unknown, and may contain unknown model parameters, $\boldsymbol{p}$. The objective is to recover these functions directly from observational data by representing each operator with a separate neural network, while simultaneously learning the latent solution $u_\theta(x,t)$.

Recovering governing operators, however, presents a fundamentally different challenge from approximating the solution itself. While sufficiently expressive neural networks can often interpolate the observed data accurately, successful operator recovery requires simultaneously balancing data fidelity, physical consistency, optimisation stability, model expressivity, and the information contained within the available observations. Small changes in architectural or optimisation choices may therefore leave the learned solution largely unchanged while substantially altering the recovered mechanisms. Users must choose the depth and width of multiple neural networks, tune learning rates, determine the relative weighting of competing loss terms, and specify sampling strategies for PDE residuals and boundary conditions~\cite{jagtap2020locally}. Increasing network expressivity may reduce approximation bias but also increase susceptibility to noise and optimisation instability~\cite{urban2025unveiling}. Likewise, over-weighting the PDE residual may enforce physical consistency at the expense of data fidelity~\cite{becerra2026role}, whereas under-weighting it may produce excellent interpolation while failing to recover the correct governing operators.

Sensitivity to the aforementioned features is therefore particularly important because, in BINN models, the recovered operators are often the primary scientific output rather than the learned solution itself. Consequently, practical identifiability becomes a central challenge. Even when a governing equation is structurally identifiable in principle~\cite{browning2023structural,loman2026structural}, sparse or weakly informative datasets may permit multiple operator combinations that generate almost indistinguishable solution dynamics. Despite the growing number of successful BINN applications, relatively little attention has been devoted to understanding how architectural design, optimisation choices, and data informativeness jointly determine the reliability of operator recovery.

As a result, the aim of this study is to identify the principles that govern reliable mechanistic inference using BINNs for a range of canonical advection--diffusion--reaction (ADR) equations commonly used in biology. To achieve this, we perform a systematic empirical investigation of architectural and optimisation choices across a suite of canonical one-dimensional advection--diffusion--reaction PDE models spanning linear diffusion, nonlinear diffusion, reaction--diffusion, and nonlinear advection. These benchmark problems probe increasing operator complexity under levels of noise and data sparsity representative of biological experiments, allowing us to examine how network architecture, learning rate, loss weighting, and batch size influence both solution reconstruction and operator recovery.

Specifically, we introduce a general ADR--BINN framework in which the solution and each constitutive operator are represented by separate neural networks, providing a consistent reference architecture for systematic evaluation. Through controlled numerical experiments, we characterise the competing trade-offs between model expressivity, optimisation, physical consistency, and data informativeness that govern reliable operator recovery. We identify practical diagnostics for recognising under-fitting, over-fitting, unstable optimisation, and operator non-identifiability, and translate these observations into evidence-based recommendations for applying BINNs to biological systems. Rather than identifying isolated optimal hyperparameter values, we demonstrate that reliable mechanistic inference is achieved by appropriately balancing these competing objectives, providing both practical guidance and a conceptual framework for understanding when BINNs succeed and why they fail. In each case we assume \textit{a priori} knowledge of which of the mechanisms is present (out of a subset of diffusion, advection and growth), leaving questions of model selection for future work. The remainder of this article is organised as follows. In the Methods section, we introduce the benchmark PDE models, describe the ADR--BINN framework, and present the network architecture, loss formulation, and training protocol. We then systematically investigate the influence of architectural and optimisation choices on operator recovery before concluding with practical recommendations, limitations, and future directions for reliable mechanistic inference using BINNs.


\section*{Methods}

In this section, we describe the models and equation learning methods used throughout this work. We begin by introducing a suite of mathematical models that span increasing levels of mechanistic complexity, ranging from purely diffusive dynamics to systems with nonlinear diffusion, advection and reaction processes. These models generate synthetic datasets that are later used to assess the ability of BINNs to recover both governing operators and solution dynamics from noisy observations. We then present the BINN framework employed throughout this study.


\subsection*{Mathematical models}

In order to evaluate the ability of BINNs to recover the correct underlying mechanisms and reproduce the results from data, we consider a family of one-dimensional advection--diffusion--reaction PDEs of the form of Eq.~\eqref{Eq:PDE} with $x\in[-7, 7] \,\mathrm{and}\, t\in[0, t_\mathrm{end}]$, where $t_\mathrm{end}$ varies between models. In the context of this work, $D\left(u\right)$ is the density-dependent diffusivity, $V(u)$ is the density-dependent advective velocity, and $G(u)$ is the density-dependent reaction term. To ensure that the problems we consider are well-posed, we also prescribe the initial condition $u(x, 0)$, and homogeneous Neumann boundary conditions so that
\begin{equation}
\label{Eq:BC}
    \left.\frac{\partial u}{\partial x}\right|_{x=-7,\,7} = 0.
\end{equation}

The examples considered in this work are intended as representative benchmark problems, rather than an exhaustive list of possible biological models that can be investigated using BINNs. As such, the models are arranged in order of increasing complexity, allowing us to systematically investigate how accurately BINNs recover operators as additional nonlinear mechanisms are introduced. Where possible, the models are constructed to have analytic solutions in order to avoid introducing numerical artefacts associated with numerical solutions into the training data. A summary of the models is provided in Table~\ref{tab:model_summary}.


\begin{table}[ht]
\centering
\small
\renewcommand{\arraystretch}{1.25}
\begin{tabular}{lcccc}
\toprule
\textbf{Model} & \textbf{Diffusion $D(u)$} & \textbf{Advection $V(u)$} & \textbf{Reaction $G(u)$} & \textbf{Analytical solution} \\
\midrule
Diffusion equation & Constant & -- & -- & Yes \\
Porous--medium equation & Nonlinear & -- & -- & Yes \\
Burgers' equation & Linear & Nonlinear & -- & No \\
Linear diffusion--nonlinear advection & Linear & Nonlinear & -- & No \\
Nonlinear ADR equation & Nonlinear & Nonlinear & Linear & No \\
\bottomrule
\end{tabular}
\caption{Summary of the mathematical models used to generate synthetic training data. The test cases are arranged in order of increasing mechanistic complexity. ADR, advection--diffusion--reaction.}
\label{tab:model_summary}
\end{table}


\paragraph*{Diffusion equation.} The simplest test case is the classical diffusion (heat) equation~\cite{rubinow1973mathematical}, with constant diffusivity ($D(u)=1/2$, $V(u)=0$ and $G(u)=0$):
\begin{equation}
\label{eq:diff}
\dfrac{\partial u}{\partial t} = \frac{1}{2} \dfrac{\partial ^2 u}{\partial x^2}.
\end{equation}
We specify an initial condition of the form
\begin{equation}
\label{eq:IC_diff}
u(x,0)=\frac{1}{2}\cos\!\left(\frac{\pi x}{7}\right)+\frac{1}{2},
\end{equation}
so that the system, with the homogeneous boundary conditions specified in Eq.~\eqref{Eq:BC}, has an explicit analytical solution given by~\cite{bluman1969general}
\begin{equation}
u(x,t)=\frac12\cos\!\left(\frac{\pi x}{7}\right)
\exp\!\left[-\frac12\left(\frac{\pi}{7}\right)^2 t\right]+\frac12. \label{eq:diff_ana}
\end{equation}


\paragraph*{Porous--medium equation.} We next consider a nonlinear porous--medium equation ($D(u)=3u^2$, $V(u)=0$ and $G(u)=0$):
\begin{equation}
\label{eq:PM}
\dfrac{\partial u}{\partial t} =\dfrac{\partial}{\partial x}\left(3u^2\dfrac{\partial u}{\partial x}\right). 
\end{equation}
The Barenblatt--Kompaneets--Zeldovich similarity solution to Eq.~\eqref{eq:PM} is given by
\begin{equation}
u(x,t)=t^{-1/4}
\sqrt{
\max\!\left(
\frac{M}{\pi\sqrt3}-\frac{x^2}{12\,\sqrt{t}},
\,0
\right)}, \label{eq:sim_sol}
\end{equation}
where $M=3$ is the initial mass of this solution profile~\cite{barenblatt1952some, zel1950towards}.
The analytical solution in Eq.~\eqref{eq:sim_sol} becomes singular as \(t \to 0^+\) where it approaches a delta-like profile. As such, we choose
\begin{equation}
\label{eq:PM_IC}
u(x,0)=2^{1/4}
\sqrt{
\max\!\left(
\frac{3}{\pi\sqrt3}-\frac{\sqrt{2}x^2}{12},
\,0
\right)},
\end{equation}
so that the solution to Eq.~\eqref{eq:PM} with boundary conditions given by Eq.~\eqref{Eq:BC} and initial condition given by Eq.~\eqref{eq:PM_IC} is
\begin{equation}
u(x,t)=(t+\tfrac12)^{-1/4}
\sqrt{
\max\!\left(
\frac{M}{\pi\sqrt3}-\frac{x^2}{12\,\sqrt{t+1/2}},
\,0
\right)}.
\label{eq:PM-sol}
\end{equation}


\paragraph*{Burgers' equation.} We next consider Burgers' equation~\cite{bonkile2018systematic}, a typical nonlinear advection--diffusion equation that, with $D(u)=1/100$, $V(u)=3u/10$ and $G(u)=0$, takes the form
\begin{equation}
\dfrac{\partial u}{\partial t} = \frac{1}{100} \dfrac{\partial^2 u}{\partial x^2}-\dfrac{\partial}{\partial x}\left(\frac{3}{10}u^2\right).\label{eq:burger}
\end{equation}
We solve Eq.~\eqref{eq:burger} with the initial condition
\begin{equation}
\label{eq:expIC}
u(x,0)=\exp\!\left(-\frac{x^2}{2}\right),
\end{equation}
and boundary conditions as in Eq.~\eqref{Eq:BC}. Numerical solutions of this equation are generated using the methods outlined in the Supporting Material.


\paragraph*{Linear diffusion--nonlinear advection equation.} We next consider a model with linear diffusion and nonlinear advection ($D(u)=1/100$, $V(u)=3u(1-u)/5$,  $G(u)=0$):
\begin{equation}
\label{eq:DA}
\dfrac{\partial u}{\partial t} = \dfrac{1}{100} \dfrac{\partial^2 u}{\partial x^2}-\dfrac{\partial}{\partial x}\!\left(\frac35\,u^2(1-u)\right),
\end{equation}
with the initial condition given by Eq.~\eqref{eq:expIC}. Numerical solutions of this equation are generated using the methods outlined in the Supporting Material.


\paragraph*{Nonlinear reaction--diffusion--advection equation.} Finally, we consider a model exhibiting nonlinear diffusion, advection and reaction processes with $D(u)=u^5/10$, $V(u)=u(1-u)$ and $G(u)=u/2$:
\begin{equation}
\label{eq:NL_ADR}
\dfrac{\partial u}{\partial t}=\dfrac{\partial}{\partial x}\!\left(\frac{1}{10}u^5\dfrac{\partial u}{\partial x}-u^2(1-u)\right)+\frac{1}{2}u,
\end{equation}
with the initial condition
\begin{equation}
u(x,0)=
\min\!\left\{
\frac{1}{10}\exp\!\left[-\frac{(x+2)^2}{2\left(\tfrac{7}{10}\right)^2}\right] +\frac{7}{100}\exp\!\left[-\frac{(x-1)^2}{2\left(\tfrac{7}{10}\right)^2}\right],\,1 \right\}, \label{eq:ic-nl}
\end{equation}
and boundary conditions as in Eq.~\eqref{Eq:BC}. Numerical solutions of this equation are generated using the methods outlined in the Supporting Material.


\paragraph*{Generating noisy observations.} For each underlying mathematical model, we assume that the quantity of interest $u(x,t)$, is observed at $N_\mathrm{data}$ points $(x_i,t_i)\in[-7, 7]\times[0,t_{\mathrm{end}}]$ for $i=1,\ldots,N_{\mathrm{data}}$, which do not need to be uniformly distributed in either space or time. The corresponding observations are  denoted as $u_o(x_i,t_i)$. As is standard in the BINNs literature, to mimic measurement uncertainty and intrinsic biological variability, synthetic observations are generated by perturbing the exact or numerical solutions with independent and identically distributed (i.i.d.)~Gaussian noise with zero mean and constant variance, $\sigma^2>0$. 
Specifically, we write 
\begin{equation}
\label{Eq:noise}
    u_o(x_i,t_i) = u(x_i,t_i) + \mathcal{N}(0, \sigma^2) \quad \text{for} \quad i=1,\ldots,N_{\mathrm{data}}.
\end{equation}
By default, results shown in text for this work have a constant variance of $\sigma^2=0.001$ and 71 spatial observation locations. However, results for both the noise-free case, $\sigma^2=0$, and a higher noise level of $\sigma^2=0.01$, as well as for more densely spatially sampled data (351 locations), can be found in the Supporting Material.


\subsection*{The BINN framework}


\begin{figure}[htbp]
    \centering
    \includegraphics[width=\linewidth]{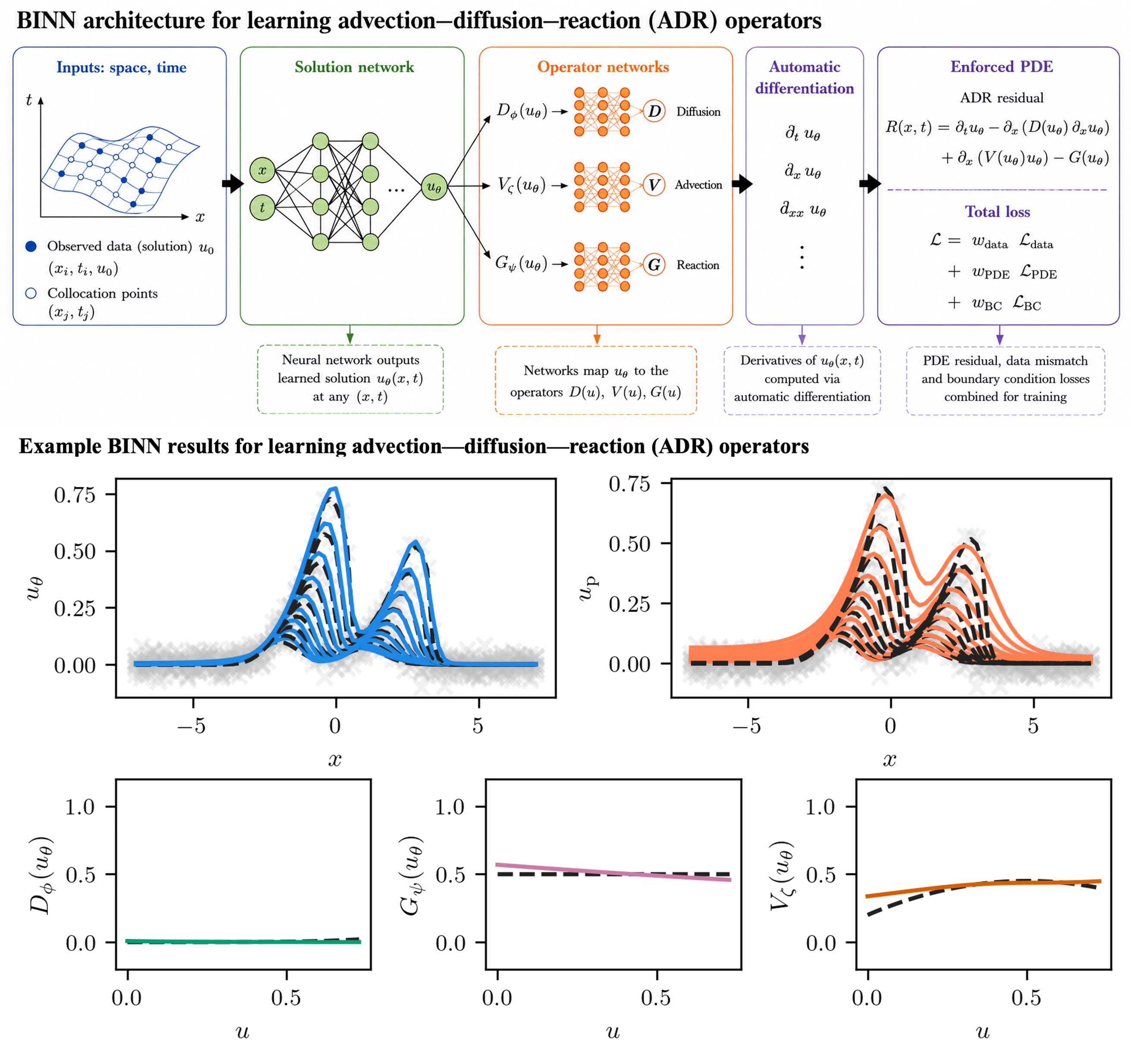}
    \caption{Overview of the ADR--BINN framework and representative output. Top: Schematic of the BINN architecture used to learn the constitutive operators in the general advection--diffusion--reaction equation (Eq.~\eqref{Eq:PDE}). The solution network $u_\theta(x,t)$ is trained from noisy observations and evaluated at randomly sampled collocation points. The predicted solution is then passed to the operator networks $D_\phi(u)$, $V_\zeta(u)$, and $G_\psi(u)$, with automatic differentiation used to compute the PDE residual. The data, PDE, and boundary-condition losses are combined to form the total loss used for joint optimisation of all four networks. Bottom: Representative ADR--BINN outputs and reconstruction for the nonlinear reaction--diffusion--advection benchmark (Eq.~\eqref{eq:NL_ADR}) with additive Gaussian noise of variance $\sigma^2=0.001$. Shown are the learned solution $u_\theta(x,t)$ (blue), and the forward-predicted solution $u_\text{p}(x,t)$ (orange) plotted at the temporal resolution described in Table~\ref{tab:time_resolution}, and the learned constitutive functions $D_\phi(u)$ (green), $G_\psi(u)$ (purple), and $V_\zeta(u)$ (brown). Dashed black curves denote the true solution or constitutive functions, coloured lines show the output from one single BINN initialisation. The forward-predicted solution is obtained by numerically solving Eq.~\eqref{Eq:PDE} using the learned operators and learned initial condition. More details can be found in the Supporting Material. The ADR--BINN hyperparameters take the default values given in Table~\ref{tab:BINN parameter}.}
    \label{fig:1-schematic}
\end{figure}


Existing applications of BINNs have focused primarily on reaction--diffusion systems~\cite{elmarakeby2021biologically, hartman2023interpreting, Lagergren:2020:BIN, Nardini:2020:LEF}.  However, in applied mathematics problems, we are often interested in learning PDEs that include advective transport. Consequently, the BINN formulation considered in this work seeks to recover (where present) the diffusion operator, $D\left(u\right)$, advection operator,  $V\left(u\right)$, and reaction operator, $G(u)$, simultaneously from the observation data, $u_o(x_i,t_i)$. To this end, the BINN model employed consists of four multi-layer perceptrons (MLPs). The first MLP $u_{\theta}$, estimates the smooth underlying latent solution $u(x,t)$ of Eq.~\eqref{Eq:PDE}. The second MLP, $D_{\phi}$, learns the diffusivity $D(u)$, whilst the third, $V_{\zeta}$, learns the advection velocity $V(u)$, and the fouth, $G_{\psi}$, learns the reaction term $G(u)$. Each one of these MLPs has its own network parameters, $\boldsymbol\theta$, $\boldsymbol\phi$, $\boldsymbol\zeta$, $\boldsymbol\psi$, respectively, and a schematic of the BINN model showing how these neural networks are connected is shown in Fig.~\ref{fig:1-schematic}.

All of the neural networks in the BINN are trained jointly, with automatic differentiation used to evaluate derivatives appearing in the PDE residual that imposes the governing equation as a soft constraint during optimisation. A multi-component loss function, denoted as $\mathcal{L}_{\mathrm{total}}(\boldsymbol\theta, \boldsymbol\phi, \boldsymbol\zeta, \boldsymbol\psi)$, is minimised during training using gradient descent methods in order to obtain the optimal network parameters $\boldsymbol\theta$, $\boldsymbol\phi$, $\boldsymbol\zeta$, and $\boldsymbol\psi$. The total loss function comprises a data-fitting term $\mathcal{L}_{\mathrm{data}}$, which penalises differences between neural network predictions of the latent variable $u(x,t)$ (given by $u_\theta(x,t)$) and the observed data; a PDE constraint term $\mathcal{L}_{\mathrm{PDE}}$, which penalises discrepancies between the left and right hand sides of the governing PDE (Eq.~\eqref{Eq:PDE}) when evaluated using the learned smooth solution, $u_\theta$, as well as the inferred $D_\phi$, $V_\zeta$ and $G_\psi$; and a boundary condition loss term $\mathcal{L}_{\mathrm{BC}}$, which penalises any deviations from the prescribed boundary conditions (Eq.~\eqref{Eq:BC}). The specific forms of $\mathcal{L}_{\mathrm{data}}$, $\mathcal{L}_{\mathrm{PDE}}$, $\mathcal{L}_{\mathrm{BC}}$, and $\mathcal{L}_{\mathrm{total}}$ are described below.


\paragraph{The data-fitting loss.} Given the assumption of i.i.d.~Gaussian noise, Eq.~\eqref{Eq:noise}, the likelihood of the observations is given by
\begin{equation*}
    p\left(\{u_o(x_i,t_i)\}\mid \boldsymbol\theta,\sigma^2\right)
=\prod_{i=1}^{N_{\mathrm{data}}}\frac{1}{\sqrt{2\pi\sigma^2}}
\exp\!\left(-\frac{\left(u_o(x_i,t_i)-u_\theta(x_i,t_i)\right)^2}{2\sigma^2}\right).
\end{equation*}
Maximising this likelihood with respect to the network parameters $\boldsymbol\theta$ is equivalent to minimising the sum of squared residuals. As such, we define the data‑fitting loss as
\begin{equation}
\label{Eq:dataloss}
    \mathcal L_{\mathrm{data}}(\boldsymbol\theta)
=\frac{1}{N_{\mathrm{data}}}\sum_{i=1}^{N_{\mathrm{data}}}\big(u_\theta(x_i,t_i)-u_o(x_i,t_i)\big)^2.
\end{equation}
Minimising $\mathcal L_{\mathrm{data}}(\boldsymbol{\theta})$ guides $u_\theta$ to match the observed data, $u_o$, at the training data points. 
Alternative assumptions regarding the structure of the observation noise can be incorporated through the relevant likelihood functions, or learned simultaneously using further neural networks during BINNs training~\cite{crossley2026likelihood}.


\paragraph{The PDE loss.} To enforce the governing PDE (Eq.~\eqref{Eq:PDE}) across the entire spatio-temporal domain, and not just at the measured training data points, during each round of training we randomly sample $N_{\mathrm{PDE}}$ grid points $\left(x^{\mathrm{PDE}}_i, t^{\mathrm{PDE}}_i\right) \in (x_{\min},x_{\max})\times[0,t_{\mathrm{end}}]$, for $i=1,\ldots,N_\text{PDE}$. At each sampled point, the PDE residual is evaluated by substituting the MLP approximations $u_\theta$, $D_{\phi}$, $V_{\zeta}$ and $G_{\psi}$ into Eq.~\eqref{Eq:PDE}. The PDE constraint loss $\mathcal{L}_{\mathrm{PDE}}$ is thus defined as
\begin{align}
\label{Eq:PDEloss}
\mathcal{L}_{\mathrm{PDE}}(\boldsymbol\theta, \boldsymbol\phi, \boldsymbol\zeta, \boldsymbol\psi) &= \frac{1}{N_{\mathrm{PDE}}} \sum_{i=1}^{N_{\mathrm{PDE}}} 
\Bigg( 
    \frac{\partial u_{\theta}\left(x^{\mathrm{PDE}}_i, t^{\mathrm{PDE}}_i\right)}{\partial t}- u_{\theta}\left(x^{\mathrm{PDE}}_i, t^{\mathrm{PDE}}_i\right) \,
      G_{\psi}\left(u_{\theta}\left(x^{\mathrm{PDE}}_i, t^{\mathrm{PDE}}_i\right)\right)\notag \\
&  \qquad 
    - \frac{\partial}{\partial x}
    \Bigg[
        D_{\phi}\left(u_{\theta}\left(x^{\mathrm{PDE}}_i, t^{\mathrm{PDE}}_i\right)\right)
        \frac{\partial u_{\theta}\left(x^{\mathrm{PDE}}_i, t^{\mathrm{PDE}}_i\right)}{\partial x}    -u_{\theta}\left(x^{\mathrm{PDE}}_i, t^{\mathrm{PDE}}_i\right)V_{\zeta}\left(u_{\theta}\left(x^{\mathrm{PDE}}_i, t^{\mathrm{PDE}}_i\right)\right)
    \Bigg] \Bigg)^{2}.
\end{align}
Randomly resampling  collocation points during training provides a stochastic approximation to enforcing the PDE throughout the spatio-temporal domain, rather than only at the observation locations.

\paragraph{The boundary condition loss.} Unlike existing PINNs methods~\cite{liu2022unified}, we further incorporate the imposed (homogeneous Neumann) boundary conditions (Eq.~\eqref{Eq:BC}) as an extra constraint into the neural network training procedure\footnote{We note that a more appropriate boundary condition might be to impose zero flux boundary conditions of the form $J(u)=D(u)\partial{u}/\partial{x}-uV(u)=0$ at each boundary. However this would introduce additional complexity, that is outside the scope of this work, by coupling the solution and operator networks at the boundary.}. To do this, we randomly sample $N_{\mathrm{BC}}$ time points in $[0, t_{\mathrm{end}}]$ on each spatial boundary ($x_\text{min}=-7$, $x_\text{max}=7$), denoted $t^{x_{\min}}_i, t^{x_{\max}}_i$, and penalise any differences between the learned and known dynamics. We therefore define the boundary condition constraint loss, $\mathcal{L}_{\mathrm{BC}}$, as
\begin{align}
\label{Eq:BCloss}
    \mathcal{L}_{\mathrm{BC}}\left(\boldsymbol\theta\right) = \frac{1}{N_{\mathrm{BC}}}\sum_{i=1}^{N_{\mathrm{BC}}}\left(\frac{\partial u_{\theta}\left(x_{\min}, t^{x_{\min}}_i\right)}{\partial x}\right)^2 + \frac{1}{N_{\mathrm{BC}}}\sum_{i=1}^{N_{\mathrm{BC}}}\left(\frac{\partial u_{\theta}\left(x_{\max}, t^{x_{\max}}_i\right)}{\partial x}\right)^2,
\end{align}
where the time samples at $x_{\min}$ and $x_{\max}$ are drawn independently.


\paragraph{The total loss function.} Combining Eqs.~\eqref{Eq:dataloss}--\eqref{Eq:BCloss}, we define the total loss function $\mathcal{L}_{\mathrm{total}}$ as
\begin{align}
\label{Eq:totalloss}
    \mathcal{L}_{\mathrm{total}}(\boldsymbol{\theta}, \boldsymbol{\phi}, \boldsymbol{\zeta}, \boldsymbol{\psi}):= w_\mathrm{data}\mathcal{L}_{\mathrm{data}}(\boldsymbol{\theta})+
    w_\mathrm{PDE}\mathcal{L}_{\mathrm{PDE}}(\boldsymbol{\theta},\boldsymbol{\phi},\boldsymbol{\zeta},\boldsymbol{\psi})+ w_\mathrm{BC}\mathcal{L}_{\mathrm{BC}}(\boldsymbol{\theta}).
\end{align}
The weights $w_\mathrm{data}\geq0$, $w_\mathrm{PDE}\geq0$, and $w_\mathrm{BC}\geq0$ balance fidelity to the observations against consistency with the governing PDE and boundary conditions. While additional loss terms have previously been used to impose biological constraints on the latent solution or constitutive operators, such as positivity, monotonicity, or boundedness~\cite{crossley2026likelihood, Lagergren:2020:BIN}, these were not required for accurate reconstruction in the present work.

The PDE and boundary conditions are evaluated at randomly sampled collocation points, allowing the mechanistic model to constrain the learned solution throughout the spatio-temporal domain, including where observations are sparse or absent. Resampling these points during training provides a stochastic approximation to the corresponding domain-wide losses without requiring a fixed, densely resolved grid, retaining computational efficiency for larger or higher-dimensional problems.

Importantly, the observations are also incorporated through a soft data-fitting constraint rather than requiring $u_\theta$ to interpolate every measurement exactly. The solution network therefore balances agreement with potentially noisy observations against consistency with the governing PDE and boundary conditions, which acts as an implicit regularisation mechanism and promotes a smooth approximation to the latent solution $u(x,t)$. This is particularly important for constitutive operator inference because the PDE residual depends on spatial and temporal derivatives of $u_\theta$; suppressing local fluctuations due to measurement noise therefore provides more reliable derivatives from which to infer the underlying constitutive relationships.

\paragraph{Mechanistic prediction.} To distinguish between accuracy in the interpolation for the latent solution and mechanistic operator recovery, we evaluate two distinct quantities throughout this work. The learned solution, denoted $u_\theta(x,t)$, is the direct output of the neural network approximating the state variable during training. In contrast, the predicted solution, denoted $u_\mathrm{p}(x,t)$, is obtained by numerically solving the governing PDE (Eq.~\eqref{Eq:PDE}) using the learned operator networks $D_\phi$, $V_\zeta$, and $G_\psi$ and the initial condition from the learned latent solution, $u_\theta(x,0)$. Full details regarding the numerical implementation of the numerical method used to solve for the predicted solution, $u_\mathrm{p}(x,t)$, can be found in the Supporting Material. Comparing $u_\theta$ and $u_\mathrm{p}$ allows us to assess whether the inferred operators generate dynamics consistent with the data beyond direct neural interpolation: even if $u_\theta$ fits the observed data closely, agreement between $u_\mathrm{p}$ and the data is not guaranteed, and it is in fact this match that provides stronger evidence that the underlying mechanisms have been correctly identified. For the model in Fig.~\ref{fig:1-schematic} we find that, whilst the latent solution $u(x,t)$ is inferred well by $u_\theta(x,t)$ throughout the domain, the predicted solution $u_\mathrm{p}(x,t)$ does not match the data well for regions of space where the true solution was in fact zero. The cause for this can be seen in the plots of the learned operators, where the largest deviations in both the advection and growth functions are found around $u=0$.


\paragraph*{The training procedure and hyperparameters.} Implementing the BINN used in this work requires choice of a range of different network architectures and hyperparameters. To keep the benchmarking fair and simple, no advanced network architectures (such as recurrent neural networks~\cite{liang2024physics}), training schedulers~\cite{muller2023achieving}, activation functions~\cite{bischof2025multi, jagtap2020locally, zhang2025simple, xiang2022self}, or modified back-propagation methods are employed~\cite{wang2023expert}. Each MLP simply consists of $h_j$ hidden layers with $n_j$ nodes, where $j=u,D,G,$ or $V$, depending on whether referencing $u_\theta$, $D_\phi$, $V_\zeta$, or $G_\psi$, respectively.
In between each hidden layer, a \texttt{tanh} activation function is employed ($f^h_j=\texttt{tanh}$ for $j=u,D,V,G$), and in the final output layer, the \texttt{SoftplusReLU} activation function ($f^o_j$ for $j=u,D,V,G$) is instead used. Definitions of all of the key variables used in the BINN are provided in Table~\ref{tab:BINN variable}.


\begin{table}[htbp]
\centering
\small
\renewcommand{\arraystretch}{1.25}

\begin{tabular}{lll}
\toprule
\textbf{Group} & \textbf{Description} & \textbf{Explanation} \\
\midrule

\multirow{6}{*}{\textbf{MLP}}
& $u_\theta$ & MLP for density $u$ \\
& $D_\phi$ & MLP for diffusivity $D(u)$ \\
& $V_\zeta$ & MLP for advection velocity $V(u)$ \\
& $G_\psi$ & MLP for reaction term $G(u)$ \\
& $f^h_{u,D,V,G}$ &
Activation function between hidden layers
(\texttt{tanh}) \\
& $f^o_{u,D,V,G}$ &
Output-layer activation function
(\texttt{SoftplusReLU}\footnotemark[1]) \\

\midrule

\multirow{3}{*}{\shortstack[l]{\textbf{Training}\\\textbf{data points}}}
& $(x_i,t_i)$ &
Training data used to compute $\mathcal{L}_{\mathrm{data}}$ \\
& $(x_i^{\mathrm{PDE}},t_i^{\mathrm{PDE}})$ &
Randomly sampled collocation points used to compute
$\mathcal{L}_{\mathrm{PDE}}$ \\
& $(x_{\min,\max},t_i^{\min,\max})$ &
Randomly sampled boundary points used to compute
$\mathcal{L}_{\mathrm{BC}}$ \\

\midrule

\multirow{4}{*}{\shortstack[l]{\textbf{Loss}\\\textbf{functions}}}
& $\mathcal{L}_{\mathrm{data}}(\boldsymbol\theta)$ &
Data-fitting loss \\
& $\mathcal{L}_{\mathrm{PDE}}(\boldsymbol\theta,\boldsymbol\phi,\boldsymbol\zeta,\boldsymbol\psi)$ &
PDE residual loss \\
& $\mathcal{L}_{\mathrm{BC}}(\boldsymbol\theta)$ &
Boundary-condition loss \\
& $\mathcal{L}_{\mathrm{total}}(\boldsymbol\theta,\boldsymbol\phi,\boldsymbol\zeta,\boldsymbol\psi)$ &
Total loss function \\

\bottomrule
\end{tabular}

\caption{Variables used throughout the ADR--BINN framework. \texttt{SoftplusReLU}$(x)=\log(1+e^x)$ for $x<20$, and
$\texttt{SoftplusReLU}(x)=\max(0,x)$ otherwise.}

\label{tab:BINN variable}

\end{table}


For each benchmark problem, the simulated data are split into training, validation, and test sets in an 80:10:10 ratio. 
Training is performed using a mini-batch strategy with batch size $B$, whereby the data-fitting loss is evaluated on randomly sampled subsets of the training observations of size $B$ at each optimisation step. This reduces the computational cost of each parameter update and introduces additional stochasticity into the optimisation process, which can improve training efficiency and help avoid poor local minima. Training is performed on the CPU using an \texttt{Adam} optimizer with learning rate $r$, and the pseudo-code for the training procedure over $N_{\mathrm{epoch}}$ epochs is provided in Algorithm~\ref{alg:Vanilla}, along with the default parameter values in Table~\ref{tab:BINN parameter} and the initialisation procedure for an MLP (either $u_{\theta}$, $D_{\phi}$, $V_{\zeta}$ or $G_{\psi}$) in Algorithm~\ref{alg:initialisation}. The validation phase, which involves evaluating the loss function at each epoch on the validation data and recording the total losses for this data (essentially repeating lines 4 to 14 of Algorithm~\ref{alg:Vanilla}), is omitted for brevity. Following standard practice, the weights of each node in $u_{\theta}$ are randomly sampled from a Xavier normal distribution~\cite{glorot2010understanding}, and the biases are initialised to zero. In order to quantify the uncertainty in the BINN output, we run Algorithm~\ref{alg:Vanilla} for $N_{\mathrm{rep}}$ independent realisations to evaluate whether the network initialisation affects the training outcomes. Exemplar results can be seen for the nonlinear reaction--advection--diffusion equation in Fig.~\ref{fig:1-schematic}. 


\begin{algorithm}[hbtp]
\caption{BINN training algorithm}
\label{alg:Vanilla}
\KwIn{\\Parameters in Table \ref{tab:BINN parameter};\\
Training data $u_{\text{train}}(x_i, t_i)$ and $(x_i, t_i)\in[x_{\min}, x_{\max}]\times[0, t_{\mathrm{end}}]$ with $i = 1, \ldots, N_{\mathrm{data}}$.
}

\vspace{\baselineskip}

Initialise the MLPs $u_{\theta}$, $D_{\phi}$, $V_{\zeta}$ and $G_{\psi}$ according to Algorithm \ref{alg:initialisation}. \\
Specify the optimiser for $\boldsymbol\theta$, $\boldsymbol\phi$, $\boldsymbol\zeta$, and $\boldsymbol\psi$ as \texttt{Adam} with learning rate $r$.
 
\vspace{\baselineskip}

\For{$p = 1$ \KwTo $N_{\mathrm{epoch}}$}
{
\For{$q = 1$ \KwTo $N_{\mathrm{data}}/B$}
{

Randomly sample training data without replacement to get a batch $\{(x_i, t_i)\}$ with size $B$. \\
Forward $\{(x_i, t_i)\}$ into $u_\theta$.\\
Compute the data-fitting loss $\mathcal{L}_{\mathrm{data}}(\boldsymbol\theta)$ according to Equation (\ref{Eq:dataloss}).

\vspace{\baselineskip}

Randomly sample $N_{\mathrm{PDE}}$ grid points in the spatio-temporal domain $(x_{\min}, x_{\max})\times [0, t_{\mathrm{end}}]$, denoted as $\{\left(x^{\mathrm{PDE}}_{i}, t^{\mathrm{PDE}}_i\right)\}$ with $i=1, \ldots, N_{\mathrm{PDE}}$. \\
Forward $\{\left(x^{\mathrm{PDE}}_{i}, t^{\mathrm{PDE}}_i\right)\}$ into $u_\theta$. Then forward $u_\theta$ into $D_\phi$, $V_\zeta$ and $G_\psi$.\\
Compute the PDE constraint loss $\mathcal{L}_{\mathrm{{PDE}}}(\boldsymbol\theta, \boldsymbol\phi, \boldsymbol\zeta, \boldsymbol\psi)$ according to Equation (\ref{Eq:PDEloss}).

\vspace{\baselineskip}

Randomly sample $N_{\mathrm{BC}}$ 
grid points in the temporal domain $[0, t_{\mathrm{end}}]$
at each boundary $x_{\min}$ and 
$x_{\max}$, denoted as $\{\left(x_{\mathrm{\min}}, t^{x_{\min}}_i\right)\}$ and $\{\left(x_{\mathrm{\max}}, t^{x_{\max}}_i\right)\}$, respectively, with $i=1, \ldots, N_{\mathrm{BC}}$.\\
Forward $\{\left(x_{\min}, t^{x_{\min}}_i\right)\}$ and $\{\left(x_{\max}, t^{x_{\max}}_i\right)\}$ into $u_\theta$.\\
Compute the boundary condition constraint loss $\mathcal{L}_{\mathrm{{BC}}}(\boldsymbol\theta)$ according to Equation (\ref{Eq:BCloss}).

\vspace{\baselineskip}

Compute the total loss function $\mathcal{L}_{\mathrm{{total}}}(\boldsymbol\theta, \boldsymbol\phi, \boldsymbol\zeta, \boldsymbol\psi)$ according to Equation (\ref{Eq:totalloss}).

\vspace{\baselineskip}

Back-propagate to update $\boldsymbol\theta$, $\boldsymbol\phi$, $\boldsymbol\zeta$, and $\boldsymbol\psi$ thus $u_{\theta}$, $D_{\phi}$, $V_{\zeta}$, and $G_{\psi}$, respectively:
$
    \boldsymbol\theta \leftarrow \boldsymbol\theta - r \times \nabla_{\theta} \mathcal{L}_{\mathrm{total}}; \quad \boldsymbol\phi \leftarrow \boldsymbol\phi - r \times \nabla_{\phi} \mathcal{L}_{\mathrm{total}}; $
    
    $\boldsymbol\zeta \leftarrow \boldsymbol\zeta - r \times \nabla_{\zeta} \mathcal{L}_{\mathrm{total}};\quad \boldsymbol\psi \leftarrow \boldsymbol\psi - r \times \nabla_{\psi} \mathcal{L}_{\mathrm{total}}.
$\footnotemark[1]
}
}
\Return{$u_{\theta}$, $D_{\phi}$, $V_{\zeta}$, $G_{\psi}$} 
\end{algorithm}


\footnotetext[1]{$\nabla_{\boldsymbol\theta, \boldsymbol\phi, \boldsymbol\zeta, \boldsymbol\psi}$ denotes the gradient with respect to the network parameters $\boldsymbol\theta$ of $u_{\theta}$, $\boldsymbol\phi$ of $D_{\phi}$, $\boldsymbol\zeta$ of $V_{\zeta}$, and $\boldsymbol\psi$ of $G_{\psi}$, respectively.}


\begin{table}[htbp]
\centering
\small
\renewcommand{\arraystretch}{1.25}

\begin{tabular}{p{3.2cm}llc}
\toprule
\textbf{Group} &
\textbf{Parameter} &
\textbf{Description} &
\textbf{Default} \\
\midrule

\multirow{6}{*}{\textbf{MLP}}
& $\boldsymbol{\theta}$ & Parameters of the solution network $u_\theta$ & N/A \\
& $\boldsymbol{\phi}$ & Parameters of the diffusivity network $D_\phi$ & N/A \\
& $\boldsymbol{\zeta}$ & Parameters of the advection network $V_\zeta$ & N/A \\
& $\boldsymbol{\psi}$ & Parameters of the reaction network $G_\psi$ & N/A \\
& $h_{u,D,V,G}$ & Hidden layers in each network & 4 \\
& $n_{u,D,V,G}$ & Nodes per hidden layer in each network & 32 \\

\midrule

\multirow{3}{*}{\shortstack[l]{\textbf{Training data}\\\textbf{and collocation points}}}
& $B$ &
Mini-batch size &
40 \\
& $N_{\mathrm{PDE}}$ &
Number of PDE collocation points &
40 \\
& $N_{\mathrm{BC}}$ &
Number of boundary collocation points &
40 \\

\midrule

\multirow{3}{*}{\textbf{Loss weights}}
& $w_{\mathrm{data}}$ &
Weighting of $\mathcal{L}_{\mathrm{data}}$ &
1 \\
& $w_{\mathrm{PDE}}$ &
Weighting of $\mathcal{L}_{\mathrm{PDE}}$ &
0.1 \\
& $w_{\mathrm{BC}}$ &
Weighting of $\mathcal{L}_{\mathrm{BC}}$ &
0.001 \\

\midrule

\textbf{Optimisation}
& $r$ &
Learning rate &
0.0005 \\

\midrule

\multirow{2}{*}{\textbf{Training}}
& $N_{\mathrm{epoch}}$ &
Training epochs &
4000 \\
& $N_{\mathrm{rep}}$ &
Independent BINN initialisations &
10 \\

\bottomrule
\end{tabular}

\caption{Default hyperparameters used throughout the ADR--BINN framework. Network parameters are learned during training; the remaining default values were determined from preliminary numerical investigations.}
\label{tab:BINN parameter}

\end{table}


\begin{algorithm}[hbtp]
\caption{Initialisation and the forward pass of the MLP $j$, where $j=u,D,V$ or $G$ depending on whether it is $u_{\theta}$, $D_{\phi}$, $V_{\zeta}$ or $G_{\psi}$.}
\label{alg:initialisation}
\KwIn{\\$h_j$: number of hidden layers; \\
$n_j$: number of nodes per hidden layer;\\
$f_j^h$: activation function between hidden layers; \\
$f_j^o$: activation function for the output layer; \\
$\left(x_i, t_i\right)$: training data points.}
\vspace{\baselineskip}

\Comment{Build network:}
Construct a fully connected MLP with $h_j$ hidden linear layers, each containing $n_j$ nodes. The hidden layers use the activation function $f_j^h$, and the output layer employs the activation function $f_j^o$.
\vspace{\baselineskip}

\Comment{Parameter initialisation for MLP $j$:}
\For{each layer in the MLP corresponding to $j$}{
    Set all biases in the current layer to zero;\\
    Initialise all weights in the current layer by sampling from a Xavier normal distribution.
}
\vspace{\baselineskip}

\Comment{Forward pass:}
\Return{MLP output for $(x_i, t_i)$}
\end{algorithm}


\paragraph*{Data and code availability.}
The code used to generate the data and implement the BINNs in this work can be found online at \texttt{github.com/YuanYIN99/ADR\_BINNs.git}.
The simulated data can be found online at \\\texttt{huggingface.co/datasets/IamYuanYinOxford/ADR\_BINNs\_data}, or by following the link in the GitHub repository.


\section*{Results}

We evaluated the performance of BINNs across a suite of canonical one-dimensional PDE models spanning diffusion, reaction--diffusion, advection--diffusion, and transport processes, where all benchmark problems are particular cases of the general advection--diffusion--reaction formulation in Eq.~\eqref{Eq:PDE}, with model-specific constitutive functions and parameters. The benchmark problems described in the \emph{Mathematical models} section of the Methods were selected to probe the ability of BINNs to recover the underlying constitutive relationships and governing operators from sparse spatio-temporal data under varying degrees of nonlinearity and model complexity. An example output for a BINN learning the governing equations from data of a model involving reaction, advection and diffusion processes can be seen in Fig.~\ref{fig:1-schematic}.

Across the benchmark PDEs, the reconstructions shown in this work are primarily demonstrative: they illustrate that BINNs can, under favourable conditions, recover good estimates of the true underlying constitutive functions. At the same time, the process of generating these results exposed several practical sensitivities, including dependence on network architecture, optimisation hyperparameters, and data informativeness. In particular, sub-optimal choices can lead to over-fitting, unstable optimisation, or apparent non-identifiability. In the following sections, we quantify these effects and extract diagnostics that help indicate when a BINN model output, or the resultant PDE solution, is likely to be reliable.


\begin{figure}[htbp]
    \centering
    \includegraphics[width=\linewidth]{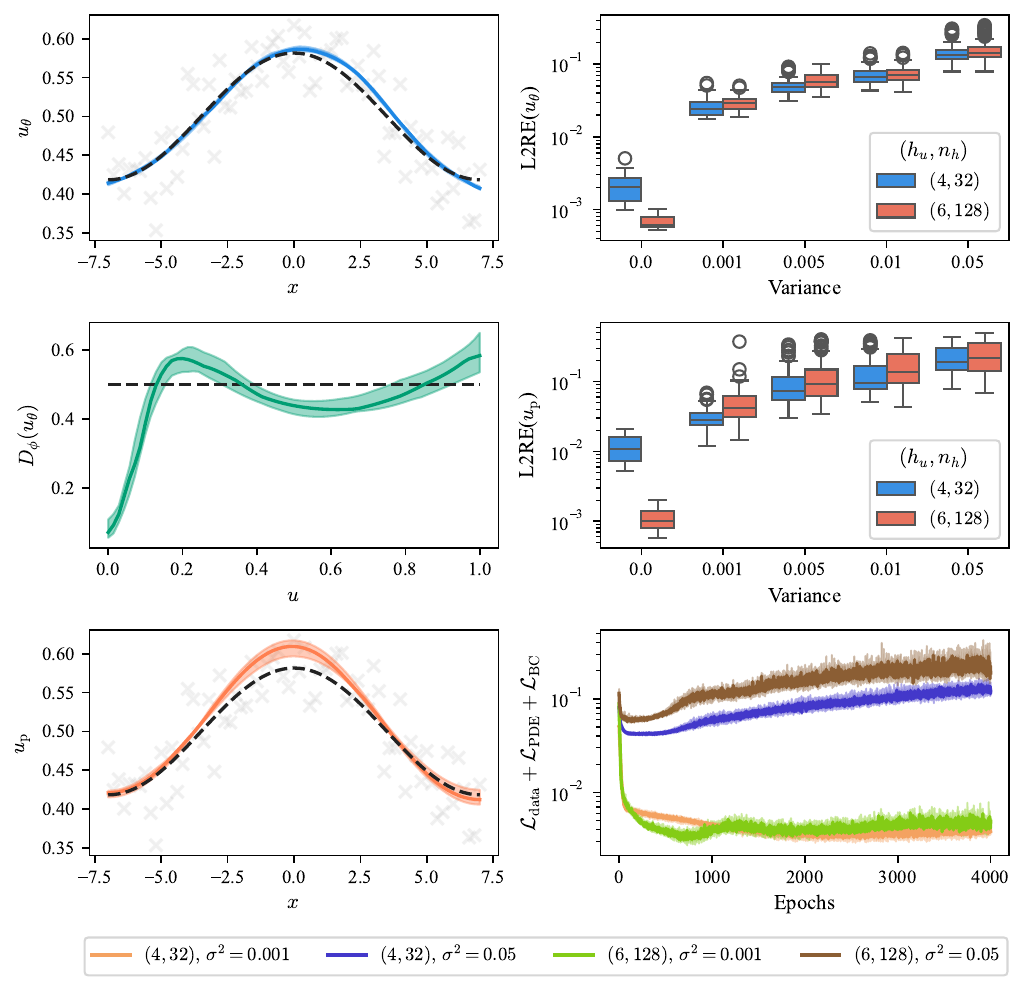}
    \caption{Effect of additive Gaussian noise on BINN performance in the linear diffusion equation (Eq.~\eqref{eq:diff}). 
    Left: Learned solution $u_\theta(x,t)$ (top), learned diffusivity $D_\phi(u)$ (middle), and forward-predicted solution $u_\text{p}(x,t)$ (bottom, plotted for $t=t_\text{end}$ as specified in Table~\ref{tab:time_resolution}). Results are for a representative ADR--BINN trained with additive Gaussian noise of variance $\sigma^2=0.001$ and $(h_u, n_h)=(4,32)$. Dashed black lines denote the true solution or constitutive function, coloured lines denote the mean over ten independent BINN initialisations, and shaded regions indicate $\pm1$ standard deviation.
    Right: Relative $L^2$ errors in the learned solution $u_\theta(x,t)$ (top) and predicted solution $u_p(x,t)$ (middle), as a function of the observation noise variance for the architectures $(h_u,n_u)=(4,32)$ and $(6,128)$. Boxplots summarise ten BINN initialisations for each of ten independently generated datasets. The botto plot shows the total validation loss during training for representative combinations of architecture and noise levels. Unless stated otherwise, all remaining ADR--BINN hyperparameters take the default values given in Table~\ref{tab:BINN parameter}.}
    \label{fig:noise}
\end{figure}


\subsection*{BINNs are robust to moderate observation noise}

Biological data are rarely noise-free, and are often sparse in both space and time. We therefore first examined how additive observation noise affects the ability of a BINN model to recover both the solution and the governing operator. To do so, we trained a BINN on synthetic data generated from the classical linear diffusion equation (Eq.~\eqref{eq:diff}) on \(x\in[-7,7]\) with homogeneous Neumann boundary conditions (Eq.~\eqref{Eq:BC}) and the smooth initial condition in Eq.~\eqref{eq:IC_diff}. For each noise level, we assessed three quantities: the learned neural-network solution \(u_\theta(x,t)\), the learned diffusivity \(D_\phi(u)\), and the forward-predicted solution \(u_\mathrm{p}(x,t)\), obtained by numerically solving Eq.~\eqref{Eq:PDE} using \(D(u)=D_\phi(u)\), the learned initial condition \(u_\theta(x,0)\), and the prescribed boundary conditions.

Fig.~\ref{fig:noise} shows that BINN performance degrades gradually as the variance, $\sigma^2$, of the additive noise increases, but remains robust for low-to-moderate noise levels. The relative error in the learned solution \(u_\theta\) increases with noise variance for both solution-network architectures considered. However, for nonzero noise levels, the shallower architecture, \((h_u,n_u)=(4,32)\), gives slightly lower and less variable errors than the larger architecture, \((h_u,n_u)=(6,128)\). In the noise-free case this trend is reversed, with the larger network giving the lowest interpolation error. This behaviour is consistent with the larger network having greater capacity to interpolate clean data, but also greater susceptibility to fitting noise once measurement error is introduced. A similar pattern is observed for the forward-predicted solution \(u_\mathrm{p}\): errors in \(u_\mathrm{p}\) increase with the noise level, and the simpler architecture again performs at least as well as the larger architecture for noisy data. This comparison is important because \(u_\mathrm{p}\) is not obtained directly from the solution network, but from a numerical solve using the learned diffusivity. Agreement between \(u_\mathrm{p}\) and the true solution therefore provides a stronger test of whether the inferred operator generates the correct dynamics, rather than merely whether \(u_\theta\) interpolates the observations.

The left-hand panels of Fig.~\ref{fig:noise} show representative results for the smallest nonzero noise level, \(\sigma^2=0.001\), using the smaller solution-network architecture. The learned solution \(u_\theta(x,t_{\mathrm{end}})\) closely follows the true solution, with little variability across independent BINN initialisations. The learned diffusivity \(D_\phi(u)\) is also centred close to the true constant value over the range of \(u\)-values explored by the data. In this example, the observed solution values lie approximately in the interval \(u\in[0.4,0.6]\), and this is the region in which the learned diffusivity is most directly constrained. Outside this interval, the apparent shape of \(D_\phi(u)\) should be interpreted with caution, because those values of \(u\) are weakly informed, or not informed at all, by the training data. This explains why the learned diffusivity can vary substantially over the full plotted interval \(u\in[0,1]\), even though the forward-predicted solution remains accurate.

The validation loss curves provide an additional diagnostic for the effect of noise. For \(\sigma^2=0.001\), the validation losses decrease and then remain relatively stable, indicating that the BINN has identified a smooth solution and operator consistent with the data. By contrast, for the largest noise level, \(\sigma^2=0.05\), the validation losses increase during later epochs for both architectures. This behaviour is consistent with over-fitting: the solution network  \(u_\theta\) begins to fit local fluctuations in the noisy observations rather than the smooth latent solution. The effect is particularly concerning for operator recovery, because small noise-driven oscillations in \(u_\theta\) can be amplified through the PDE residual and lead to biased or oscillatory estimates of \(D(u)\).

Together, these results show that BINNs can recover the solution dynamics and the underlying diffusion operator from moderately noisy data, but that increasing network expressivity is not automatically beneficial. For noisy biological data, simpler architectures can provide useful implicit regularisation, improving robustness without sacrificing predictive accuracy. In practice, we therefore recommend monitoring both the relative errors between the observation data and \(u_\theta\) or \(u_\mathrm{p}\), together with validation loss trajectories and the range of \(u\)-values covered by the data. Operator estimates should be interpreted primarily over the observed range of \(u\), and extrapolation beyond that range should be treated as unreliable unless additional data or constraints are available.

The conclusions drawn from Fig.~\ref{fig:noise} remain qualitatively unchanged when the spatial sampling density is increased. Supporting Material Fig.~\ref{fig:SI-noise} presents the same analysis using five times as many observations in the spatial dimension. As expected, the additional observations reduce the errors in both the learned solution, $u_\theta$, and the forward-predicted solution, $u_\mathrm{p}$, across all noise levels. The reconstruction of the diffusivity, $D_\phi(u)$, is also improved, exhibiting both reduced variability between independent BINN initialisations and closer agreement with the true constant diffusivity over the range of $u$-values sampled by the data. Outside the observed range, however, the learned diffusivity remains comparatively uncertain, reflecting the limited information available to constrain the operator. The dependence on network architecture is likewise consistent with the results in Fig.~\ref{fig:noise}. Increasing the observation noise leads to larger reconstruction errors for both architectures, while the larger solution network, $(h_u,n_u)=(6,128)$, again produces more variable operator estimates than the smaller network, \((h_u,n_u)=(4,32)\). The increased variability, accompanied by the same late-stage increase in the validation loss, indicates that the additional network capacity allows the model to over-fit noisy observations rather than recover the underlying smooth dynamics.

Overall, these findings demonstrate that increasing the quantity of data improves both solution and operator recovery, but does not remove the need for appropriate regularisation of the BINN. Reliable operator identification depends not only on the number of observations, but also on the range of solution values and spatio-temporal gradients represented in the data. Additional measurements improve reconstruction because they provide denser coverage of the evolving solution and therefore constrain the governing operators over a broader range of values. In the following sections, we investigate how architectural and optimisation choices influence the ability of BINNs to exploit the available information for robust operator recovery.


\subsection*{Intermediate learning rates balance efficient optimisation with stable operator recovery}

The learning rate controls the size of each optimisation step taken during gradient descent. If the learning rate is too small, the optimisation progresses only slowly through the loss landscape and may fail to converge within the available training budget. Conversely, if it is too large, successive parameter updates can overshoot ``good'' minima, leading to unstable optimisation or even divergence. Consequently, selecting an appropriate learning rate is essential for reliable operator recovery using BINNs. As demonstrated in Fig.~\ref{fig:noise}, validation loss curves provide a useful diagnostic for assessing optimisation behaviour, identifying both under- and over-fitting. Here we investigate how these diagnostics can be used to identify an appropriate learning rate.

\begin{figure}[htbp]
    \centering
    \includegraphics[width=\linewidth]{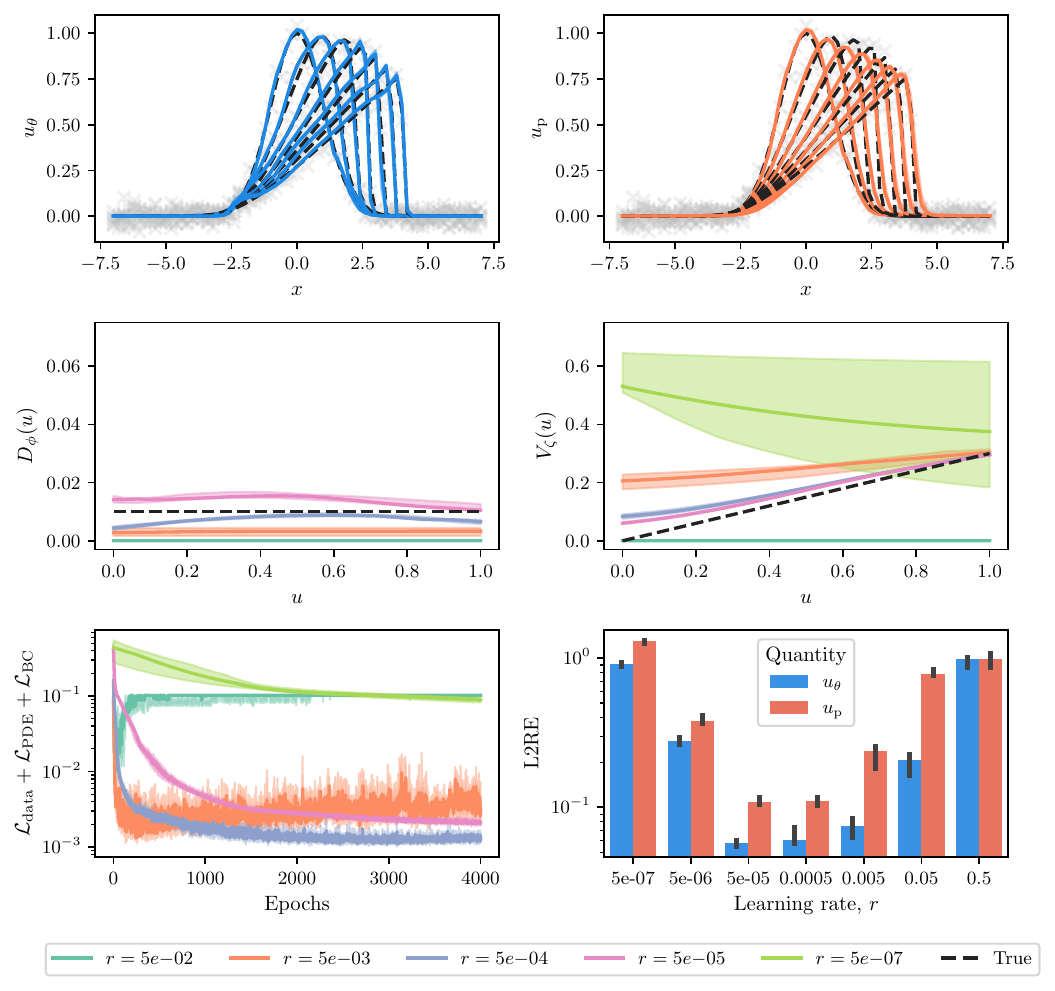}
    \caption{Effect of the learning rate, $r$, on BINNs performance on data simulated from the Burgers' equation~\eqref{eq:burger} with additive Gaussian noise of variance $\sigma^2=0.001$. 
    Top: Learned solution $u_\theta(x,t)$ (left) and forward-predicted solution $u_\text{p}(x,t)$ (right) plotted at the time resolution described in Table~\ref{tab:time_resolution}. 
    Middle: Learned diffusivity $D_\phi(u)$ (left) and advection velocity $V_\zeta(u)$ (right).
    Results correspond to the representative learning rate $r=5\times10^{-4}$ (top); dashed black curves denote the true operators and shaded regions indicate $\pm1$ standard deviation over ten independent ADR--BINN initialisations.
    Bottom: Total validation loss (left) and relative $L^2$ errors in $u_\theta$ and $u_p$ (right) for various learning rates, $r$. Unless stated otherwise, all remaining ADR--BINN hyperparameters take the default values given in Table~\ref{tab:BINN parameter}.}
    \label{fig:learning-rate}
\end{figure}

To do so, we trained a BINN on synthetic data generated from Burgers' equation, Eq.~\eqref{eq:burger}, a nonlinear advection--diffusion equation that provides a challenging benchmark for learning state-dependent advective fluxes. Fig.~\ref{fig:learning-rate} summarises the performance of the BINN model over a range of learning rates. We find that the choice of learning rate has a pronounced effect on both optimisation behaviour and the accuracy of the recovered solution and governing operators.


The validation loss curves (Fig.~\ref{fig:learning-rate}, bottom left) reveal three distinct optimisation regimes. For very small learning rates, such as $r=5\times10^{-7}$, the validation loss decreases only slowly throughout training and the BINN has not converged after 4000 epochs, indicating that the optimiser is unable to efficiently explore the loss landscape. Slow convergence, accompanied by relatively large errors in both the learned solution, $u_\theta$, and the forward-predicted solution, $u_\mathrm{p}$ (Fig.~\ref{fig:learning-rate}, bottom right), are characteristic of under-fitting. At the opposite extreme, learning rates of order $10^{-2}$ or larger give rise to unstable optimisation. The validation losses cease to decrease smoothly, and the resulting errors in both $u_\theta$ and $u_\mathrm{p}$ increase substantially. In these cases, the optimiser takes parameter updates that are too large to allow it to settle into a ``good'' minimum of the loss landscape, leading to poor recovery of both the governing operators and the solution dynamics.

Between these two extremes lies a broad intermediate regime in which optimisation is both stable and efficient. Across all benchmark problems considered in this study, the lowest reconstruction errors are consistently obtained for learning rates of approximately $r\approx10^{-4}$. In Fig.~\ref{fig:learning-rate}, both the learned solution, $u_\theta$, and the forward-predicted solution, $u_\mathrm{p}$, achieve their lowest relative errors for $r=5\times10^{-5}$ and $5\times10^{-4}$, where the validation loss curves also decrease smoothly and reach the lowest values. In this regime, the recovered diffusivity, $D_\phi(u)$, and advective velocity, $V_\zeta(u)$, are both closest to the true constitutive functions and exhibit relatively little variability across independent BINN initialisations. Representative reconstructions for $r=5\times10^{-4}$ (top panels of Fig.~\ref{fig:learning-rate}) illustrate the close agreement between both the learned and forward-predicted solutions and the underlying dynamics.

Taken together, these results demonstrate that the validation loss curves provide a simple and effective diagnostic for selecting an appropriate learning rate. Persistently slow decreases in the validation loss indicate that the optimiser is making insufficient progress, whereas unstable or erratic loss trajectories signal excessively large optimisation steps. Ultimately, smooth, monotonic convergence, coupled with low reconstruction errors, can be used to identify a regime in which the BINN is able to efficiently explore the loss landscape while recovering robust estimates of the governing operators.


\subsection*{Accurate operator recovery requires balancing loss contributions}

Unlike conventional neural networks, which optimise only a data-fitting objective, BINNs must simultaneously reconcile agreement with noisy observations and consistency with the governing PDE model. These competing objectives are balanced through the weighting parameters in the total loss function, Eq.~\eqref{Eq:totalloss}. Consequently, the choice of loss weightings determines the extent to which the optimisation prioritises fitting the observations or recovering a mechanistically consistent operator. Within the ADR--BINN framework, three weighting parameters, $w_{\mathrm{data}}$, $w_{\mathrm{PDE}}$, and $w_{\mathrm{BC}}$, control the relative contributions of the data-fitting, PDE residual, and boundary-condition losses, respectively (Table~\ref{tab:BINN parameter}). We investigate the influence of these weightings using synthetic data generated from the porous--medium equation (Eq.~\eqref{eq:PM}), whose nonlinear, state-dependent diffusivity provides a stringent test of operator recovery. Correctly interpolating the observed solution alone is insufficient for this problem; successful mechanistic inference additionally requires accurate reconstruction of the nonlinear diffusion operator.

Preliminary investigations established that the boundary-condition weighting, $w_{\mathrm{BC}}$, had little influence on reconstruction accuracy. Once the learned solution, $u_\theta$, accurately approximates the observations, the homogeneous Neumann boundary conditions are naturally satisfied and the associated residual becomes negligible. We therefore fix $w_{\mathrm{BC}}=0$ and focus on the balance between the data-fitting and PDE residual losses, setting, without loss of generality, $w_{\mathrm{data}}=1$ and varying $w_{\mathrm{PDE}}$ over several orders of magnitude each side.


\begin{figure}[htbp]
    \centering
    \includegraphics[width=\linewidth]{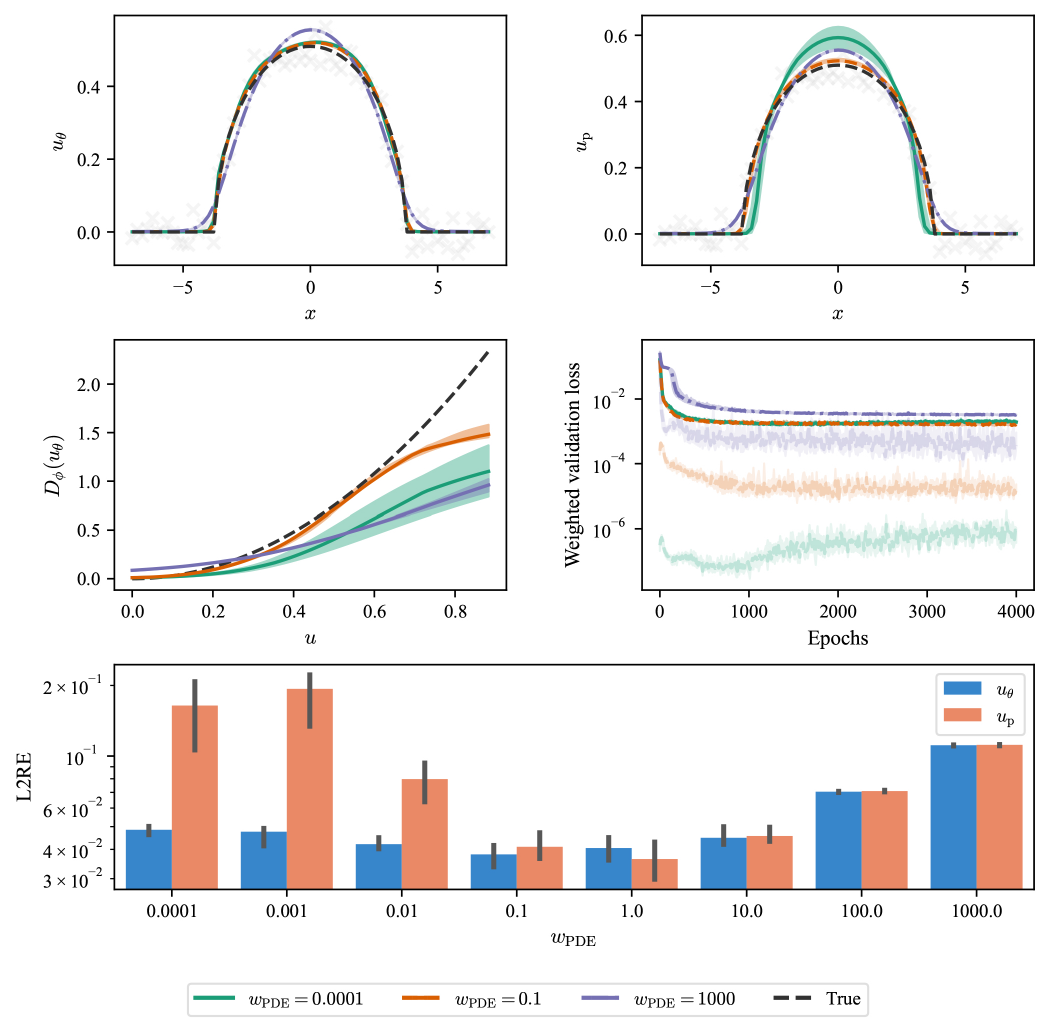}
    \caption{
    Effect of the PDE residual weighting $w_{\mathrm{PDE}}$ on BINN performance for the porous-medium equation (Eq.~\eqref{eq:PM}) with additive Gaussian noise of variance $\sigma^2=0.001$, whilst $w_{\mathrm{data}}=1$ and $w_{\mathrm{BC}}=0$. 
    Top: Learned solution $u_\theta(x,t)$ (left) and forward-predicted solution $u_p(x,t)$ (right) plotted for $t=t_\text{end}$ as specified in Table~\ref{tab:time_resolution}. 
    Middle left: Learned diffusion function $D_\phi(u)$. Representative results are shown for $w_{\mathrm{PDE}}\in\{10^{-4},10^{-1},10^3\}$; dashed black curves denote the true solution or constitutive function and shaded regions represent $\pm1$ standard deviation across ten BINN initialisations. 
    Middle right: Weighted validation losses, showing $L_{\mathrm{data}}$ (solid) and $w_{\mathrm{PDE}}L_{\mathrm{PDE}}$ (dashed) with shaded regions representing $\pm1$ standard deviation across ten BINN initialisations. 
    Bottom: Relative $L^2$ errors in $u_\theta$ and $u_\text{p}$ as $w_{\mathrm{PDE}}$ varies. 
    Unless stated otherwise, all remaining ADR--BINN hyperparameters take the default values given in Table~\ref{tab:BINN parameter}.  
    }
    \label{fig:PDEweight}
\end{figure}


Fig.~\ref{fig:PDEweight} reveals two distinct failure modes. When the PDE residual weighting, $w_{\mathrm{PDE}}$, is too small relative to the data-fitting weight, $w_{\mathrm{data}}$, the optimisation is dominated by the data-fitting objective. The learned solution, $u_\theta$, closely interpolates the observations, resulting in small relative errors (Fig.~\ref{fig:PDEweight}, bottom). However, the inferred diffusion operator, $D_\phi(u)$, fails to reproduce the underlying dynamics, leading to substantially larger errors in the forward-predicted solution, $u_\mathrm{p}$. This behaviour is particularly evident for $w_{\mathrm{PDE}}=10^{-4}$, where the reconstructed diffusivity deviates markedly from the true constitutive function despite the excellent agreement between $u_\theta$ and the observed data. The weighted validation losses explain this behaviour. For $w_{\mathrm{PDE}}=10^{-4}$, the weighted PDE residual remains more than three orders of magnitude smaller than the weighted data loss throughout training (Fig.~\ref{fig:PDEweight}, middle right). Consequently, satisfying the governing PDE contributes very little to the overall optimisation objective, and the BINN receives almost no incentive to recover the correct diffusion operator. In this regime, the BINN successfully interpolates the observations but fails to identify the underlying diffusion mechanism. This illustrates an important caveat: accurate interpolation of the data does not necessarily imply successful mechanistic recovery.

At the opposite extreme, excessively large relative PDE weightings produce the opposite failure mode. When the PDE residual weighting dominates the total loss, optimisation prioritises satisfying the governing equation before accurately representing the observed data. The learned solution therefore becomes a poor approximation to the latent dynamics, which subsequently propagates into errors in both the recovered diffusivity, $D_\phi(u)$, and the forward-predicted solution, $u_\mathrm{p}$. This behaviour is evident for $w_{\mathrm{PDE}}=10^3$, where both $u_\theta$ and $u_\mathrm{p}$ exhibit large reconstruction errors despite the strong emphasis placed on the PDE residual during training.

Between these two extremes lies a broad regime in which the competing objectives are appropriately balanced. For the porous--medium equation, the lowest reconstruction errors occur when $w_{\mathrm{PDE}}$ and $w_{\mathrm{data}}$ lie approximately between $10^{-1}$ and $10^{0}$. Within this regime, the learned solution accurately represents the observations whilst simultaneously providing sufficient information for recovery of the nonlinear diffusion operator. Consequently, both the inferred diffusivity, $D_\phi(u)$, and the forward-predicted solution, $u_\mathrm{p}$, exhibit close agreement with the true underlying dynamics.

Taken together, these results demonstrate that the PDE residual acts as a regularising constraint that guides mechanistic discovery, but only once the learned solution provides an accurate representation of the underlying dynamics. Across all experiments considered in this work, an effective balance is achieved when the data-fitting objective remains at least as influential as the PDE residual, that is, when $w_{\mathrm{data}}\gtrsim w_{\mathrm{PDE}}$. More generally, discrepancies between the errors in $u_\theta$ and $u_\mathrm{p}$ provide a powerful diagnostic of mechanistic failure: good agreement in $u_\theta$ accompanied by poor agreement in $u_\mathrm{p}$ indicates that the BINN has learned to interpolate the observations without recovering the correct governing operator.


\subsection*{Moderately sized operator networks are sufficient for accurate operator recovery}

The previous section demonstrated that accurate operator recovery depends on balancing the competing objectives in the BINN loss function. Once this balance has been established, a second question naturally arises: how expressive do the operator neural networks, $D_\phi$, $V_\zeta$ and $g_\psi$, themselves need to be? While increasing network depth and width increases the representational capacity of the neural networks, it also introduces additional trainable parameters that may increase computational cost, reduce reproducibility, and promote over-fitting. We therefore investigate the extent to which successful operator recovery requires highly expressive operator networks.


\begin{figure}[htbp]
    \centering
    \includegraphics[width=\linewidth]{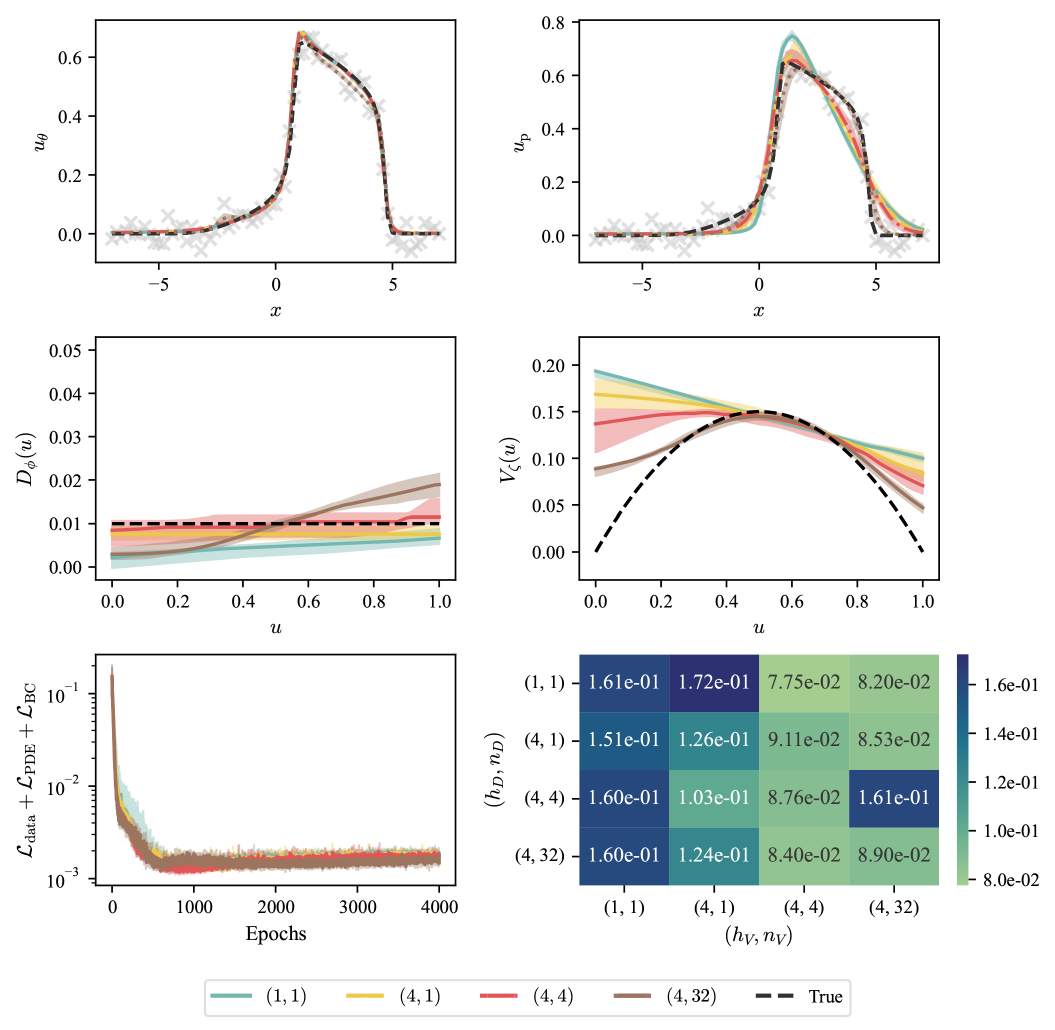}
    \caption{
    Effect of operator network architecture on ADR--BINN performance for the linear diffusion--nonlinear advection equation (Eq.~\eqref{eq:DA}) with additive Gaussian noise of variance $\sigma^2=0.001$. Top: Learned solution $u_\theta(x,t)$ (left) and forward-predicted solution $u_\text{p}(x,t)$ (right). 
    Middle: Learned constitutive functions $D_\phi(u)$ (left) and $V_\zeta(u)$ (right) plotted for $t=t_\text{end}$ as specified in Table~\ref{tab:time_resolution}. Results are shown for operator network architectures $(h_{D,V},n_{D,V})\in\{(1,1),(4,1),(4,4),(4,32)\}$; dashed black curves denote the true solution or operators, and shaded regions indicate $\pm1$ standard deviation over ten independent ADR--BINN initialisations. Note that when there is only one hidden layer, an \texttt{Identity} activation function is used in the input layer and a \texttt{Linear} activation function, where $\texttt{Linear}(x;W,b)= xW^{\top}+b$,
     where $W$ is the weight matrix and $b$ is the bias vector, is used for the output layer.
     Bottom: Total validation loss during training for each operator network architecture (left). Mean relative $L^2$ error in the forward-predicted solution $u_\text{p}$ for all sixteen combinations of diffusion and advection network architectures. Unless stated otherwise, all remaining ADR--BINN hyperparameters take the default values given in Table~\ref{tab:BINN parameter}.  }
    \label{fig:Networks}
\end{figure}


To address this question, we trained an ADR--BINN on synthetic data generated from a linear diffusion--nonlinear advection equation, Eq.~\eqref{eq:DA}, while systematically varying the architectures of the diffusion and advection networks, $D_\phi$ and $V_\zeta$. This benchmark provides a useful test because one constitutive relationships is nonlinear, requiring the operator networks to capture a non-trivial functional form, whilst the other is described by a simple constant value. 
Fig.~\ref{fig:Networks} shows that the learned solution $u_\theta$ is relatively insensitive to the particular architecture, with the different latent solutions remaining broadly similar. The forward-predicted solution $u_\mathrm{p}$ follows the true solution near the central peak for most architectures, although noticeable discrepancies occur away from the peak, particularly on the descending branch. In contrast, the recovered constitutive functions show larger and more systematic differences. The learned diffusivities remain close to an approximately constant value over much of the domain, but do not consistently reproduce the true dependence on $u$. The advection functions capture the broad concave shape of the target relationship near its maximum, while deviating substantially near the ends of the input range. Thus, even in the case of more expressive network architecture, such as $(4,32)$, which provides a visually plausible solution reconstruction, we find that it does not necessarily imply accurate recovery of the underlying operators.
An example of overly restrictive network architectures can be seen when considering the smallest example in Fig.~\ref{fig:Networks}, where$(h_{D,V},n_{D,V})=(1,1)$. In this case, small architectures produce overly restrictive approximations of both the diffusivity and advection functions and consequently, the recovered operators are biased, leading to comparatively large errors in the forward-predicted solution, $u_\mathrm{p}$. 




The effect of increasing the operator neural network capacity is, however, not a simple monotonic improvement. The validation loss curves decrease rapidly during the initial stages of training and then level off at similar values for the different architectures, with small fluctuations persisting during later epochs. The forward predicted solution error as shown in the bottom right plot in Fig.~\ref{fig:Networks} likewise shows no clear ordering with network size: some larger architectures perform comparably to, or worse than, smaller ones, depending on the combination of diffusion and advection networks. The $(4,32)$ architecture, for example, does not provide a consistent improvement over $(4,4)$; its learned diffusivity departs more strongly from the nearly constant target at larger $u$, while its advection function remains similar in overall shape but still differs from the true curve near the boundaries. These results indicate that, for this benchmark problem and dataset, additional depth or width in the neural networks does not reliably improve either predictive accuracy or constitutive-function recovery. Moderate architectures appear sufficient to reach the observed validation loss plateau, but the plots do not support the stronger conclusion that the true operators are accurately recovered or that the remaining differences are definitively caused by over-fitting. In fact, users should take care not to employ overly complex network architectures, as this could lead to over-fitting to the data,  as would be indicated by an increasing gradient over epochs in the loss functions.

Taken together, these results suggest that successful operator recovery requires only sufficient network capacity to represent the complexity of the underlying constitutive functions, rather than maximising the size of the operator networks. Beyond this point, increasing network expressivity yields diminishing returns, increasing computational cost and reducing robustness without materially improving mechanistic inference. For the benchmark problems considered here, moderately sized operator networks therefore provide a practical balance between representational flexibility, computational efficiency, and reproducibility. Accordingly, when implementing BINNs with unknown constitutive operators, we recommend users to evaluate a range of increasingly expressive network architectures and select a level of complexity beyond which further increases yield only limited improvements.


\subsection*{Batch size governs the trade-off between optimisation efficiency and reconstruction accuracy}

The previous sections have shown that successful operator recovery depends on appropriate optimisation and network design. A further practical consideration is how the optimisation is performed. Rather than evaluating the loss using the entire training dataset at every optimisation step, ADR--BINNs employ mini-batch gradient descent, in which the gradient is estimated from a randomly selected subset of the data of size $B$ \cite{goyal2017accurate,goodfellow2016deep,keskar2016large}. The batch size therefore determines the stochasticity of the optimisation process. Small batches produce noisy gradient estimates that encourage exploration of the loss landscape, whereas large batches yield smoother, more deterministic updates. This introduces a trade-off between optimisation efficiency, computational cost, and the robustness of the recovered operators.


\begin{figure}[htbp]
    \centering
    \includegraphics[width=\linewidth]{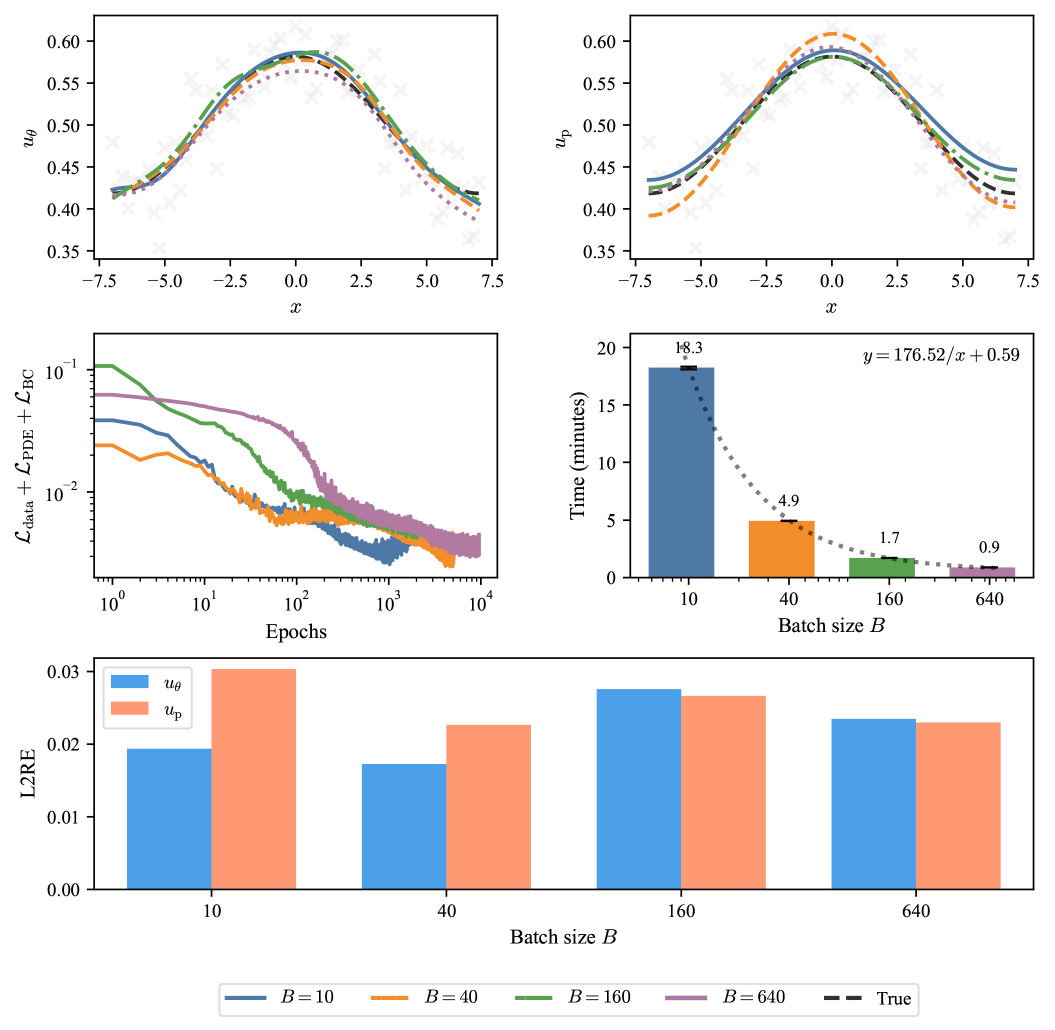}
    \caption{
    Effect of mini-batch size, $B$, on ADR--BINN optimisation and reconstruction accuracy for the diffusion equation (Eq.\eqref{eq:diff}) with additive Gaussian noise of variance $\sigma^2=0.001$.
    Top: Learned solution $u_\theta(x,t)$ (left) and forward-predicted solution $u_\text{p}(x,t)$ (right) plotted for $t=t_\text{end}$ as specified in Table~\ref{tab:time_resolution}. Middle: Total validation loss during training for $B=\{10,40,160,640\}$ (left). Median wall-clock training time over ten repetitions for $4000$ epochs as a function of batch size; the dashed line shows the fitted inverse relationship (right). Bottom: Relative $L^2$ errors in $u_\theta$ and $u_\text{p}$ after a fixed training budget of five minutes. Unless stated otherwise, $N_{\mathrm{PDE}}=N_{\mathrm{BC}}=40$ and all remaining hyperparameters take the default values given in Table~\ref{tab:BINN parameter}.}
    \label{fig:batch}
\end{figure}


To investigate this trade-off, we trained an ADR--BINN on synthetic data generated from the diffusion equation, Eq.~\eqref{eq:diff}, using a range of mini-batch sizes. Fig.~\ref{fig:batch} compares the resulting optimisation behaviour, computational cost, and reconstruction accuracy. Small batch sizes, such as $B=10$, initially reduce the validation loss rapidly (Fig.~\ref{fig:batch}, middle left). The stochasticity in the gradient estimates enables the optimiser to explore the loss landscape effectively, helping it escape shallow local minima and avoid premature convergence. However, because many more parameter updates are required to process the complete training dataset, this improvement comes at the expense of substantially increased computational cost. Across 4000 training epochs, the median wall-clock training time scales approximately as $T\propto1/B$ (Fig.~\ref{fig:batch}, middle right), consistent with the expected computational complexity of mini-batch optimisation. At the opposite extreme, very large batch sizes ($B=640$) substantially reduce gradient stochasticity, producing smoother validation loss trajectories after the initial stages of training. However, the optimisation explores the loss landscape less effectively and consequently makes slower progress per unit computational time. Although each optimisation step is cheaper, the reduced efficiency in exploration results in poorer reconstruction accuracy within a fixed training budget.

These competing effects give rise to an intermediate regime in which optimisation efficiency and reconstruction accuracy are simultaneously maximised. For a fixed training budget of five minutes, intermediate batch sizes ($B=40$--$160$) consistently yielded the smallest errors in both the learned solution, $u_\theta$, and the forward-predicted solution, $u_\mathrm{p}$ (Fig.~\ref{fig:batch}, bottom). In this regime, the optimiser benefits from sufficient stochasticity to efficiently explore the loss landscape while remaining computationally efficient enough to perform many optimisation steps within the available time. We note that batch size also influences the reproducibility of operator recovery. While small batches often converge rapidly, they also exhibit greater variability between independent BINN initialisations, indicating increased sensitivity to stochastic optimisation noise. Intermediate batch sizes substantially reduce this variability while maintaining high reconstruction accuracy, producing more consistent estimates of the underlying operators across repeated training runs.

Taken together, these results demonstrate that batch size should be viewed as an optimisation parameter rather than simply a computational one. Successful operator recovery depends on balancing gradient stochasticity with computational efficiency: too little stochasticity limits exploration of the loss landscape, whereas too much reduces reproducibility and increases computational cost. Across the benchmark problems considered here, intermediate batch sizes provide the most favourable compromise between optimisation efficiency, reconstruction accuracy, and robustness of the recovered constitutive operators.


\section*{Discussion}

Biologically-informed neural networks models seek to bridge mechanistic modelling and data-driven inference by embedding governing differential equations directly into neural network training. The hybrid formulation enables the recovery of interpretable operators from sparse and noisy biological data, offering an attractive alternative to purely data-driven or fully mechanistic approaches. However, our results demonstrate that successful operator recovery is not an automatic consequence of employing the BINNs framework. Rather, reliable mechanistic inference emerges only when the network architecture, optimisation strategy, and available data are appropriately balanced. Through a systematic investigation across a suite of canonical advection--diffusion--reaction models, we have identified both the regimes in which BINNs reliably recover governing operators and the failure modes that arise when these competing requirements are not adequately balanced.

Perhaps the most important finding of this study is that accurate interpolation should not be conflated with successful mechanistic inference. Throughout this work, we distinguished between the learned neural network solution, $u_\theta$, and the forward-predicted solution, $u_\mathrm{p}$, generated by numerically solving the governing PDE using the recovered operators. This distinction proved crucial. In several experiments, the learned solution closely matched the observations while the forward-predicted solution remained inaccurate, demonstrating that excellent interpolation alone does not imply that the correct governing mechanisms have been recovered. Instead, agreement between $u_\mathrm{p}$ and the observed dynamics provides a much stronger indication that the inferred constitutive operators capture the underlying biological processes. We therefore advocate that future applications of BINNs routinely assess both quantities, rather than relying solely on the accuracy of the learned neural network solution.

A second consistent theme emerging from our results is that each of the principal BINN hyperparameters balances competing objectives during optimisation. Learning rates mediate a trade-off between efficient exploration of the loss landscape and optimisation stability. Loss weightings balance fidelity to the observed data against consistency with the governing PDE. Network architecture governs the balance between representational flexibility and robustness to noisy observations, while batch size controls the trade-off between stochastic exploration of the loss landscape and computational efficiency. Although these hyperparameters are often tuned independently in practice, our results suggest that they should instead be viewed collectively as mechanisms for balancing competing aspects of the optimisation problem. Successful operator recovery therefore depends less on identifying a single optimal hyperparameter value than on identifying a regime in which these competing objectives remain appropriately balanced.

These observations also highlight that the principal limitation in operator recovery is often not the expressivity of the neural networks themselves, but rather the information contained within the available data. Across all benchmark problems considered here, increasing network complexity beyond a moderate level rarely improved reconstruction accuracy once the underlying constitutive relationships could already be represented. Instead, the dominant factors limiting recovery of operators are the observation noise and the range of solution values represented in the data. This finding is particularly encouraging for biological applications, where data are frequently sparse, noisy, and only partially informative. It suggests that increasing model complexity is unlikely to compensate for limited data informativeness, and may instead reduce robustness through over-fitting or unstable optimisation.

In experimental applications, the true governing operators are unknown and operator reconstruction errors cannot be evaluated directly. Under these circumstances, practitioners should instead monitor quantities that indicate whether the optimisation is behaving appropriately. Stable validation loss trajectories, close agreement between the learned solution and forward simulations generated using the recovered operators, and consistency across different neural network initialisations all provide practical evidence that the inferred mechanisms are likely to be reliable. Conversely, discrepancies between $u_\theta$ and $u_\mathrm{p}$, increasing validation losses during training, or substantial variability between repeated optimisations with different initialisations provide early indications that the recovered operators should be interpreted with caution. These diagnostics offer practical guidance for applying BINNs to experimental datasets where the ground truth is unavailable.

Taken together, our findings suggest several practical recommendations for using BINN models to study biological systems. Moderately expressive network architectures should generally be preferred over highly over-parameterised models unless there is clear evidence of systematic under-fitting. Data-fitting should remain at least as influential as the PDE residual throughout optimisation, either through iterative training strategies~\cite{Nardini:2020:LEF} or by ensuring that $w_{\mathrm{data}}\gtrsim w_{\mathrm{PDE}}$. Intermediate learning rates and batch sizes consistently provide the most favourable balance between optimisation efficiency, reconstruction accuracy, and robustness. More generally, we recommend that practitioners use optimisation diagnostics to guide model refinement, increasing architectural complexity only when simpler models consistently fail to capture the observed dynamics.

Several limitations of this study should also be acknowledged. All benchmark problems were based on synthetic data generated from one-dimensional PDE models with known ground truth, enabling controlled comparisons but inevitably simplifying the complexity of real biological systems. Extensions to higher-dimensional domains, such as in~\cite{lavery2026physicsinformedneuralnetworksbiological}, heterogeneous media, partially observed states, or coupled multi-species systems may introduce additional challenges and are beyond the scope of the present work. Furthermore, uncertainty in the recovered operators was assessed through repeated network initialisations rather than formal Bayesian uncertainty quantification, and we did not investigate adaptive loss re-weighting, adaptive sampling strategies, or alternative optimisation algorithms that may further improve convergence and stability~\cite{mcclenny2023self, wang2022and, wu2023comprehensive, daw2022rethinking, gao2023failure}. By deliberately focusing on a comparatively simple and reproducible BINN framework, however, we were able to isolate the influence of architecture, optimisation, and data informativeness without introducing additional methodological confounders.

More broadly, this work contributes to the ongoing effort to make neural equation learning methods reliable tools for scientific discovery. For biological modelling, predictive accuracy alone is rarely sufficient; the ultimate objective is to recover mechanisms that are both interpretable and reproducible. By identifying the competing trade-offs that govern optimisation, demonstrating how these trade-offs influence mechanistic recovery, and proposing practical diagnostics for assessing reliability when ground truth is unavailable, we hope this work provides a useful foundation for the broader application of BINN models in biology. Future developments that integrate principled experimental design, formal uncertainty quantification, adaptive optimisation strategies, and symbolic simplification of recovered operators will further strengthen the ability of BINN models to bridge mechanistic theory and data-driven inference. More generally, our results suggest that reliable mechanistic inference with BINNs is achieved not by maximising network complexity or optimisation effort, but by appropriately balancing model expressivity, optimisation, and data informativeness.


\section*{Acknowledgments}

R.M.C. acknowledges support from the Engineering and Physical Sciences Research Council (EP/Z534870/1). R.E.B. acknowledges support of a grant from the Simons Foundation (MP-SIP-00001828). Y.Y. acknowledges support from the Engineering and Physical Sciences Research Council (EP/W524311/1). ChatGPT (v5.5) was used to aid in the production of the top panel in Fig.~\eqref{fig:1-schematic}, and in assisting with proofreading and language editing prior to publication---this service was used strictly to improve text clarity and readability. For the purpose of Open Access, the authors have applied a CC BY public copyright licence to any Author Accepted Manuscript (AAM) version arising from this submission.


\bibliographystyle{unsrt}
\bibliography{refs}


\pagebreak

\begin{center}
\section*{Supporting Material}
\end{center}

\subsection*{Synthetic data generation}

Synthetic training data were generated by solving each benchmark PDE using the constitutive functions and parameter values listed in Table~\ref{tab:BINN parameter}. Collectively, the benchmark suite spans both linear and nonlinear diffusion, advection, and reaction terms, enabling a systematic assessment of the ability of ADR--BINNs to recover constitutive relationships of increasing complexity.

Analytical solutions were used whenever available. Specifically, the diffusion and porous--medium benchmark datasets were generated directly from their analytical solutions presented in the main text (Eqs.~\eqref{eq:diff_ana} and \eqref{eq:PM-sol}). The remaining benchmark problems were solved numerically using standard spatial discretisation techniques and time integration methods implemented in Python. Unless otherwise stated, all simulations were performed on the spatial domain $x\in[-7,7],$ using a uniform mesh with spacing $\Delta x=0.04$, corresponding to $351$ spatial nodes. In the sparsely sampled data in the main text, there are only $71$ spatial nodes, selected by subsampling every 5 points. Homogeneous Neumann boundary conditions (Eq.~\eqref{Eq:BC}) were imposed throughout. Temporal sampling differed between benchmark problems and is summarised in Table~\ref{tab:time_resolution}.


\begin{table}[htbp]
\centering
\small
\renewcommand{\arraystretch}{1.25}

\begin{tabular}{p{2.0cm}p{7.2cm}ccc}
\toprule
\textbf{PDE} &
\textbf{Model} &
\textbf{$N_t$} &
\textbf{Time interval} &
\textbf{$\Delta t$} \\
\midrule
Eq.~\eqref{eq:diff}
& Diffusion equation
& 7
& $[0,18]$
& 3 \\

Eq.~\eqref{eq:PM}
& Porous--medium equation
& 7
& $[0,4]$
& $2/3$ \\

Eq.~\eqref{eq:burger}
& Burgers' equation
& 7
& $[0,10]$
& $5/3$ \\

Eq.~\eqref{eq:DA}
& Linear diffusion--nonlinear advection equation
& 7
& $[0,18]$
& 3 \\

Eq.~\eqref{eq:NL_ADR}
& Nonlinear reaction--diffusion--advection equation
& 9
& $[0,4]$
& 0.5 \\

\bottomrule
\end{tabular}

\caption{Temporal discretisations used to generate the synthetic datasets for each benchmark PDE.}
\label{tab:time_resolution}

\end{table}


\paragraph{Nonlinear advection--diffusion equation.} The nonlinear reaction--diffusion--advection problem (Eq.~\eqref{eq:NL_ADR}) was solved using a method-of-lines approach. Let $u_i(t)\approx u(x_i,t),$ for $i=0, \dots, 70$ denote the semi-discrete solution at node \(x_i\). Spatial derivatives were approximated using second-order centred finite differences, such that
\begin{equation*}
(u_x)_i
\approx
\frac{u_{i+1}-u_{i-1}}{2\Delta x}.
\end{equation*}
The nonlinear flux, $F=D(u)u_x-uV(u),$
was evaluated pointwise using the prescribed constitutive relationships, $D(u)=0.1u^5$, and $V(u)=u(1-u)$, before its divergence was approximated by
\begin{equation*}
(\partial_xF)_i
\approx
\frac{F_{i+1}-F_{i-1}}
{2\Delta x}.
\end{equation*}
The reaction term was evaluated pointwise using the prescribed growth function, $G(u)=-0.5$, yielding the semi-discrete system
\begin{equation*}
\frac{\mathrm du_i}{\mathrm dt}
=
(\partial_xF)_i
+
u_iG(u_i).
\end{equation*}
Homogeneous Neumann boundary conditions were imposed using mirrored ghost nodes. The resulting ordinary differential equation system was integrated using the adaptive Runge--Kutta solver \texttt{RK45} implemented in SciPy.


\paragraph{Burgers' equation.} The Burgers' equation problem (Eq.~\eqref{eq:burger}) was solved using a Fourier pseudospectral method \cite{trefethen2000spectral}. The semi-discrete solution was represented on the uniform spatial mesh, and first and second derivatives were computed spectrally using fast Fourier transforms,
\begin{equation*}
\widehat{u_x}=ik\widehat{u},
\qquad
\widehat{u_{xx}}=-k^2\widehat{u},
\end{equation*}
where \(k\) denotes the discrete wave numbers. Nonlinear products were evaluated in physical space before transforming back to Fourier space for differentiation. This yielded the semi-discrete system
\begin{equation*}
\frac{\mathrm du_i}{\mathrm dt}
=
-\mu u_i(u_x)_i
+
D(u_{xx})_i,
\end{equation*}
which was integrated using \texttt{scipy.integrate.odeint}.


\paragraph{Linear diffusion--nonlinear advection equation.} The linear diffusion--nonlinear advection problem (Eq.~\eqref{eq:DA}) was solved using continuous piecewise-linear finite elements on the uniform spatial mesh. The weak formulation of the governing equation produced the semi-discrete system
\begin{equation*}
M\dot{\mathbf U}=-K\mathbf U+\mathbf C(\mathbf U),
\end{equation*}
where \(M\) and \(K\) denote the standard finite-element mass and stiffness matrices, respectively, and \(\mathbf C\) represents the nonlinear advection contribution. The nonlinear term was evaluated using two-point Gaussian quadrature on each element. The resulting stiff system of ordinary differential equations was integrated using the implicit Radau solver implemented in SciPy.


\paragraph{Noise model.} To emulate measurement noise commonly encountered in biological experiments, i.i.d.~additive Gaussian noise was introduced after generation of the noiseless solution. Specifically,
\begin{equation}
u_{\mathrm{obs}}(x_i,t_i)
=
u(x_i,t_i)
+
\eta,
\end{equation}
where
\[
\eta\sim\mathcal N(0,\sigma^2).
\]
Three noise levels were considered,
\[
\sigma^2
\in
\left\{
0,\,
10^{-3},\,
10^{-2}
\right\},
\]
and the random seed was fixed throughout to ensure reproducibility.


\section*{Numerical reconstruction using learned constitutive functions.}

To assess whether the learned constitutive relationships accurately capture the underlying transport and reaction mechanisms, we reconstructed solutions by numerically solving Eq.~\eqref{Eq:PDE} using the learned constitutive functions, $D_{\phi}(u)$, $V_{\zeta}(u)$, and $G_{\psi}(u)$, together with the initial condition predicted by the solution network, $u_\theta(x,0)$, and the homogeneous Neumann boundary conditions given by Eq.~\eqref{Eq:BC}. The reconstructed solution was evaluated on the same spatial mesh and at the same temporal sampling points used to generate the synthetic data. A conservative finite-volume discretisation was employed. Let \(x_0<x_1<\cdots<x_{N-1}\) denote the computational mesh, with interface locations
\[
x_{i+\frac12}=\frac{x_i+x_{i+1}}{2},
\]
and let $u_i(t)\approx u(x_i,t)$ denote the semi-discrete approximation at node \(x_i\). The governing equation is written in conservative form,
\begin{equation*}
u_t
=
\partial_xF
+
uG(u),
\end{equation*}
where the total flux is
\begin{equation*}
F=D(u)u_x-uV(u).
\end{equation*}
The diffusive and advective contributions to the flux were approximated separately before being combined at each cell interface. The learned diffusivity was first evaluated at the nodal values, $D_i=D_\phi(u_i)$, and linearly interpolated to the interfaces using arithmetic averaging,
\begin{equation*}
D_{i+\frac12}
=
\frac{D_i+D_{i+1}}{2}.
\end{equation*}
The solution gradient at each interface was approximated using a centred finite difference,
\begin{equation*}
(u_x)_{i+\frac12}
\approx
\frac{u_{i+1}-u_i}
{\Delta x_{i+\frac12}},
\end{equation*}
giving the diffusive flux as
\begin{equation*}
F^{\mathrm{diff}}_{i+\frac12}
=
D_{i+\frac12}
\frac{u_{i+1}-u_i}
{\Delta x_{i+\frac12}}.
\end{equation*}
Similarly, the learned velocity was evaluated at the nodes, $V_i=V_\zeta(u_i)$, and averaged to obtain the interface velocity,
\begin{equation*}
V_{i+\frac12}
=
\frac{V_i+V_{i+1}}{2}.
\end{equation*}
To improve numerical stability of the advection term, a first-order upwind approximation was employed. 
The upwind state was defined as
\begin{equation*}
u^{\mathrm{up}}_{i+\frac12}
=
\begin{cases}
u_i, & V_{i+\frac12}\ge0,\\[0.5em]
u_{i+1}, & V_{i+\frac12}<0,
\end{cases}
\end{equation*}
yielding the advective flux
\begin{equation*}
F^{\mathrm{adv}}_{i+\frac12}
=
V_{i+\frac12}
u^{\mathrm{up}}_{i+\frac12}.
\end{equation*}
The total interface flux is therefore
\begin{equation*}
F_{i+\frac12}
=
F^{\mathrm{diff}}_{i+\frac12}
-
F^{\mathrm{adv}}_{i+\frac12}.
\end{equation*}
The spatial divergence of the flux was approximated by
\begin{equation*}
\left(\partial_xF\right)_i
\approx
\frac{F_{i+\frac12}-F_{i-\frac12}}
{\Delta x},
\end{equation*}
where \(\Delta x\) denotes the uniform width between nodes on the computational mesh. 
The learned reaction term was evaluated pointwise as
$G_i=G_\psi(u_i).$
The resulting semi-discrete system is therefore given by
\begin{equation*}
\frac{\mathrm du_i}{\mathrm dt}
=
\frac{F_{i+\frac12}-F_{i-\frac12}}
{\Delta x}
+
u_iG_i,
\qquad
i=0,\ldots,N-1.
\end{equation*}

Homogeneous Neumann boundary conditions were imposed using mirrored ghost nodes~\cite{morton_numerical_2005}, $u_{-1}=u_1$ and $u_N=u_{N-2}$, which ensure that the centred finite difference approximation satisfies $u_x(-7,t)=u_x(7,t)=0$. The interface fluxes at the domain boundaries were then evaluated using the extended solution vector including the ghost nodes.

Finally, the resulting semi-discrete system was integrated over the prescribed time interval using the adaptive Runge--Kutta solver \texttt{RK45} implemented in SciPy. The reconstructed solution, $u_{\mathrm{p}}(x,t)$, was subsequently compared directly with the true underlying solution generated from the known constitutive functions. Agreement between the two provides an independent assessment of whether the learned constitutive relationships faithfully reproduce the underlying system dynamics.


\subsection*{Results for noise-free datasets}

\begin{figure}[htbp]
    \centering
    \includegraphics[width=\linewidth]{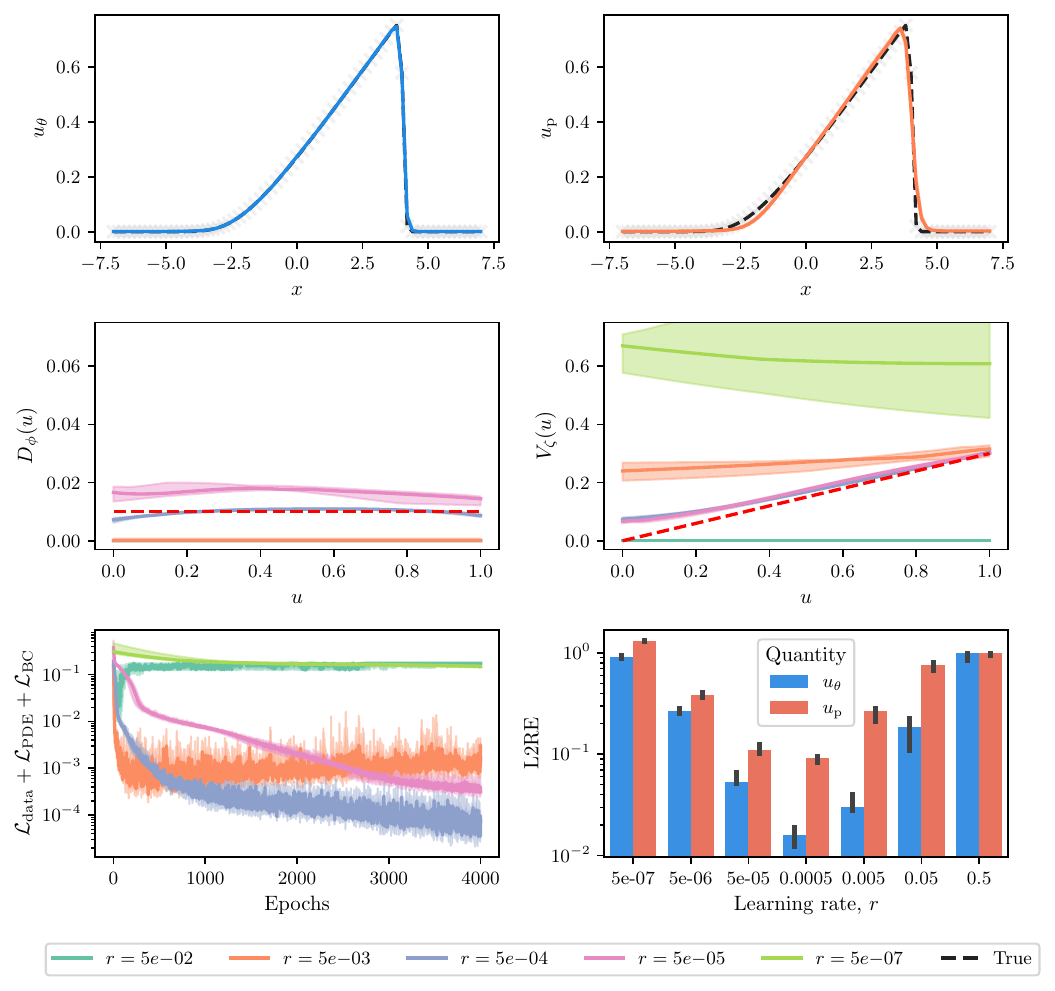}
     \caption{Effect of the learning rate, $r$, on BINNs performance on data simulated from Burgers' equation (Eq.~\eqref{eq:burger}) with no noise. 
    Top: Learned solution $u_\theta(x,t)$ (left) and forward-predicted solution $u_\text{p}(x,t)$ (right) plotted for $t=t_\text{end}$ as specified in Table~\ref{tab:time_resolution}. 
    Middle: Learned diffusivity $D_\phi(u)$ (left) and advection velocity $V_\zeta(u)$ (right).
    Results correspond to the representative learning rate $r=5\times10^{-4}$ (top); dashed black curves denote the true operators and shaded regions indicate $\pm1$ standard deviation over ten independent ADR--BINN initialisations.
    Bottom: Total validation loss (left) and relative $L^2$ errors in $u_\theta$ and $u_p$ (right) for various learning rates, $r$. Unless stated otherwise, all remaining ADR--BINN hyperparameters take the default values given in Table~\ref{tab:BINN parameter}.}
    \label{fig:0-r}
\end{figure}

\begin{figure}[htbp]
    \centering
    \includegraphics[width=\linewidth]{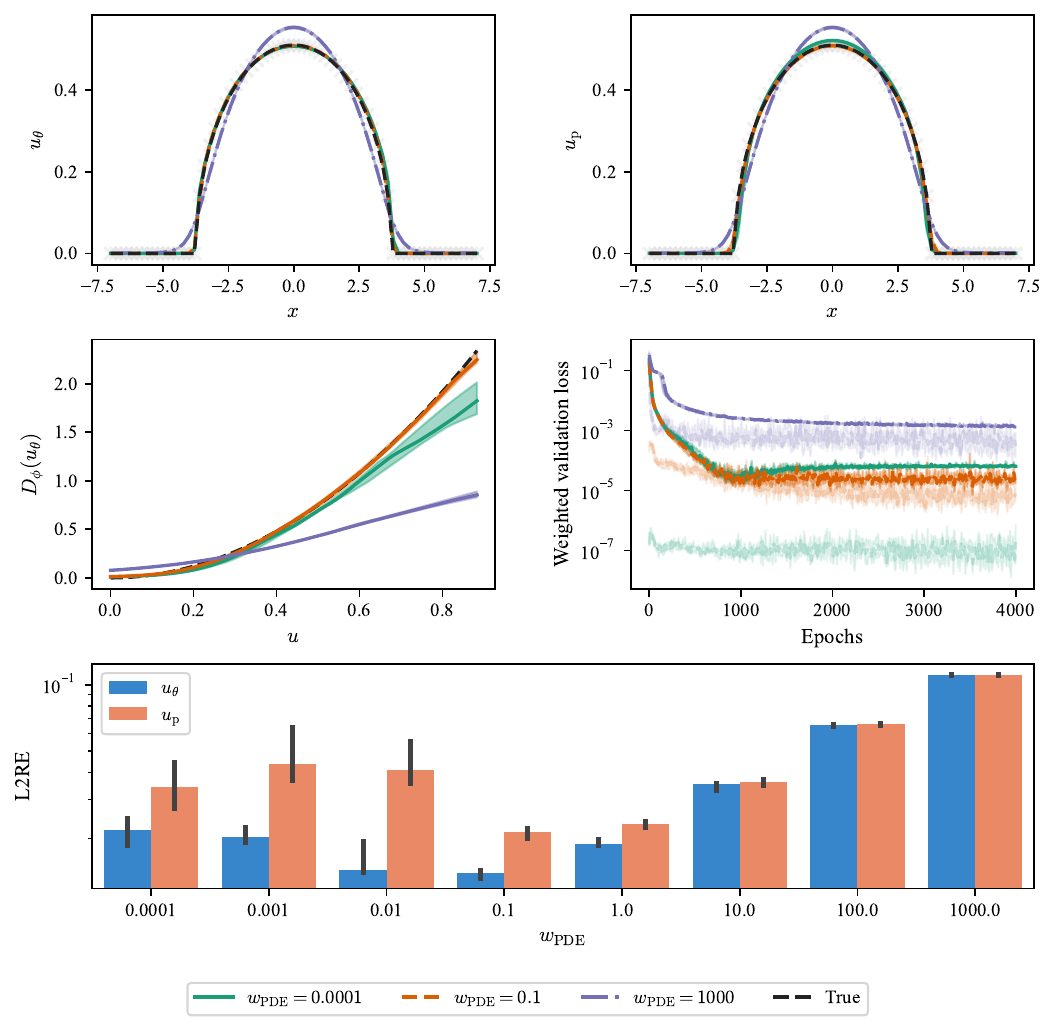}
    \caption{
    Effect of the PDE residual weighting $w_{\mathrm{PDE}}$ on BINN performance for the porous-medium equation (Eq.~\eqref{eq:PM}) with no noise, whilst $w_{\mathrm{data}}=1$ and $w_{\mathrm{BC}}=0$. 
    Top: Learned solution $u_\theta(x,t)$ (left) and forward-predicted solution $u_p(x,t)$ (right) plotted for $t=t_\text{end}$ as specified in Table~\ref{tab:time_resolution}. 
    Middle left: Learned diffusion function $D_\phi(u)$. Representative results are shown for $w_{\mathrm{PDE}}\in\{10^{-4},10^{-1},10^3\}$; dashed black curves denote the true solution or constitutive function and shaded regions represent $\pm1$ standard deviation across ten BINN initialisations. 
    Middle right: Weighted validation losses, showing $L_{\mathrm{data}}$ (solid) and $w_{\mathrm{PDE}}L_{\mathrm{PDE}}$ (dashed) with shaded regions representing $\pm1$ standard deviation across ten BINN initialisations. 
    Bottom: Relative $L^2$ errors in $u_\theta$ and $u_\text{p}$ as $w_{\mathrm{PDE}}$ varies. 
    Unless stated otherwise, all remaining ADR--BINN hyperparameters take the default values given in Table~\ref{tab:BINN parameter}.  
    }
    \label{fig:0-w}
\end{figure}

\begin{figure}[htbp]
    \centering
    \includegraphics[width=\linewidth]{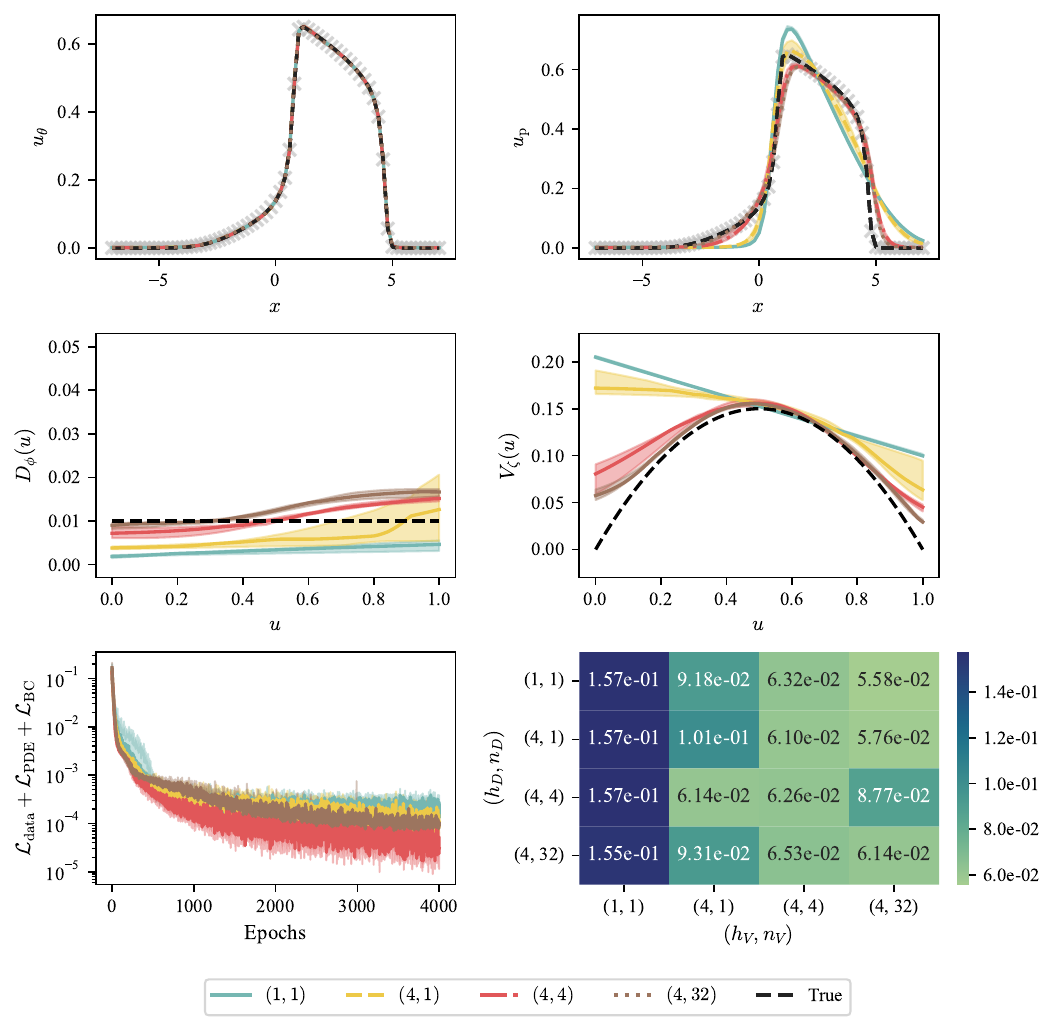}
    \caption{
    Effect of operator network architecture on ADR--BINN performance for the linear diffusion--nonlinear advection equation (Eq.~\eqref{eq:DA}) with no noise. Top: Learned solution $u_\theta(x,t)$ (left) and forward-predicted solution $u_\text{p}(x,t)$ (right) plotted for $t=t_\text{end}$ as specified in Table~\ref{tab:time_resolution}. 
    Middle: Learned constitutive functions $D_\phi(u)$ (left) and $V_\zeta(u)$ (right). Results are shown for operator network architectures $(h_{D,V},n_{D,V})\in\{(1,1),(4,1),(4,4),(4,32)\}$; dashed black curves denote the true solution or operators, and shaded regions indicate $\pm1$ standard deviation over ten independent ADR--BINN initialisations. Note that when there is only one hidden layer, an \texttt{Identity} activation function is used in the input layer and a \texttt{Linear} activation function, where $\texttt{Linear}(x;W,b)= xW^{\top}+b$,
    where $W$ is the weight matrix and $b$ is the bias vector, is used for the output layer.
    Bottom: Total validation loss during training for each operator network architecture (left). Mean relative $L^2$ error in the forward-predicted solution $u_\text{p}$ for all sixteen combinations of diffusion and advection network architectures. Unless stated otherwise, all remaining ADR--BINN hyperparameters take the default values given in Table~\ref{tab:BINN parameter}.}
    \label{fig:0-network}
\end{figure}

\begin{figure}[htbp]
    \centering
    \includegraphics[width=\linewidth]{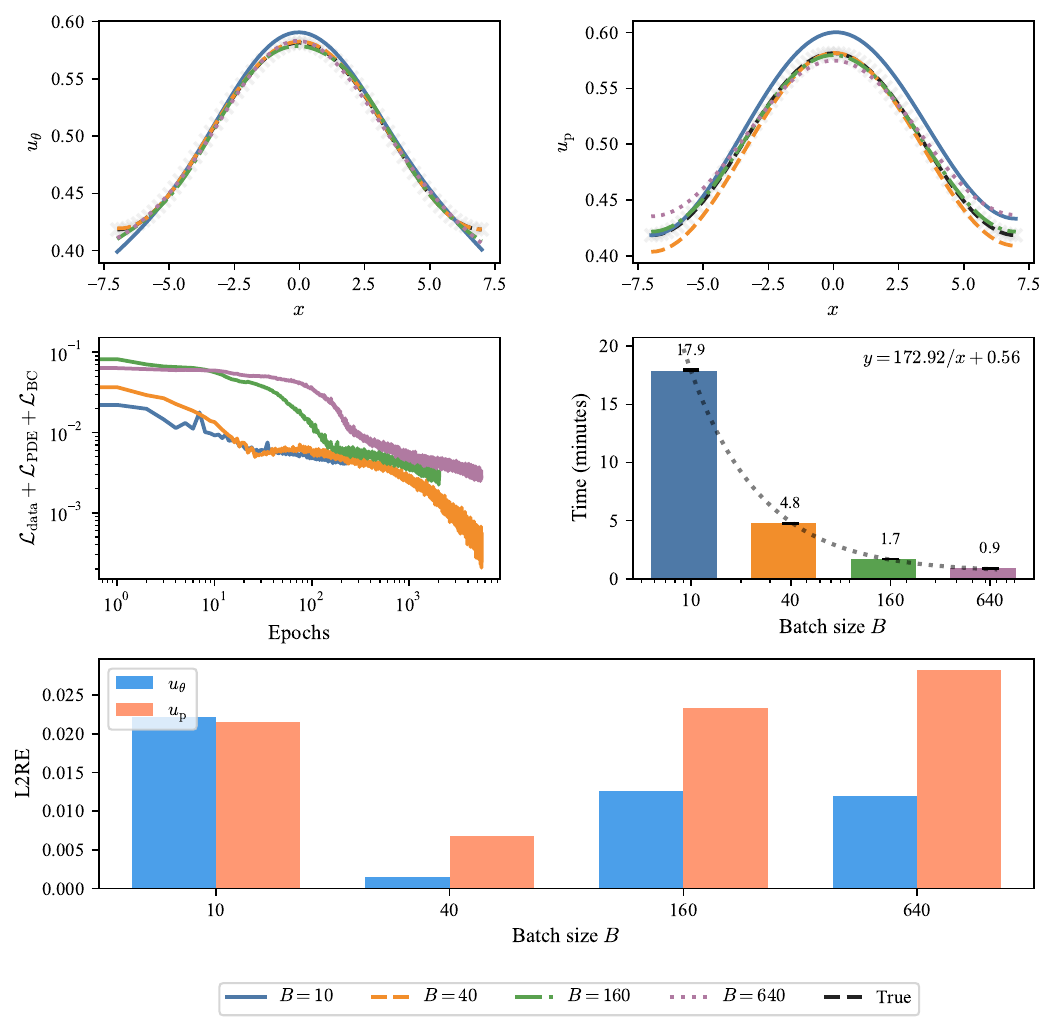}
    \caption{
    Effect of mini-batch size, $B$, on ADR--BINN optimisation and reconstruction accuracy for the diffusion equation (Eq.\eqref{eq:diff}) with no  noise.
    Top: Learned solution $u_\theta(x,t)$ (left) and forward-predicted solution $u_\text{p}(x,t)$ (right) plotted for $t=t_\text{end}$ as specified in Table~\ref{tab:time_resolution}. Middle: Total validation loss during training for $B=\{10,40,160,640\}$ (left). Median wall-clock training time over ten repetitions for $4000$ epochs as a function of batch size; the dashed line shows the fitted inverse relationship (right). Bottom: Relative $L^2$ errors in $u_\theta$ and $u_\text{p}$ after a fixed training budget of five minutes. Unless stated otherwise, $N_{\mathrm{PDE}}=N_{\mathrm{BC}}=40$ and all remaining hyperparameters take the default values given in Table~\ref{tab:BINN parameter}. }
    \label{fig:0-B}
\end{figure}

Figs.~\ref{fig:0-r}--\ref{fig:0-B} present the same analyses as Figs.~\ref{fig:learning-rate}--\ref{fig:batch} in the main text, but for synthetic datasets without additive observation noise ($\sigma^2=0$). Overall, the qualitative conclusions remain unchanged.  Intermediate learning rates continue to provide the most reliable optimisation behaviour (Fig.~\ref{fig:0-r}), balanced PDE and data-loss weightings remain essential for accurate operator recovery (Fig.~\ref{fig:0-w}), moderately expressive operator networks remain sufficient to recover the underlying constitutive relationships (Fig.~\ref{fig:0-network}), and intermediate mini-batch sizes continue to provide the best balance between optimisation efficiency and computational cost (Fig.~\ref{fig:0-B}).

The absence of observation noise primarily improves the quantitative accuracy of both the learned solution, $u_\theta$, and the forward-predicted solution, $u_{\mathrm p}$, across all experiments. In particular, the learned constitutive functions exhibit reduced variability between independent ADR--BINN initialisations, and the validation losses decrease more smoothly throughout training. Consequently, the operator estimates generally lie closer to the true constitutive relationships than in the corresponding noisy experiments presented in the main text and the next section of the Supporting Material. These results demonstrate that the trends identified throughout the manuscript are not artefacts of noisy observations, but rather reflect intrinsic properties of ADR--BINN optimisation and architecture.


\clearpage

\subsection*{Results for highly noisy datasets}

\begin{figure}[htbp]
    \centering
    \includegraphics[width=\linewidth]{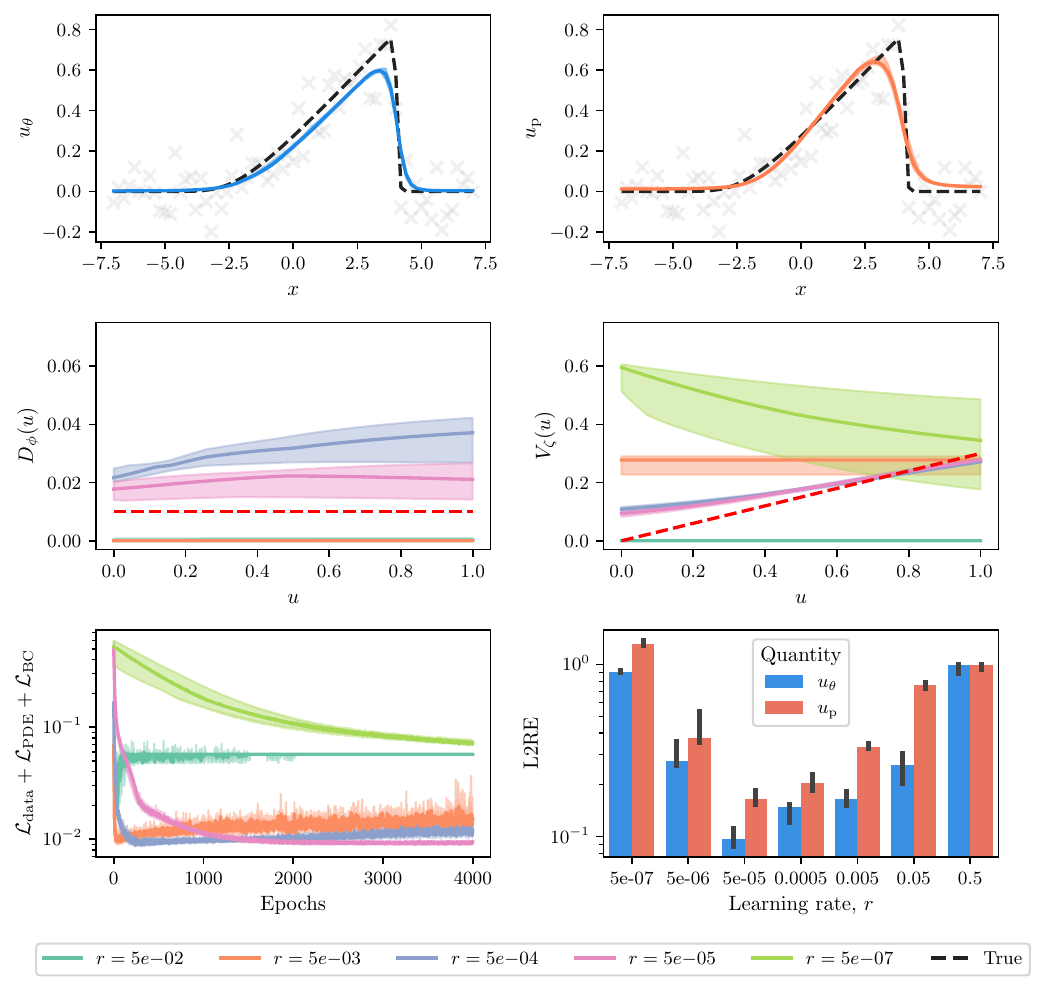}
     \caption{Effect of the learning rate, $r$, on BINNs performance on data simulated from Burgers' equation (Eq.~\eqref{eq:burger}) with additive Gaussian noise of variance $\sigma^2=0.01$. 
    Top: Learned solution $u_\theta(x,t)$ (left) and forward-predicted solution $u_\text{p}(x,t)$ (right) plotted for $t=t_\text{end}$ as specified in Table~\ref{tab:time_resolution}. 
    Middle: Learned diffusivity $D_\phi(u)$ (left) and advection velocity $V_\zeta(u)$ (right).
    Results correspond to the representative learning rate $r=5\times10^{-4}$ (top); dashed black curves denote the true operators and shaded regions indicate $\pm1$ standard deviation over ten independent ADR--BINN initialisations.
    Bottom: Total validation loss (left) and relative $L^2$ errors in $u_\theta$ and $u_p$ (right) for various learning rates, $r$. Unless stated otherwise, all remaining ADR--BINN hyperparameters take the default values given in Table~\ref{tab:BINN parameter}.}
    \label{fig:high-r}
\end{figure}

\begin{figure}[htbp]
    \centering
    \includegraphics[width=\linewidth]{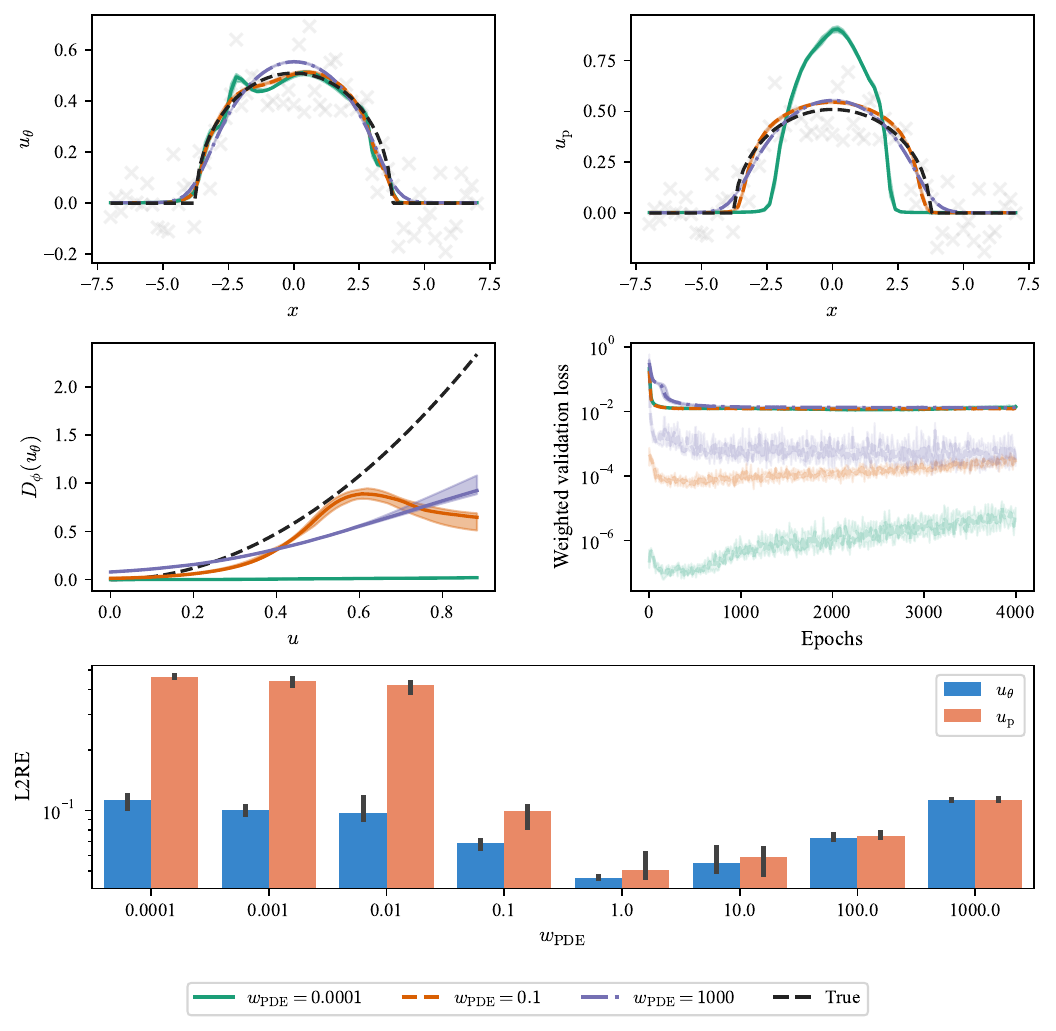}
    \caption{
    Effect of the PDE residual weighting $w_{\mathrm{PDE}}$ on BINN performance for the porous-medium equation (Eq.~\eqref{eq:PM}) with additive Gaussian noise of variance $\sigma^2=0.01$, whilst $w_{\mathrm{data}}=1$ and $w_{\mathrm{BC}}=0$. 
    Top: Learned solution $u_\theta(x,t)$ (left) and forward-predicted solution $u_p(x,t)$ (right) plotted for $t=t_\text{end}$ as specified in Table~\ref{tab:time_resolution}. 
    Middle left: Learned diffusion function $D_\phi(u)$. Representative results are shown for $w_{\mathrm{PDE}}\in\{10^{-4},10^{-1},10^3\}$; dashed black curves denote the true solution or constitutive function and shaded regions represent $\pm1$ standard deviation across ten BINN initialisations. 
    Middle right: Weighted validation losses, showing $L_{\mathrm{data}}$ (solid) and $w_{\mathrm{PDE}}L_{\mathrm{PDE}}$ (dashed) with shaded regions representing $\pm1$ standard deviation across ten BINN initialisations. 
    Bottom: Relative $L^2$ errors in $u_\theta$ and $u_\text{p}$ as $w_{\mathrm{PDE}}$ varies. 
    Unless stated otherwise, all remaining ADR--BINN hyperparameters take the default values given in Table~\ref{tab:BINN parameter}.  
    }
    \label{fig:high-w}
\end{figure}

\begin{figure}[htbp]
    \centering
    \includegraphics[width=\linewidth]{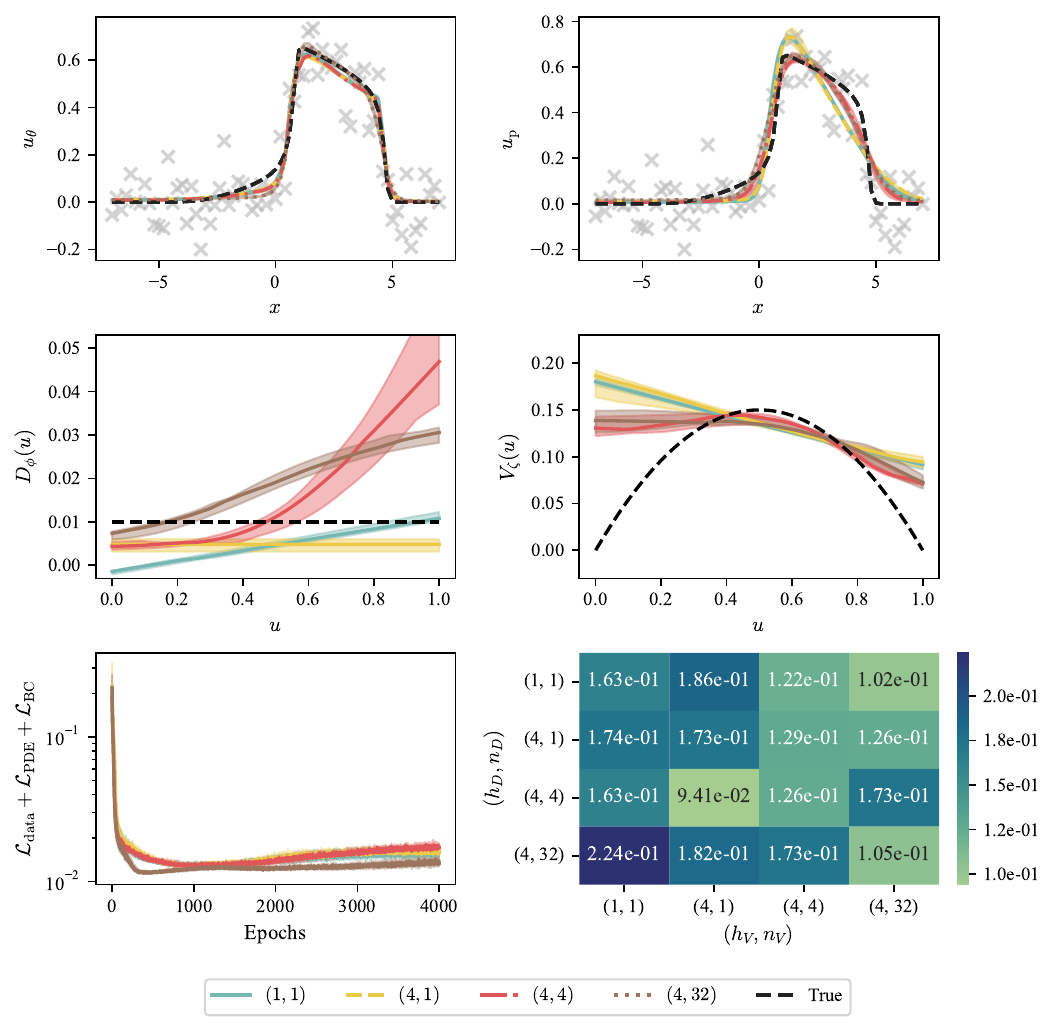}
    \caption{
    Effect of operator network architecture on ADR--BINN performance for the linear diffusion--nonlinear advection equation (Eq.~\eqref{eq:DA}) with additive Gaussian noise of variance $\sigma^2=0.01$. Top: Learned solution $u_\theta(x,t)$ (left) and forward-predicted solution $u_\text{p}(x,t)$ (right) plotted for $t=t_\text{end}$ as specified in Table~\ref{tab:time_resolution}. 
    Middle: Learned constitutive functions $D_\phi(u)$ (left) and $V_\zeta(u)$ (right). Results are shown for operator network architectures $(h_{D,V},n_{D,V})\in\{(1,1),(4,1),(4,4),(4,32)\}$; dashed black curves denote the true solution or operators, and shaded regions indicate $\pm1$ standard deviation over ten independent ADR--BINN initialisations. Note that when there is only one hidden layer, an \texttt{Identity} activation function is used in the input layer and a \texttt{Linear} activation function, where $\texttt{Linear}(x;W,b)= xW^{\top}+b$,
     where $W$ is the weight matrix and $b$ is the bias vector, is used for the output layer.
     Bottom: Total validation loss during training for each operator network architecture (left). Mean relative $L^2$ error in the forward-predicted solution $u_\text{p}$ for all sixteen combinations of diffusion and advection network architectures. Unless stated otherwise, all remaining ADR--BINN hyperparameters take the default values given in Table~\ref{tab:BINN parameter}.  }
    \label{fig:high-network}
\end{figure}

\begin{figure}[htbp]
    \centering
    \includegraphics[width=\linewidth]{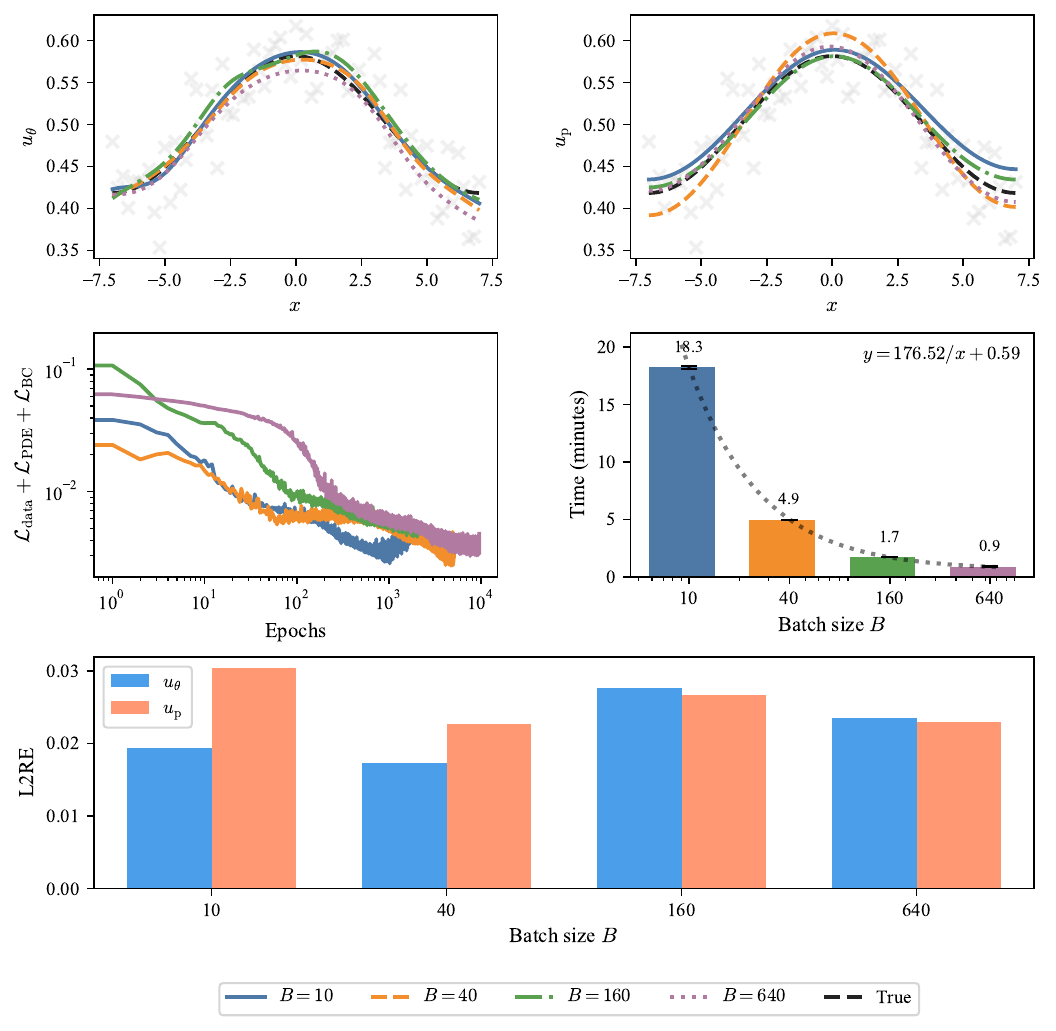}
    \caption{
    Effect of mini-batch size, $B$, on ADR--BINN optimisation and reconstruction accuracy for the diffusion equation (Eq.~\eqref{eq:diff}) with additive Gaussian noise of variance $\sigma^2=0.01$.
    Top: Learned solution $u_\theta(x,t)$ (left) and forward-predicted solution $u_\text{p}(x,t)$ (right) plotted for $t=t_\text{end}$ as specified in Table~\ref{tab:time_resolution}. Middle: Total validation loss during training for $B=\{10,40,160,640\}$ (left). Median wall-clock training time over ten repetitions for $4000$ epochs as a function of batch size; the dashed line shows the fitted inverse relationship (right). Bottom: Relative $L^2$ errors in $u_\theta$ and $u_\text{p}$ after a fixed training budget of five minutes. Unless stated otherwise, $N_{\mathrm{PDE}}=N_{\mathrm{BC}}=40$ and all remaining hyperparameters take the default values given in Table~\ref{tab:BINN parameter}. }
    \label{fig:high-B}
\end{figure}

Figs.~\ref{fig:high-r}--\ref{fig:high-B} repeat the analyses presented in the main text using a larger observation noise variance of $\sigma^2=0.01$. 
As expected, increasing the observation noise reduces the overall accuracy of both the learned solutions and recovered constitutive functions.
Nevertheless, the principal conclusions of the main text remain unchanged. Across all four investigations, intermediate learning rates (Fig.~\ref{fig:high-r}), balanced loss weightings (Fig.~\ref{fig:high-w}), moderately expressive operator networks (Fig.~\ref{fig:high-network}), and intermediate batch sizes (Fig.~\ref{fig:high-B}) continue to provide the most reliable reconstructions. The increased noise primarily manifests as greater variability between independent ADR--BINN initialisations, wider uncertainty bands surrounding the learned constitutive functions, and larger reconstruction errors for both $u_\theta$ and $u_{\mathrm p}$. The validation losses also exhibit increased stochastic fluctuations throughout training, reflecting the greater difficulty of distinguishing the underlying dynamics from noisy observations. Despite these quantitative reductions in performance, the same optimisation and architectural recommendations identified in the main manuscript remain robust under substantially noisier datasets.


\clearpage

\subsection*{Results for densely sampled datasets}

\begin{figure}[htbp]
    \centering
    \includegraphics[width=\linewidth]{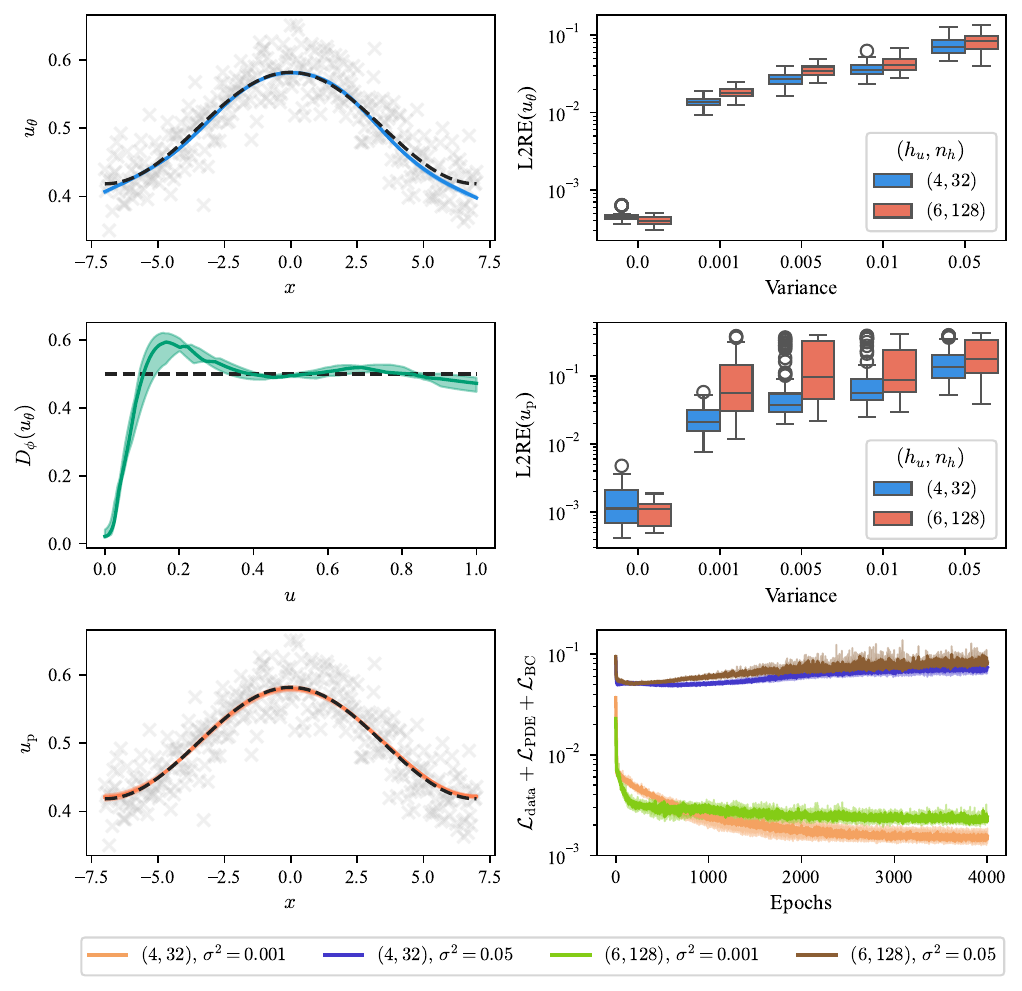}
    \caption{Effect of additive Gaussian noise on BINN performance in the linear diffusion equation (Eq.~\eqref{eq:diff}) with five times as many spatial data points as in Fig.~\ref{fig:noise}. 
    Left: Learned solution $u_\theta(x,t)$ (top), middle left: learned diffusivity $D_\phi(u)$, and bottom left: forward-predicted solution $u_\text{p}(x,t)$ plotted for $t=t_\text{end}$ as specified in Table~\ref{tab:time_resolution} for a representative ADR--BINN trained with additive Gaussian noise of variance $\sigma^2=0.001$ and $(h_u, n_h)=(4,32)$. 
    Dashed black lines denote the true solution or constitutive function, coloured lines denote the mean over ten independent BINN initialisations, and shaded regions indicate $\pm1$ standard deviation.
    Right: Relative $L^2$ errors in the learned solution $u_\theta$ (top) and predicted solution $u_p$ (middle), as a function of the observation noise variance for the architectures $(h_u,n_u)=(4,32)$ and $(6,128)$. Boxplots summarise ten BINN initialisations for each of ten independently generated datasets.
    Bottom right shows the total validation loss during training for representative combinations of architecture and noise levels. Unless stated otherwise, all remaining ADR--BINN hyperparameters take the default values given in Table~\ref{tab:BINN parameter}.}
    \label{fig:SI-noise}
\end{figure}

\begin{figure}[htbp]
    \centering
    \includegraphics[width=\linewidth]{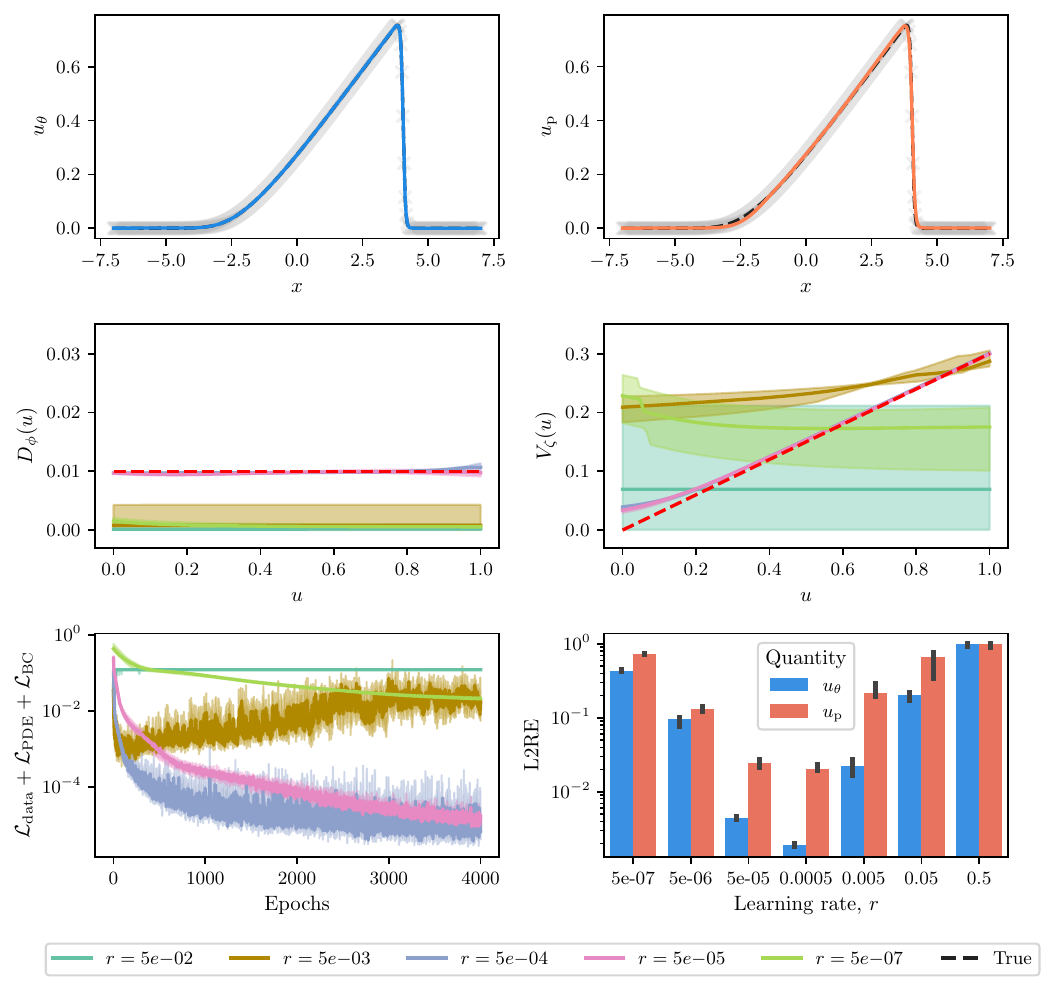}
    \caption{Effect of the learning rate, $r$, on BINNs performance on data simulated from Burgers' equation (Eq.~\eqref{eq:burger}) with no noise and five times as many spatial data points as Fig.~\ref{fig:0-r}. 
    Top: Learned solution $u_\theta(x,t)$ (left) and forward-predicted solution $u_\text{p}(x,t)$ (right) plotted for $t=t_\text{end}$ as specified in Table~\ref{tab:time_resolution}. 
    Middle: Learned diffusivity $D_\phi(u)$ (left) and advection velocity $V_\zeta(u)$ (right).
    Results correspond to the representative learning rate $r=5\times10^{-4}$ (top); dashed black curves denote the true operators and shaded regions indicate $\pm1$ standard deviation over ten independent ADR--BINN initialisations.
    Bottom: Total validation loss (left) and relative $L^2$ errors in $u_\theta$ and $u_p$ (right) for various learning rates, $r$. Unless stated otherwise, all remaining ADR--BINN hyperparameters take the default values given in Table~\ref{tab:BINN parameter}.}
    \label{fig:5-0-r}
\end{figure}

\begin{figure}[htbp]
    \centering
    \includegraphics[width=\linewidth]{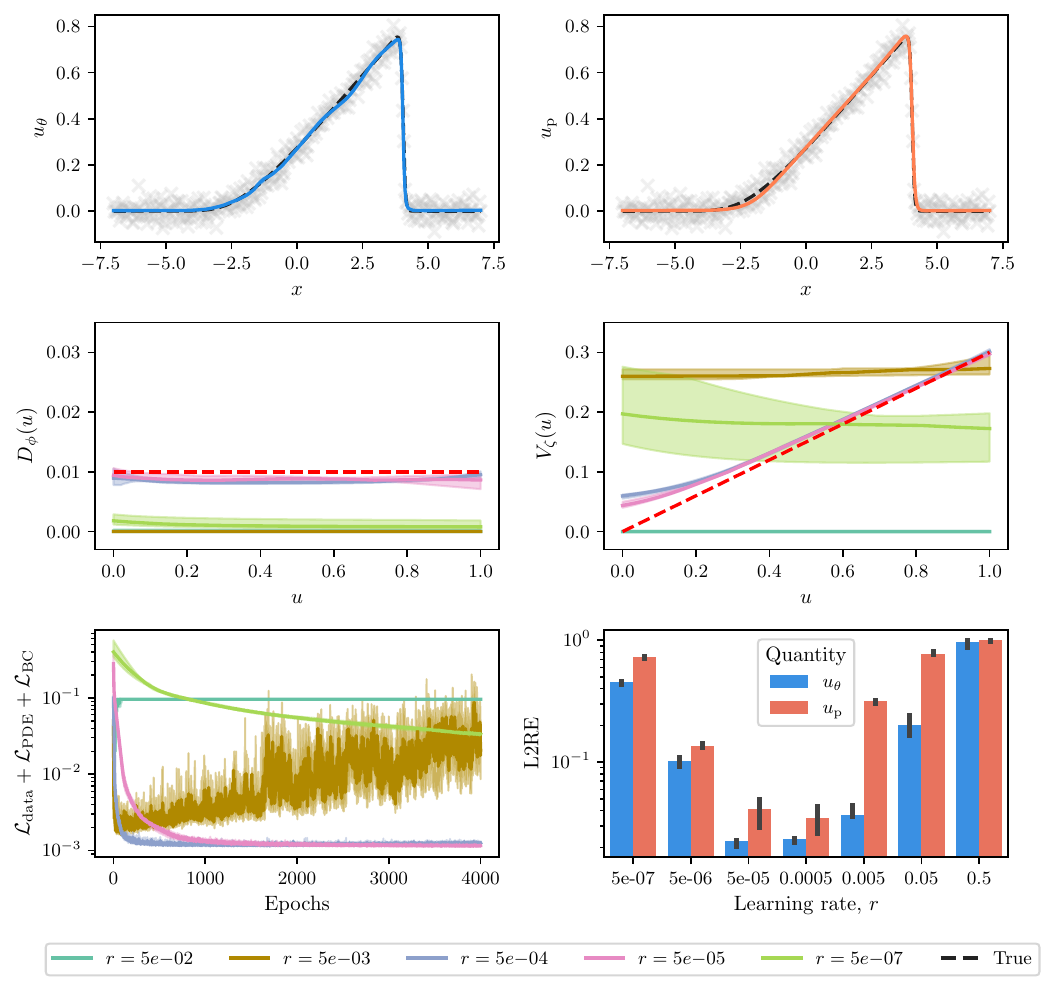}
    \caption{Effect of the learning rate, $r$, on BINNs performance on data simulated from Burgers' equation (Eq.~\eqref{eq:burger}) with additive Gaussian noise of variance $\sigma^2=0.001$ and five times as many data points as Fig.~\ref{fig:learning-rate}. 
    Top: Learned solution $u_\theta(x,t)$ (left) and forward-predicted solution $u_\text{p}(x,t)$ (right) plotted for $t=t_\text{end}$ as specified in Table~\ref{tab:time_resolution}. 
    Middle: Learned diffusivity $D_\phi(u)$ (left) and advection velocity $V_\zeta(u)$ (right).
    Results correspond to the representative learning rate $r=5\times10^{-4}$ (top); dashed black curves denote the true operators and shaded regions indicate $\pm1$ standard deviation over ten independent ADR--BINN initialisations.
    Bottom: Total validation loss (left) and relative $L^2$ errors in $u_\theta$ and $u_p$ (right) for various learning rates, $r$. Unless stated otherwise, all remaining ADR--BINN hyperparameters take the default values given in Table~\ref{tab:BINN parameter}.}
    \label{fig:5-0.001-r}
\end{figure}

\begin{figure}[htbp]
    \centering
    \includegraphics[width=\linewidth]{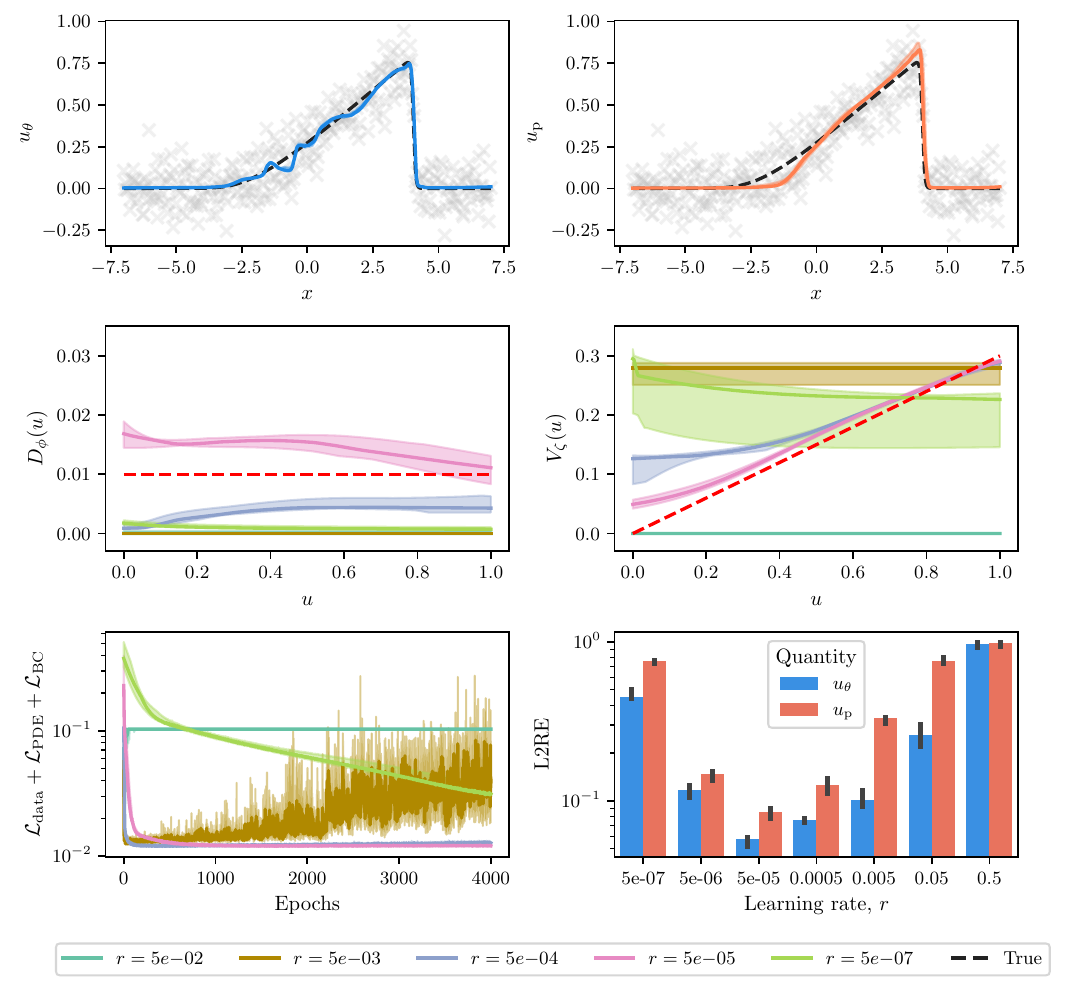}
    \caption{Effect of the learning rate, $r$, on BINNs performance on data simulated from  Burgers' equation (Eq.~\eqref{eq:burger}) with additive Gaussian noise of variance $\sigma^2=0.01$ and five times as many spatial data points as Fig.~\ref{fig:high-r}. 
    Top: Learned solution $u_\theta(x,t)$ (left) and forward-predicted solution $u_\text{p}(x,t)$ (right) plotted for $t=t_\text{end}$ as specified in Table~\ref{tab:time_resolution}. 
    Middle: Learned diffusivity $D_\phi(u)$ (left) and advection velocity $V_\zeta(u)$ (right).
    Results correspond to the representative learning rate $r=5\times10^{-4}$ (top); dashed black curves denote the true operators and shaded regions indicate $\pm1$ standard deviation over ten independent ADR--BINN initialisations.
    Bottom: Total validation loss (left) and relative $L^2$ errors in $u_\theta$ and $u_p$ (right) for various learning rates, $r$. Unless stated otherwise, all remaining ADR--BINN hyperparameters take the default values given in Table~\ref{tab:BINN parameter}.}
    \label{fig:5-0.01-r}
\end{figure}

\begin{figure}[htbp]
    \centering
    \includegraphics[width=\linewidth]{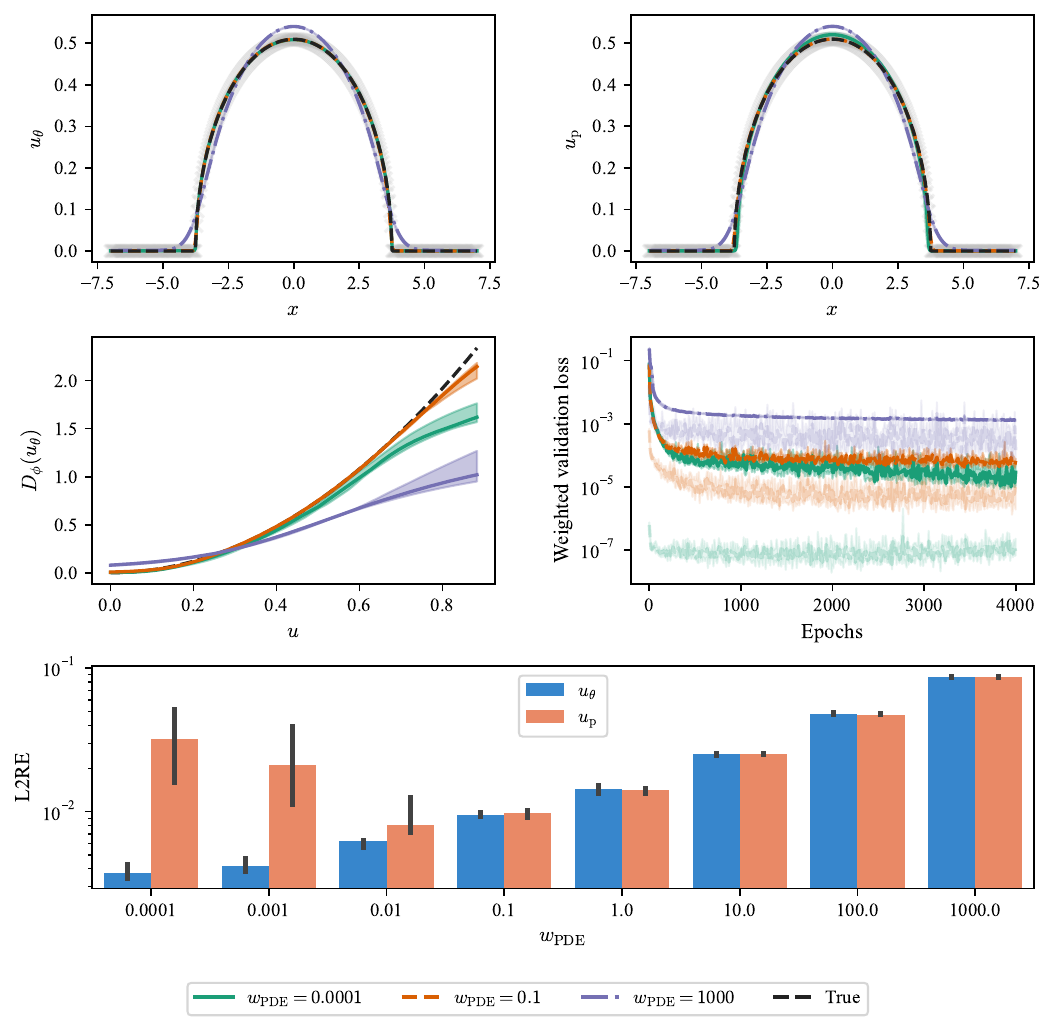}
    \caption{
    Effect of the PDE residual weighting $w_{\mathrm{PDE}}$ on BINN performance for the porous-medium equation (Eq.~\eqref{eq:PM}) with no noise and five times as many spatial data points as Fig.~\ref{fig:0-w}, whilst $w_{\mathrm{data}}=1$ and $w_{\mathrm{BC}}=0$. 
    Top: Learned solution $u_\theta(x,t)$ (left) and forward-predicted solution $u_p(x,t)$ (right) plotted for $t=t_\text{end}$ as specified in Table~\ref{tab:time_resolution}. 
    Middle left: Learned diffusion function $D_\phi(u)$. Representative results are shown for $w_{\mathrm{PDE}}\in\{10^{-4},10^{-1},10^3\}$; dashed black curves denote the true solution or constitutive function and shaded regions represent $\pm1$ standard deviation across ten BINN initialisations. 
    Middle right: Weighted validation losses, showing $L_{\mathrm{data}}$ (solid) and $w_{\mathrm{PDE}}L_{\mathrm{PDE}}$ (dashed) with shaded regions representing $\pm1$ standard deviation across ten BINN initialisations. 
    Bottom: Relative $L^2$ errors in $u_\theta$ and $u_\text{p}$ as $w_{\mathrm{PDE}}$ varies. 
    Unless stated otherwise, all remaining ADR--BINN hyperparameters take the default values given in Table~\ref{tab:BINN parameter}.  
    }
    \label{fig:5-0-w}
\end{figure}

\begin{figure}[htbp]
    \centering
    \includegraphics[width=\linewidth]{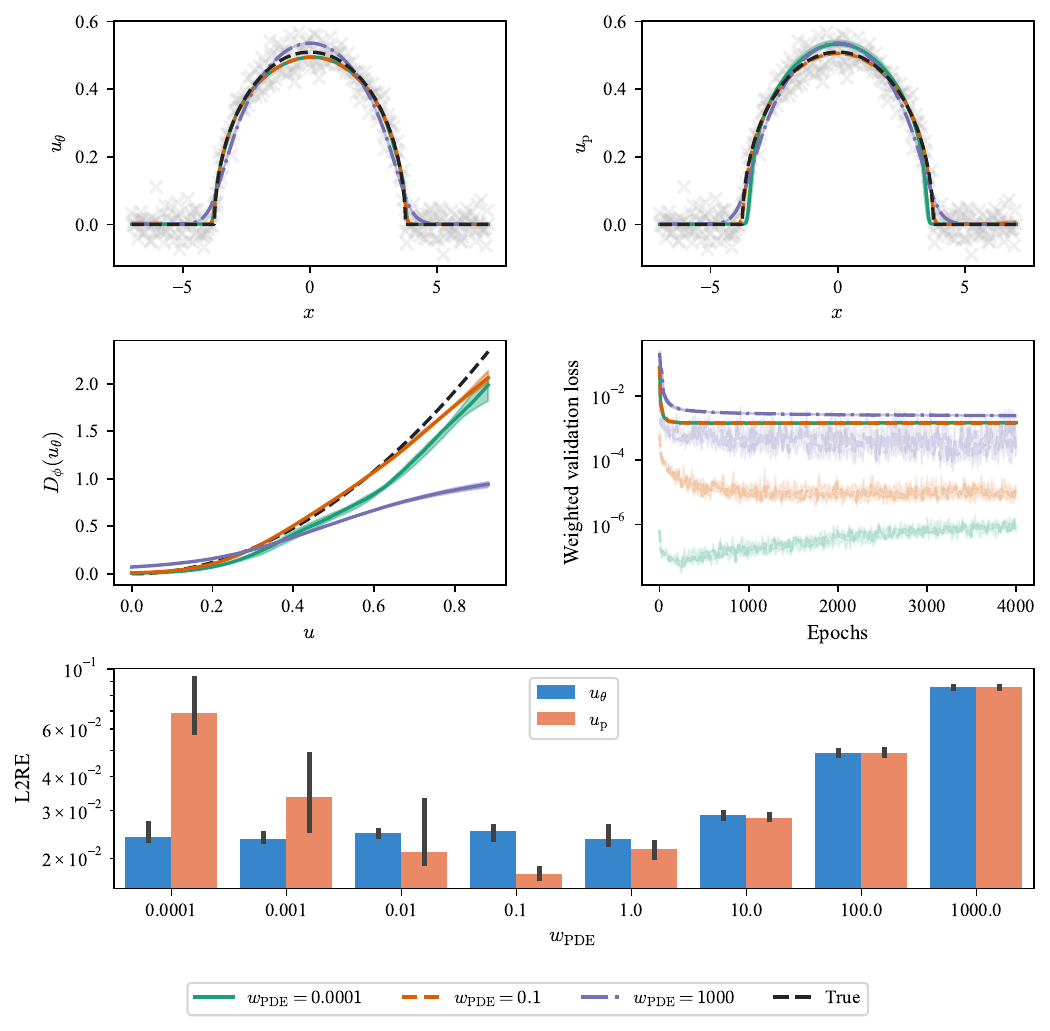}
\caption{
    Effect of the PDE residual weighting $w_{\mathrm{PDE}}$ on BINN performance for the porous-medium equation (Eq.~\eqref{eq:PM}) with additive Gaussian noise of variance $\sigma^2=0.001$ and five times as many spatial data points as Fig.~\ref{fig:PDEweight}, whilst $w_{\mathrm{data}}=1$ and $w_{\mathrm{BC}}=0$. 
    Top: Learned solution $u_\theta(x,t)$ (left) and forward-predicted solution $u_p(x,t)$ (right) plotted for $t=t_\text{end}$ as specified in Table~\ref{tab:time_resolution}. 
    Middle left: Learned diffusion function $D_\phi(u)$. Representative results are shown for $w_{\mathrm{PDE}}\in\{10^{-4},10^{-1},10^3\}$; dashed black curves denote the true solution or constitutive function and shaded regions represent $\pm1$ standard deviation across ten BINN initialisations. 
    Middle right: Weighted validation losses, showing $L_{\mathrm{data}}$ (solid) and $w_{\mathrm{PDE}}L_{\mathrm{PDE}}$ (dashed) with shaded regions representing $\pm1$ standard deviation across ten BINN initialisations. 
    Bottom: Relative $L^2$ errors in $u_\theta$ and $u_\text{p}$ as $w_{\mathrm{PDE}}$ varies. 
    Unless stated otherwise, all remaining ADR--BINN hyperparameters take the default values given in Table~\ref{tab:BINN parameter}.  
    }
    \label{fig:5-0.001-w}
\end{figure}

\begin{figure}[htbp]
    \centering
    \includegraphics[width=\linewidth]{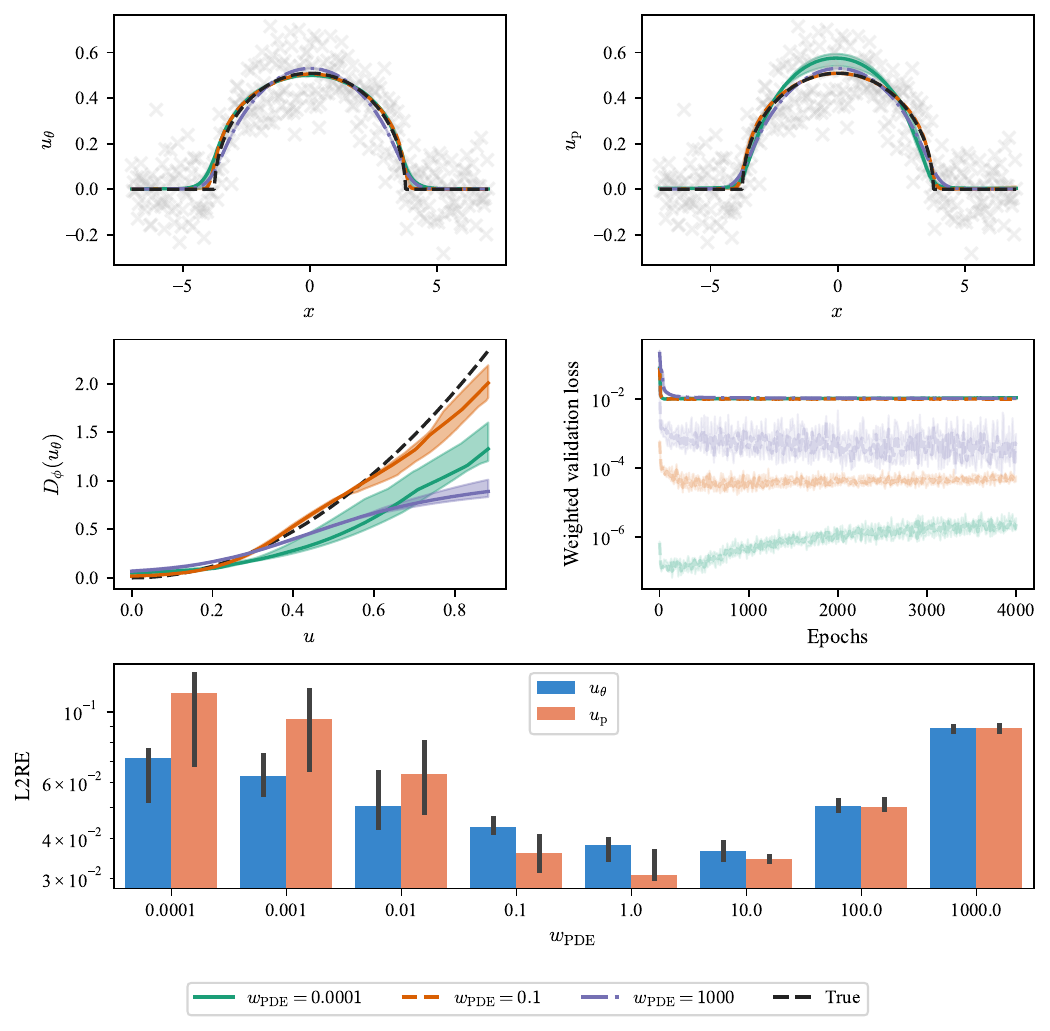}
    \caption{
    Effect of the PDE residual weighting $w_{\mathrm{PDE}}$ on BINN performance for the porous-medium equation (Eq.~\eqref{eq:PM}) with additive Gaussian noise of variance $\sigma^2=0.01$ and five times as many spatial data points as Fig.~\ref{fig:high-w}, whilst $w_{\mathrm{data}}=1$ and $w_{\mathrm{BC}}=0$. 
    Top: Learned solution $u_\theta(x,t)$ (left) and forward-predicted solution $u_p(x,t)$ (right) plotted for $t=t_\text{end}$ as specified in Table~\ref{tab:time_resolution}. 
    Middle left: Learned diffusion function $D_\phi(u)$. Representative results are shown for $w_{\mathrm{PDE}}\in\{10^{-4},10^{-1},10^3\}$; dashed black curves denote the true solution or constitutive function and shaded regions represent $\pm1$ standard deviation across ten BINN initialisations. 
    Middle right: Weighted validation losses, showing $L_{\mathrm{data}}$ (solid) and $w_{\mathrm{PDE}}L_{\mathrm{PDE}}$ (dashed) with shaded regions representing $\pm1$ standard deviation across ten BINN initialisations. 
    Bottom: Relative $L^2$ errors in $u_\theta$ and $u_\text{p}$ as $w_{\mathrm{PDE}}$ varies. 
    Unless stated otherwise, all remaining ADR--BINN hyperparameters take the default values given in Table~\ref{tab:BINN parameter}.  
    }
    \label{fig:5-0.01-w}
\end{figure}

\begin{figure}[htbp]
    \centering
    \includegraphics[width=\linewidth]{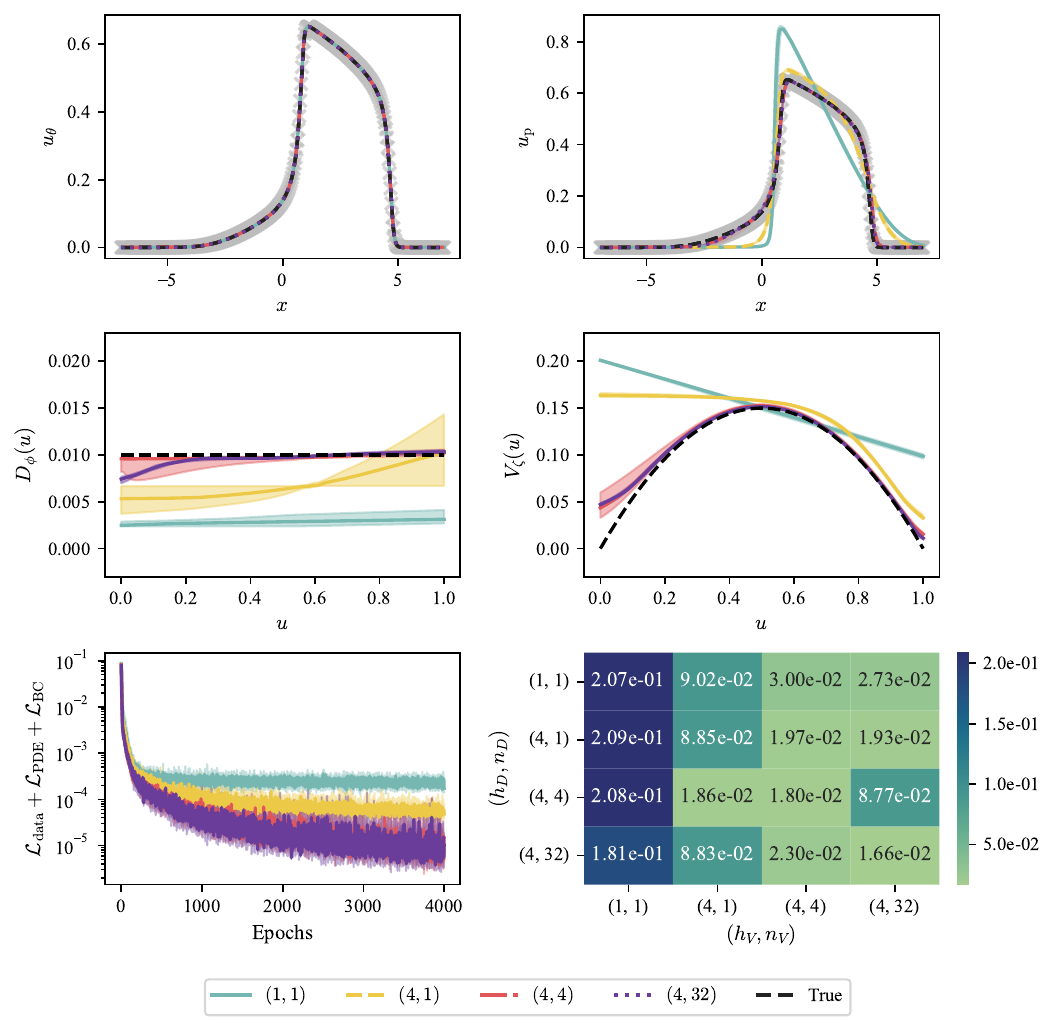}
    \caption{
    Effect of operator network architecture on ADR--BINN performance for the linear diffusion--nonlinear advection equation (Eq.~\eqref{eq:DA}) with no noise and five times as many spatial data points as Fig.~\ref{fig:0-network}. Top: Learned solution $u_\theta(x,t)$ (left) and forward-predicted solution $u_\text{p}(x,t)$ (right) plotted for $t=t_\text{end}$ as specified in Table~\ref{tab:time_resolution}. 
    Middle: Learned constitutive functions $D_\phi(u)$ (left) and $V_\zeta(u)$ (right). Results are shown for operator network architectures $(h_{D,V},n_{D,V})\in\{(1,1),(4,1),(4,4),(4,32)\}$; dashed black curves denote the true solution or operators, and shaded regions indicate $\pm1$ standard deviation over ten independent ADR--BINN initialisations. Note that when there is only one hidden layer, an \texttt{Identity} activation function is used in the input layer and a \texttt{Linear} activation function, where $\texttt{Linear}(x;W,b)= xW^{\top}+b$,
     where $W$ is the weight matrix and $b$ is the bias vector, is used for the output layer.
     Bottom: Total validation loss during training for each operator network architecture (left). Mean relative $L^2$ error in the forward-predicted solution $u_\text{p}$ for all sixteen combinations of diffusion and advection network architectures. Unless stated otherwise, all remaining ADR--BINN hyperparameters take the default values given in Table~\ref{tab:BINN parameter}.  }
    \label{fig:5-0-network}
\end{figure}

\begin{figure}[htbp]
    \centering
    \includegraphics[width=\linewidth]{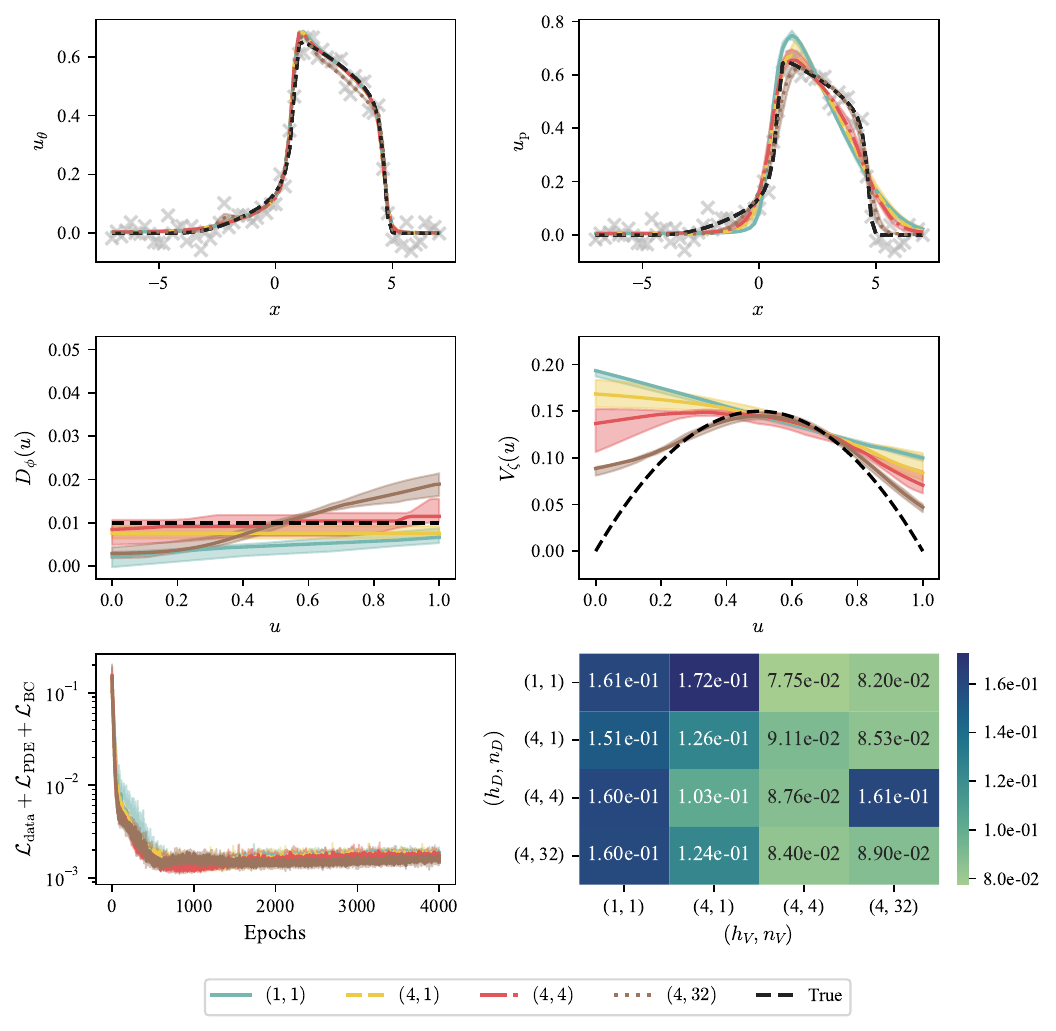}
    \caption{
    Effect of operator network architecture on ADR--BINN performance for the linear diffusion--nonlinear advection equation (Eq.~\eqref{eq:DA}) with additive Gaussian noise of variance $\sigma^2=0.001$ and five times as many spatial data points as Fig.~\ref{fig:Networks}. Top: Learned solution $u_\theta(x,t)$ (left) and forward-predicted solution $u_\text{p}(x,t)$ (right) plotted for $t=t_\text{end}$ as specified in Table~\ref{tab:time_resolution}. 
    Middle: Learned constitutive functions $D_\phi(u)$ (left) and $V_\zeta(u)$ (right). Results are shown for operator network architectures $(h_{D,V},n_{D,V})\in\{(1,1),(4,1),(4,4),(4,32)\}$; dashed black curves denote the true solution or operators, and shaded regions indicate $\pm1$ standard deviation over ten independent ADR--BINN initialisations. Note that when there is only one hidden layer, an \texttt{Identity} activation function is used in the input layer and a \texttt{Linear} activation function, where $\texttt{Linear}(x;W,b)= xW^{\top}+b$,
     where $W$ is the weight matrix and $b$ is the bias vector, is used for the output layer.
     Bottom: Total validation loss during training for each operator network architecture (left). Mean relative $L^2$ error in the forward-predicted solution $u_\text{p}$ for all sixteen combinations of diffusion and advection network architectures. Unless stated otherwise, all remaining ADR--BINN hyperparameters take the default values given in Table~\ref{tab:BINN parameter}.  }
    \label{fig:5-0.001-network}
\end{figure}

\begin{figure}[htbp]
    \centering
    \includegraphics[width=\linewidth]{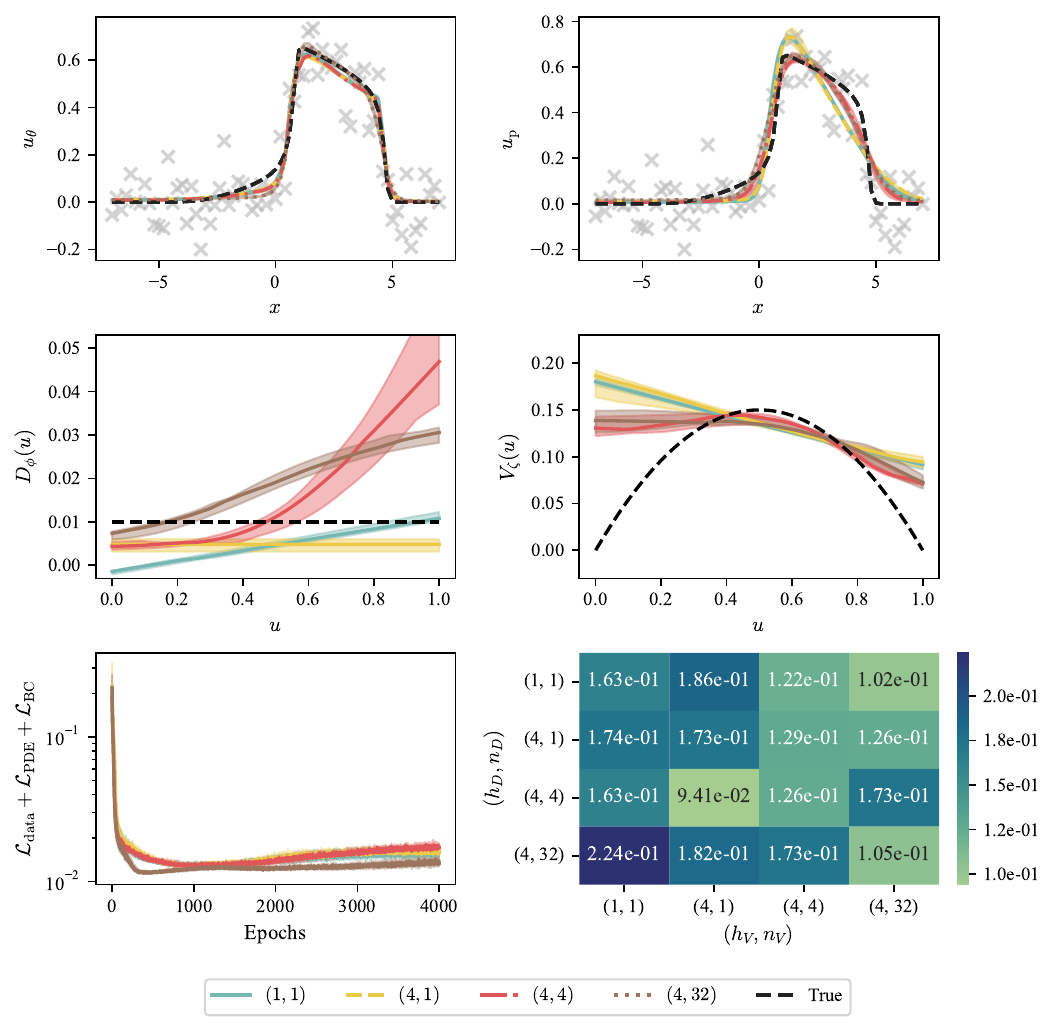}
    \caption{
    Effect of operator network architecture on ADR--BINN performance for the linear diffusion--nonlinear advection equation (Eq.~\eqref{eq:DA}) with additive Gaussian noise of variance $\sigma^2=0.01$ and five times as many spatial data points as Fig.~\ref{fig:high-network}. Top: Learned solution $u_\theta(x,t)$ (left) and forward-predicted solution $u_\text{p}(x,t)$ (right) plotted for $t=t_\text{end}$ as specified in Table~\ref{tab:time_resolution}. 
    Middle: Learned constitutive functions $D_\phi(u)$ (left) and $V_\zeta(u)$ (right). Results are shown for operator network architectures $(h_{D,V},n_{D,V})\in\{(1,1),(4,1),(4,4),(4,32)\}$; dashed black curves denote the true solution or operators, and shaded regions indicate $\pm1$ standard deviation over ten independent ADR--BINN initialisations. Note that when there is only one hidden layer, an \texttt{Identity} activation function is used in the input layer and a \texttt{Linear} activation function, where $\texttt{Linear}(x;W,b)= xW^{\top}+b$,
     where $W$ is the weight matrix and $b$ is the bias vector, is used for the output layer.
     Bottom: Total validation loss during training for each operator network architecture (left). Mean relative $L^2$ error in the forward-predicted solution $u_\text{p}$ for all sixteen combinations of diffusion and advection network architectures. Unless stated otherwise, all remaining ADR--BINN hyperparameters take the default values given in Table~\ref{tab:BINN parameter}.  }
    \label{fig:5-0.01-network}
\end{figure}

\begin{figure}[htbp]
    \centering
    \includegraphics[width=\linewidth]{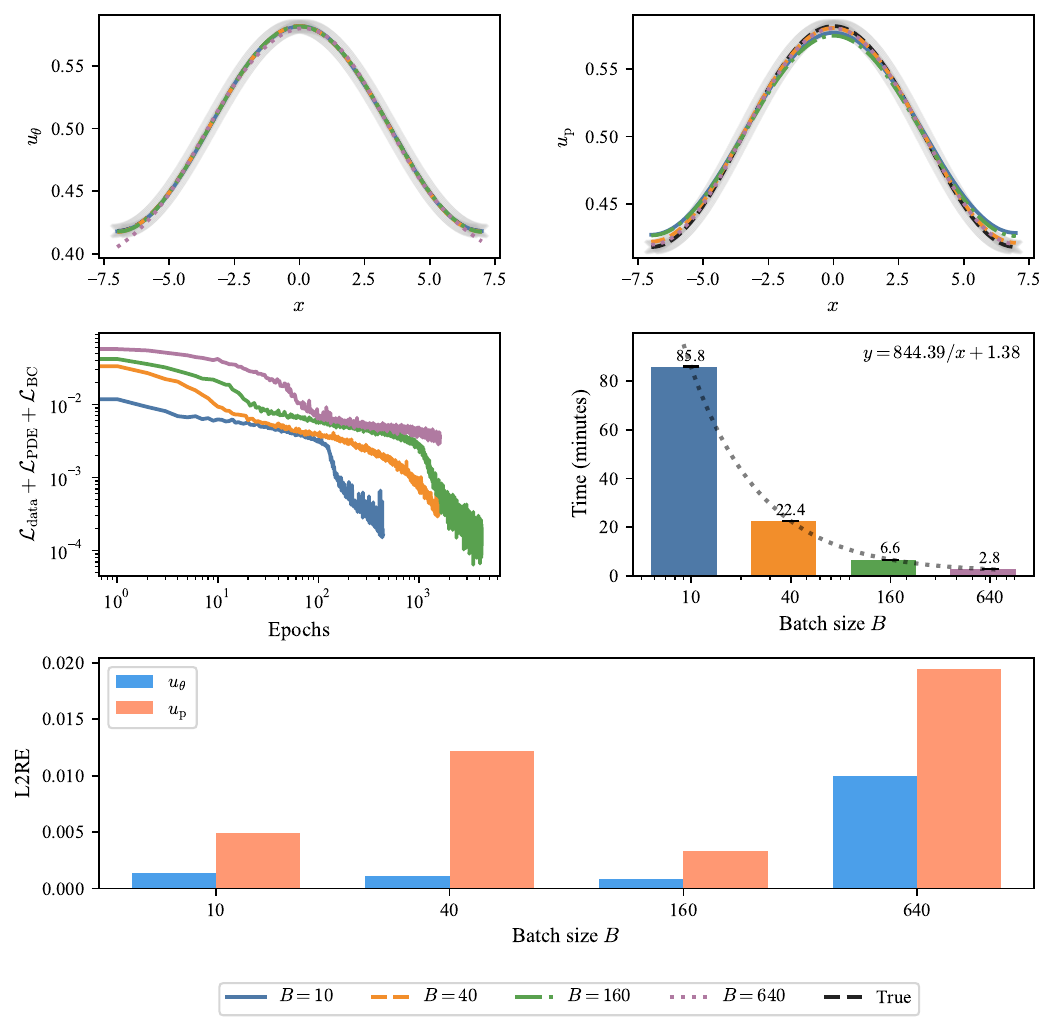}
    \caption{
    Effect of mini-batch size, $B$, on ADR--BINN optimisation and reconstruction accuracy for the diffusion equation (Eq.\eqref{eq:diff}) with no noise and five times as many spatial data points as in Fig.~\ref{fig:0-B}.
    Top: Learned solution $u_\theta(x,t)$ (left) and forward-predicted solution $u_\text{p}(x,t)$ (right) plotted for $t=t_\text{end}$ as specified in Table~\ref{tab:time_resolution}. Middle: Total validation loss during training for $B=\{10,40,160,640\}$ (left). Median wall-clock training time over ten repetitions for $4000$ epochs as a function of batch size; the dashed line shows the fitted inverse relationship (right). Bottom: Relative $L^2$ errors in $u_\theta$ and $u_\text{p}$ after a fixed training budget of five minutes. Unless stated otherwise, $N_{\mathrm{PDE}}=N_{\mathrm{BC}}=40$ and all remaining hyperparameters take the default values given in Table~\ref{tab:BINN parameter}. }
    \label{fig:5-0-B}
\end{figure}

\begin{figure}[htbp]
    \centering
    \includegraphics[width=\linewidth]{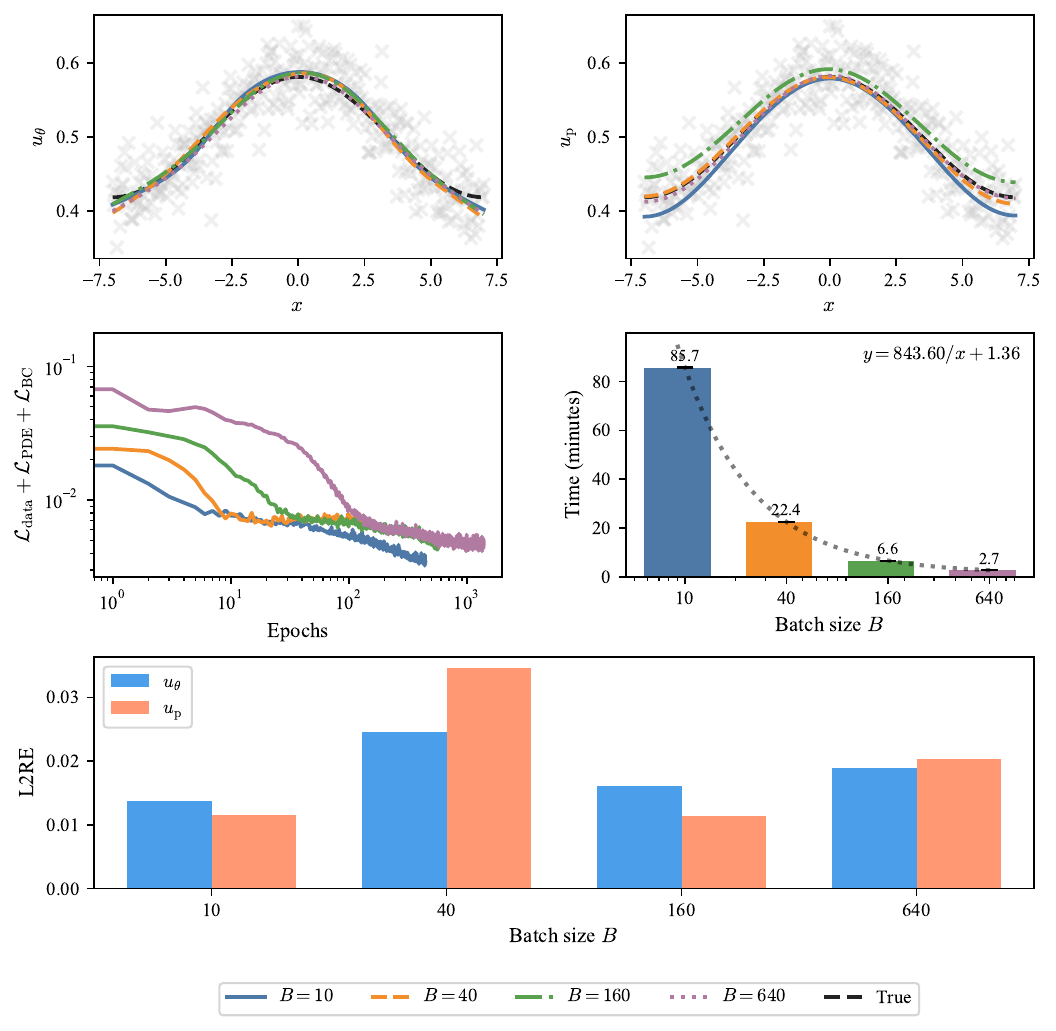}
    \caption{Effect of mini-batch size, $B$, on ADR--BINN optimisation and reconstruction accuracy for the diffusion equation (Eq.\eqref{eq:diff}) with additive Gaussian noise with variance $\sigma^2=0.001$ and five times as many spatial data points as in Fig.~\ref{fig:batch}.
    Top: Learned solution $u_\theta(x,t)$ (left) and forward-predicted solution $u_\text{p}(x,t)$ (right) plotted for $t=t_\text{end}$ as specified in Table~\ref{tab:time_resolution}. Middle: Total validation loss during training for $B=\{10,40,160,640\}$ (left). Median wall-clock training time over ten repetitions for $4000$ epochs as a function of batch size; the dashed line shows the fitted inverse relationship (right). Bottom: Relative $L^2$ errors in $u_\theta$ and $u_\text{p}$ after a fixed training budget of five minutes. Unless stated otherwise, $N_{\mathrm{PDE}}=N_{\mathrm{BC}}=40$ and all remaining hyperparameters take the default values given in Table~\ref{tab:BINN parameter}. }
    \label{fig:5-0.001-B}
\end{figure}

\begin{figure}[htbp]
    \centering
    \includegraphics[width=\linewidth]{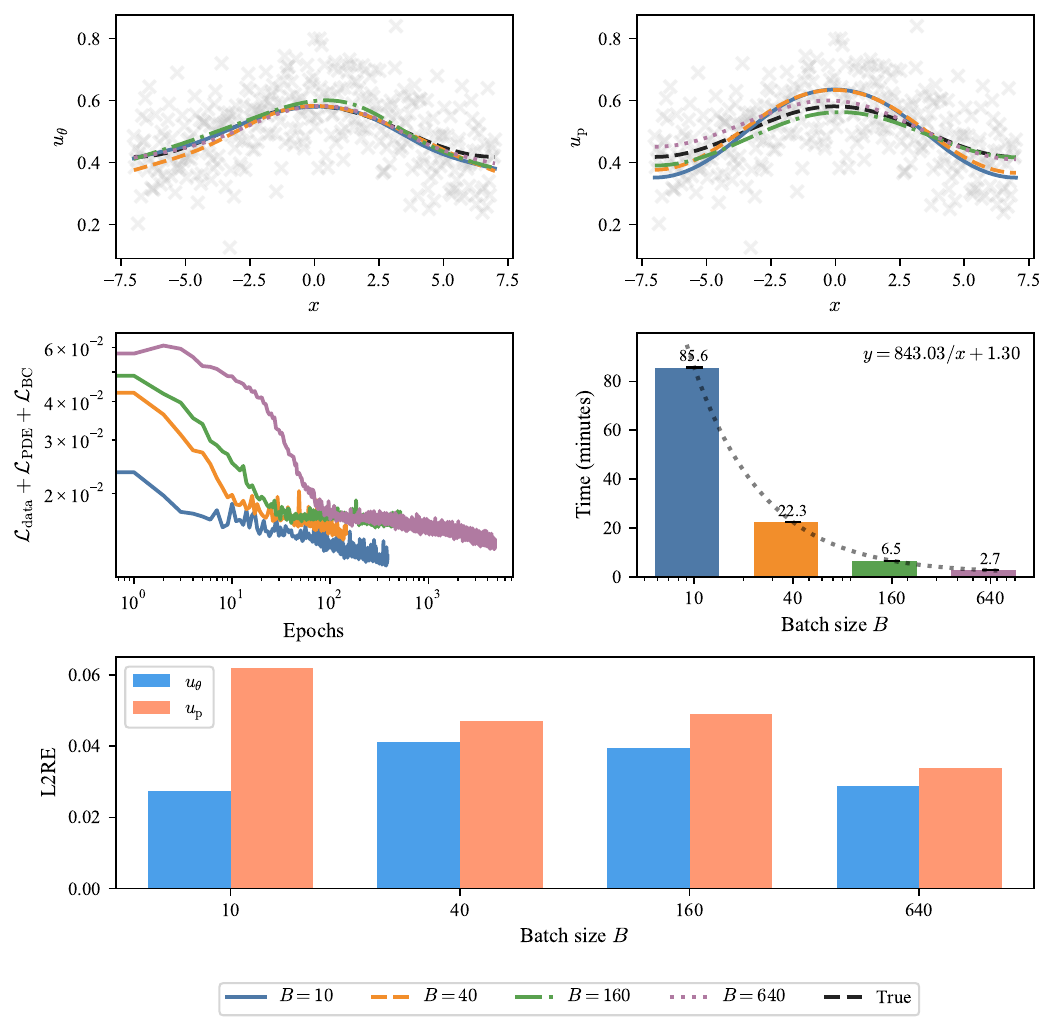}
    \caption{Effect of mini-batch size, $B$, on ADR--BINN optimisation and reconstruction accuracy for the diffusion equation (Eq.\eqref{eq:diff}) with additive Gaussian noise with variance $\sigma^2=0.01$ and five times as many spatial data points as in Fig.~\ref{fig:high-B}.
    Top: Learned solution $u_\theta(x,t)$ (left) and forward-predicted solution $u_\text{p}(x,t)$ (right) plotted for $t=t_\text{end}$ as specified in Table~\ref{tab:time_resolution}. Middle: Total validation loss during training for $B=\{10,40,160,640\}$ (left). Median wall-clock training time over ten repetitions for $4000$ epochs as a function of batch size; the dashed line shows the fitted inverse relationship (right). Bottom: Relative $L^2$ errors in $u_\theta$ and $u_\text{p}$ after a fixed training budget of five minutes. Unless stated otherwise, $N_{\mathrm{PDE}}=N_{\mathrm{BC}}=40$ and all remaining hyperparameters take the default values given in Table~\ref{tab:BINN parameter}. }
    \label{fig:5-0.01-B}
\end{figure}

Figs.~\ref{fig:SI-noise}--\ref{fig:5-0.01-B} repeat all experiments presented in the main manuscript using five times as many spatial observations. 
Increasing the sampling density consistently improves the recovery of both the governing operators and the underlying solution dynamics. Relative errors in the learned solution, $u_\theta$, and the forward-predicted solution, $u_{\mathrm p}$, decrease across almost all experiments, while the recovered constitutive functions exhibit reduced variability between independent ADR--BINN initialisations.

Importantly, increasing the quantity of data does not alter the qualitative conclusions drawn in the main manuscript. Intermediate learning rates remain optimal across all noise levels (Figs.~\ref{fig:5-0-r}--\ref{fig:5-0.01-r}), balanced weighting of the PDE and data-fitting losses continues to be required for accurate operator recovery (Figs.~\ref{fig:5-0-w}--\ref{fig:5-0.01-w}), and moderately expressive operator networks remain sufficient once they possess adequate representational capacity (Figs.~\ref{fig:5-0-network}--\ref{fig:5-0.01-network}). Similarly, intermediate mini-batch sizes continue to provide the best compromise between optimisation efficiency and computational cost (Figs.~\ref{fig:5-0-B}--\ref{fig:5-0.01-B}). The primary effect of the additional observations is therefore to reduce uncertainty in the learned constitutive functions, rather than to change the optimal architectural or optimisation choices. These experiments reinforce the conclusion that BINN performance depends more strongly on the informativeness of the available data than on simply increasing model complexity.


\end{document}